\documentclass[aps,prd,twocolumn,10pt]{revtex4-1}

\usepackage[T1]{fontenc}
\usepackage{amsmath}
\usepackage{amssymb}

\usepackage{hyperref}
\usepackage{multirow}

\begin{document}

\title{Improved predictions for magnetic moments and M1 decay widths of
heavy hadrons}

\author{V.~\v{S}imonis}

\email{vytautas.simonis@tfai.vu.lt}

\affiliation{ Vilnius University, Institute of Theoretical Physics
and Astronomy, Saul\.{e}tekio 3, LT-10257,\ Vilnius, Lithuania }

\date{\today}
\begin{abstract}
In the framework of an extended bag model the magnetic moments, M1
transition moments, and decay widths of all ground-state heavy hadrons
are calculated. For the heavy baryons containing three quarks of different
flavors the effect of hyperfine mixing of the states is taken into
account. The additional care is taken to get more accurate theoretical
estimates for the mass splittings of heavy hadrons. The use of such
improved values enables one to provide more accurate predictions for
the decay widths. These values of the hyperfine splittings between
baryons may be also useful for the further experimental searches of
new heavy hadrons. For instance, we predict $M(\Xi_{cc}^{*})=3695\pm5$ MeV.
The agreement of our results for the M1 decay rates with available
experimental data is good. We also present a wide comparison of the
predictions obtained in our work with the results obtained using various
other approaches. 
\end{abstract}

\maketitle

\section{Introduction} \label{sec_int}

The heavy hadrons are the permanent subject of interest from both
experimental and theoretical sides \cite{KR10,BeA11a,C15,CCLLZ17}.
A typical example could be recently discovered by the LHCb Collaboration
the doubly charmed baryon $\Xi_{\thinspace cc}^{++}$ \cite{AeA17b}
with the mass near 3621 MeV. From the experimental side, the long-standing
puzzle of the $\Xi_{cc}$ was at last solved. On the other side, the
utility of various theoretical approaches has been proven. The $\Xi_{\thinspace cc}^{++}$ was
found almost exactly where it was predicted to reside. The observed
value agrees well with the quark model prediction $3.61$ GeV \cite{KKP94}
and more recent estimate by Karliner and Rosner $M(\Xi_{cc})=3627\pm12$ MeV
\cite{KR14}. The calculations in the framework of the nonrelativistic
model with different potentials \cite{S96d,S96e} provide values in
the range from 3607 to 3631 MeV. The estimate obtained using the relativistic
approach in the quark-diquark approximation \cite{EFGM02} is 3620 MeV. Similar results were also obtained in the recent lattice QCD
calculations: $3603\pm31$ MeV \cite{NeA13}, $3610\pm45$ MeV \cite{BDMO14},
and $3606\pm19$ MeV \cite{AK17}.

Many heavy hadrons are still waiting to be discovered. Moreover, the
understanding of the properties of yet discovered hadrons is not complete.
Therefore the study of various properties of hadrons is important
task. Recently we have studied the magnetic properties (magnetic moments,
M1 decay rates) of ground-state mesons using the modified version
of the bag model \cite{S16a}. Now we extend our investigation to
include the magnetic properties of heavy baryons. It could be one
more step towards the comprehensive analysis of the magnetic properties
of all hadrons. We expect our approach could be suitable for this
purpose because the bag model treats the mesons and baryons on equally
the same footing. There are many other attempts devoted to the unified
treatment of mesons and baryons using various approaches \cite{L78a,R81b,BCN81,BSS02,CI86,ECAKR09,RCCR11}.
Lattice QCD and QCD sum rules can also serve as appropriate framework.

In the baryon sector several additional difficulties emerge, such
as the state mixing problem of the baryons ($\Xi_{c}$, $\Xi_{c}^{\prime}$),
($\Xi_{b}$, $\Xi_{b}^{\prime}$), ($\Xi_{cb}$, $\Xi_{cb}^{\prime}$),
and ($\Omega_{cb}$, $\Omega_{cb}^{\prime}$). Another problem is
possibly insufficient accuracy of the available theoretical predictions
for the hyperfine splittings of doubly and triply heavy baryons. To
deal with these problems is the object of the present work, too.

There is a vast number of publications devoted to the theoretical
treatment of the magnetic properties of various hadrons. The magnetic moments of the mesons
were studied in Refs.~\cite{S16a,S03a,AOS09,CGNSS95a,MS02,J03,CJ04,ddMF14,MddF15,dT17,SD17a,HP99,BM08,RBGRW11,PSRRSW13,BS13d,DEGM14,AW97,LMW08,HKLLWZ07,OKLMM15a,SDE15,LST17,GS15a,BS80,LCD15,CBCT15,BS14,KPT18},
the M1 decays of the mesons in Refs.~\cite{S16a,ABL78,HI82,KRQ87,ST87,KX92,R13,GYS14,CDN93,FR94,JMS95,OX94,CGNSS95b,J96b,J99,CJ99,C07,CJ09,HW07,HDDT78,IDS82,WMM85,SM86a,MS88,SM89,ZCT93,OH99,EGKLY80a,EQ94,EGMR08,BJ82,GOS84,ZSG91,GKLT95a,F99d,BSG02,BEH03,GVGV03,CS05,LS07,PPV10,BVM11,MBV17a,CYC12,SOEF16,DLGZ17a,DLGZ17b,ASMA17,ER93,IV95,GR01b,EFG02b,EFG03a,GI85,G04,BGS05,GM15,GM16,GMS16,RR07,RRS09,GRR17,DR11,DR12,DR13,DKR14,DS87,FM93,CDN94,A94a,DFS00,CK01c,BD94,JPT02,JMPS10a,JMMP10,SPV14,SPV16a,SPV16b,PSDR15,PDKPB16,PDKPB17,MBV17b,BGS17,DHJ93,MRMP95,KRMMP00,IKR99,ASV05,KWLC10,LNR00,L03b,BeA04b,BJV06,LPW12,P12,PS13,S80d,K80a,A84,AIP94b,ADIP96,EK85,BR85b,DN96b,ZHY97b,V08,W13b,ADMNS07,DET09,CeA11b,BH11,BS13c,DeA12a,DDKL14,OKLMM15b,LW16,CCLLYY93,CCLLYY94a,CDDGFN97,DDGNP98,NW00,LZ11,CH14,BBHMR93,IT95,DCD14},
the magnetic moments of heavy baryons in Refs.~\cite{S96d,S96e,BS80,FLNC81,DL96,AHNV07b,KDV05,DV09,DKV13,MPV08,PRV08a,PMV09,GRQR14,STR16,STRV16b,SR17a,MTV16,TMV16,JR04,FGIKLNP06,GSP16,BD83a,JR86,JB95,SDCG10,OMRS91,SW04,S94a,LMLZ17a,MLLZ17,ZHY97a,AOS02b,AAO08,AAO09a,ABS15a,CEIOT13,CEIOT14b,CEOT15,BS13a},
and the M1 decay properties of heavy baryons in Refs.~\cite{STRV16b,HDDT78,IDS82,CCLLYY93,CCLLYY94a,C97e,HM06b,BPS00,JCZ15,LMLZ18,TKO01,XWLZZ17,WYZZ17,DDSV94,MPV09,AHN10b,IKLR99b,BFGIKLO10,LWXZ17,AOS02a,AAO09b,AAS14b,AAS15b,ABS16,ASZ12,ZD99,W10a,W10b,BCEO15,BCEOT17,CCCLZ18,BS13b}.
References are representative, not necessarily exhaustive. It is interesting
to perform a detailed comparison between results obtained using various
approaches. Such comparison yields some information which can help
us to get a comprehensive picture of the contemporary capabilities
to predict the magnetic properties of the hadrons.

The plan of the paper is the following one. In the next section 
the method used for our investigation is presented, and the expressions
needed for the calculation of the magnetic observables of the hadrons
are given. The details of the derivations are skipped. In Sec.~\ref{sec_mix}
the problem of the state mixing is discussed. In Sec.~\ref{sec_hyp}
the method to improve theoretical predictions for hyperfine splittings
of doubly (and triply) heavy baryons is proposed, and the new estimates
are obtained. Subsequently these values are used in the calculations
of the decay widths of heavy baryons. In Sec.~\ref{sec_mes}
the model parameter setting procedure is discussed, and the magnetic
properties of the mesons are analyzed. Sec.~\ref{sec_bar-1} is devoted
to the study of the magnetic properties of heavy baryons. Magnetic moments,
magnetic dipole transition moments, and M1 decay widths of these baryons are calculated. 
Our predictions are compared with the results obtained using various other approaches
and with available experimental data. Finally, Sec.~\ref{sec_sum} contains
the concluding discussion.

\section{Formalism: Bag model and magnetic properties of the hadrons} \label{sec_bag}

The model used for the present investigation is the same as in our
previous work \cite{S16a} except for the specific values of the scale
factors used for the parameterization of the quark magnetic moments.
It is a version of the MIT bag model designed to reconcile the initial
model with the heavy quark spectroscopy \cite{BS04b}. In the bag
model \cite{CJJTW74,CJJT74,DJJK75,T84b} the hadrons are considered
as extended objects localized in space. The internal structure of
a particle is associated with quark and gluon fields. The model possesses
many desirable features inspired by QCD and relativity. For practical
calculations it is convenient to use the static spherical cavity approximation,
in which the bag surface is frozen. The motion of the quarks inside
the cavity of radius $R$ is described by the free Dirac equation.
The wave function of the quark in the $s$-mode is written in the
form

\begin{equation}
\Psi_{\ m}^{1/2}(r)=\frac{1}{\sqrt{4\pi}}\left(\begin{array}{c}
G(r)\\
-i(\mathbf{\sigma}\cdot\mathbf{\hat{r})}F(r)
\end{array}\right)\Phi_{\ m}^{1/2} ,\label{mbag 001}
\end{equation}
where $\mathbf{\sigma}$ are Pauli matrices, $\mathbf{\hat{r}}$ is
the unit radius-vector, $\Phi_{\ m}^{1/2}$ is usual two-component
spinor. $G(r)$ and $F(r)$ are radial parts of the upper and lower
components respectively. The eigenenergy of the quark is determined
by the boundary condition $G(R)=-F(R).$

The wave functions of the ground-state mesons are characterized by
the total spin $J$. In the baryon sector the situation is
more complex. The baryon wave function can be constructed by coupling
the spins of the first two quarks to the intermediate spin $S$ and
then adding the spin of the third quark to form the total spin $J$.

\begin{align}
\left|B_{[q_{1}q_{2}]q_{3}}\right\rangle  & =\left|(q_{1}q_{2})^{S=0}q_{3},J=\frac{1}{2}\right\rangle ,\label{mbag 003a}\\
\left|B_{\{q_{1}q_{2}\}q_{3}}\right\rangle  & =\left|(q_{1}q_{2})^{S=1}q_{3},\,J=\frac{1}{2}\right\rangle ,\label{mbag 003b}\\
\left|B^{*}\right\rangle  & =\left|(q_{1}q_{2})^{S=1}q_{3},J=\frac{3}{2}\right\rangle .\label{mbag 003c}
\end{align}

The energy  associated with a specific hadron is

\begin{equation}
E=\frac{4\pi}{3}BR^{3}+\sum\limits _{i}\varepsilon_{i}+E_{\mathrm{int}} .\label{mbag 004}
\end{equation}

The bag radius $R_{H}$ of the particular hadron is determined from
the equilibrium condition $\frac{\partial E_{H}}{R}=0$. The three
terms on the right-hand side are the so called volume energy (parameter
$B$ is the bag constant), the sum of single-particle eigenenergies,
and the quark interaction energy due to one-gluon-exchange $E_{\mathrm{int}}=E^{e}+E^{m}$.
The $E^{e}$ and $E^{m}$are the color-electric and color-magnetic
pieces of the interaction energy

\begin{equation}
E^{\mathrm{e}}=\lambda\alpha_{c}(R)\sum\limits _{j>i}I_{ij}(R),\label{mbag 005}
\end{equation}

\begin{equation}
E^{\mathrm{m}}=\alpha_{c}(R)\sum\limits _{j>i}a_{ij}K_{ij}(R),\label{mbag 006}
\end{equation}
where 
\begin{equation}
\lambda=\left\{ \begin{array}{rl}
-1 & \quad\textrm{for baryons,}\\
-2 & \quad\textrm{for mesons.}
\end{array}\right.\label{mbag 007}
\end{equation}

Parameters $a_{ij}$ that specify the color-magnetic energy for
mesons with the total spin $J$ are 
\begin{equation}
a_{ij}=\left\{ \begin{array}{rl}
-6 & \quad(J=0),\\
2 & \quad(J=1).
\end{array}\right.\label{mbag 008}
\end{equation}

The corresponding parameters for the color-magnetic energy of baryons
are presented in Table~\ref{t2.1}. Note that in the case of $J=1/2$
baryons containing three differently flavored quarks the off-diagonal
(with respect to intermediate spin $S$) matrix elements are present.
Therefore, in general case, the physical states $\left|B\right\rangle $
and $\left|B\,'\right\rangle $ are the linear combinations of
the states $\left|B[q_{1}q_{2}]q_{3}\right\rangle $, $\left|B\{q_{1}q_{2}\}q_{3}\right\rangle $
and are obtained by the diagonalization of the interaction energy matrix.

\begin{table}
\caption{\label{t2.1} Parameters which specify the color-magnetic interaction energy of
baryons.}
\begin{tabular}{ccccc}
\hline\hline  
$(S_{1},\,S_{2})$ & $J$  & $a_{12}$  & $a_{13}$  & $a_{23}$ \\
\hline 
$(0,\,0)$ & $\frac{1}{2}$  & $-3$  & $0$  & $0$ \\
$(0,\,1)$ & $\frac{1}{2}$  & $0$  & $-\sqrt{3}$  & $\sqrt{3}$\\
$(1,\,1)$ & $\frac{1}{2}$  & $1$  & $-2$  & $-2$ \\
$(1,\,1)$ & $\frac{3}{2}$  & $1$  & $1$  & $1$ \\
\hline\hline  
\end{tabular}
\end{table}

Functions $I_{ij}(R)$ and $K_{ij}(R)$ can be written as

\begin{equation}
I_{ij}(R)=\frac{2}{3}\int\limits _{0}^{R}r^{2}dr\rho_{i}^{\prime}(r)V_{j}(r,R),\label{mbag 009}
\end{equation}
\begin{equation}
K_{ij}(R)=\frac{4}{3}\int\limits _{0}^{R}r^{2}dr\mu_{i}^{\prime}(r)A_{j}(r,R),\label{mbag 010}
\end{equation}
where 
\begin{equation}
\rho_{i}^{\prime}(r)=G_{i}^{2}\left(r\right)+F_{i}^{2}\left(r\right)\label{mbag 011}
\end{equation}
is the charge density, and 
\begin{equation}
\mu_{i}^{\prime}(r)=-\frac{2r}{3}G_{i}(r)F_{i}(r)\label{mbag 012}
\end{equation}
is the scalar magnetization density of an \textit{i}-th quark. $V_{i}(r,R)$
and $A_{i}(r,R)$ are semiclassical scalar and vector potentials generated
by the \textit{i}-th quark

\begin{equation}
V_{i}(r,R)=\rho_{i}(r)\left(\frac{1}{r}-\frac{1}{R}\right)+\int\limits _{r}^{R}x^{2}dx\frac{\rho_{i}^{\prime}(x)}{x},\label{mbag 013}
\end{equation}
\begin{equation}
A_{i}(r,R)=\frac{\mu_{i}(r)}{r^{3}}+\frac{\mu_{i}(R)}{2R^{3}}+\mathrm{M}_{i}(r,R),\label{mbag 014}
\end{equation}
where 
\begin{equation}
\rho_{i}(r)=\int\limits _{0}^{r}x^{2}dx\rho_{i}^{\prime}(x),\label{mbag 015}
\end{equation}
\begin{equation}
\mu_{i}(r)=\int\limits _{0}^{r}x^{2}dx\mu_{i}^{\prime}(x),\label{mbag 016}
\end{equation}

\begin{equation}
\mathrm{M}_{i}(r,R)=\int\limits _{r}^{R}x^{2}dx\frac{\mu_{i}^{\prime}(x)}{x^{3}} .\label{mbag 017}
\end{equation}

The interaction energy depends on the effective running coupling constant
$\alpha_{\mathrm{c}}(R)$. We allow this constant to vary with the
bag radius in a manner consistent with asymptotic freedom and adopt
the parameterization proposed in Ref.~\cite{DJ80}

\begin{equation}
\alpha_{\mathrm{c}}(R)=\frac{2\pi}{9\,\ln(A+R_{0}/R)},\label{mbag 018}
\end{equation}
where$R_{0}$ is the scale parameter analogous to the QCD constant
$\Lambda^{-1}$, and the parameter $A$ is introduced in order to avoid
divergences when $R_{0}\rightarrow R$. We also introduce effective
$R$-dependent (running) quark mass 
\begin{equation}
m_{f}(R)=\widetilde{m}_{f}+\alpha_{\mathrm{c}}(R)\cdot\delta_{f},\label{mbag 019}
\end{equation}
where parameters $\widetilde{m}_{f}$ and $\delta_{f}$ define the
mass functions $m_{f}(R)$ for each quark flavor.

The mass of the hadron is defined as the corresponding bag energy
corrected for the center-of-mass motion (c.m.m.) according to the prescription

\begin{equation}
M^{2}=E^{2}-\gamma\sum\limits _{i}p_{i}^{2},\label{mbag 020}
\end{equation}
where $p_{i}=\sqrt{\varepsilon_{i}^{2}-m_{i}^{2}}$ is the momentum
of the \textit{i}-th quark, $m_{i}$ is the effective quark mass given
by Eq.~(\ref{mbag 019}), and $\gamma$ is the free parameter.

The model parameters $B$, $\gamma$, $A$, $R_{0}$, $\tilde{m}_{s}$,
$\delta_{s}$, $\tilde{m}_{c}$, $\delta_{c}$, $\tilde{m}_{b}$,
$\delta_{b}$ were taken from the Ref.~\cite{S16a} ($B=7.301\times10^{-4}~\mathrm{GeV}^{4}$,
$\gamma=1.785$, $A=0.7719$, $R_{0}=3.876~\mathrm{GeV}^{-1}$, $\widetilde{m}_{s}=0.2173~\mathrm{GeV}$,
$\delta_{s}=0.1088$~$\mathrm{GeV}$, $\widetilde{m}_{c}=1.456~\mathrm{GeV}$,
$\delta_{c}=0.1003~\mathrm{GeV}$, $\widetilde{m}_{b}=4.746~\mathrm{GeV}$,
and $\delta_{b}=0.0880~\mathrm{GeV}$). The lightest (up and down)
quarks are assumed to be massless.

The magnetic moment of a hadron $\mu(H)$ and the static transition
moment $\mu(H_{1}\rightarrow H_{2})$ can be calculated by the matrix
element of the operator $\boldsymbol{\mu}$

\begin{equation}
\mu(H)=\left\langle H\left|\boldsymbol{\mu}\right|H\right\rangle ,\label{mbag 021}
\end{equation}

\begin{equation}
\mu(H_{1}\rightarrow H_{2})=\left\langle H_{1}\left|\boldsymbol{\mu}\right|H_{2}\right\rangle ,\label{mbag 022}
\end{equation}

\begin{equation}
\boldsymbol{\mu}=\frac{1}{2}\int\mathrm{d}^{3}x\left[\mathbf{r}\times\mathbf{j}_{em}\right],\label{mbag 023}
\end{equation}
where $\mathbf{j}_{em}$ is the Dirac electromagnetic current, and
$\left|H_{i}\right\rangle $ is the hadron state of definite polarization.
After some algebra these expressions can be rewritten as

\begin{equation}
\mu=\sum\widetilde{\mu}_{i}\left\langle H_{1}\uparrow\left|e_{i}\mathbf{\sigma}_{i}\right|H_{2}\uparrow\right\rangle ,\label{mbag 024}
\end{equation}
where $e_{i}$ is the charge of the corresponding quark, and $\widetilde{\mu}_{i}$
is the reduced (charge-independent) quark magnetic moment 
\begin{equation}
\widetilde{\mu}_{i}=\int r^{2}\mathrm{d}r\frac{2r}{3}G_{i}(r)F_{i}(r),\label{mbag 025}
\end{equation}
$\mu_{i}=e_{i}\widetilde{\mu}_{i}$. For the vector meson made of
quark $q_{1}$ and antiquark $\overline{q}_{2}$ one has 
\begin{equation}
\mu=\mu_{1}+\mu_{2}=e_{1}\widetilde{\mu}_{1}+\overline{e}_{2}\widetilde{\mu}_{2},\label{mbag 026}
\end{equation}
where $e_{a}$ ($\overline{e}_{b}$) is the charge of the corresponding
quark (antiquark).

The static transition moment connecting vector and pseudoscalar mesons
is given by

\begin{equation}
\mu(V\rightarrow PS)=\mu_{1}-\mu_{2}=e_{1}\widetilde{\mu}_{1}-\overline{e}_{2}\widetilde{\mu}_{2}.\label{mbag 027}
\end{equation}

For the baryons one has (see Ref.~\cite{FLNC81}):

\begin{align}
\mu(B_{[q_{1}q_{2}]q_{3}}) & =\mu_{3},\label{mbag 028}\\
\mu(B_{\{q_{1}q_{2}\}q_{3}}) & =\frac{1}{3}(2\mu_{1}+2\mu_{2}-\mu_{3}),\label{mbag 029}\\
\mu(B^{*}) & =\mu_{1}+\mu_{2}+\mu_{3},\label{mbag 030}
\end{align}
\begin{align}
\mu(B_{\{q_{1}q_{2}\}q_{3}}\rightarrow B_{[q_{1}q_{2}]q_{3}}) & =\frac{1}{\sqrt{3}}(\mu_{2}-\mu_{1}),\label{mbag 031}\\
\mu(B^{*}\rightarrow B_{[q_{1}q_{2}]q_{3}}) & =\sqrt{\frac{2}{3}}(\mu_{1}-\mu_{2}),\label{mbag 032}\\
\mu(B^{*}\rightarrow B_{\{q_{1}q_{2}\}q_{3}}) & =\frac{\sqrt{2}}{3}(\mu_{1}+\mu_{2}-2\mu_{3}).\label{mbag 033}
\end{align}

For the M1 decay widths we use the following expressions

\begin{equation}
\Gamma(H_{1}\rightarrow H_{2})=\frac{\alpha k^{3}}{M_{P}^{2}}\frac{1}{2J+1}\mu^{2}(H_{1}\rightarrow H_{2}) \label{mbag 034}
\end{equation}
for mesons, and

\begin{equation}
\Gamma(H_{1}\rightarrow H_{2})=\frac{\alpha k^{3}}{M_{P}^{2}}\frac{2}{2J+1}\mu^{2}(H_{1}\rightarrow H_{2}) \label{mbag 035}
\end{equation}
for baryons. In these expressions  $\alpha\approx\frac{1}{137}$
is the fine structure constant, $M_{P}$ is the proton mass, 

\begin{equation}
k=(M_{1}^{2}-M_{2}^{2})/(2M_{1})  \label{mbag 039}
\end{equation}
is the photon momentum in the rest frame of a decaying particle, $J$
is the spin of decaying hadron, and $\mu(H_{1}\rightarrow H_{2})$
is the corresponding M1 transition moment expressed in nuclear magnetons
$\mu_{N}=1/(2M_{p})$. 
The transition moment depends on the momentum
of the emitted photon and can be obtained from Eqs.~(\ref{mbag 031})--(\ref{mbag 033})
by replacing static moments 
with the corresponding $k$-dependent M1 transition moments 
\begin{equation}
\widetilde{\mu}_{i}^{TR}(k)=\int r^{2}\mathrm{d}rj_{1}(kr)[G_{1i}(r)F_{2i}(r)+G_{2i}(r)F_{1i}(r)].\label{mbag 036}
\end{equation}
Indices $1$ and $2$ stand for initial and final particles, $j_{1}(kr)$
is the Bessel function. When the radii of the bags differ we choose
the smaller one as the upper limit of the integral. It can be checked,
that Eqs.~(\ref{mbag 034}), (\ref{mbag 035}) and the expression
(17) for the M1 decay width obtained in Ref.~\cite{HDDT78} are equivalent.

It is well-known that magnetic observables calculated in the bag model
should be corrected for the center-of-mass motion (c.m.m.), recoil,
and possibly other effects. We use simple prescription adopted in
Ref.~\cite{S16a} 
\begin{equation}
\mu_{L}=C_{L}\,\mu_{L}^{0},\quad\mu_{H}=C_{H}\,\mu_{H}^{0},\label{mbag 037}
\end{equation}
where $C_{L}$, $C_{H}$ are the model parameters and $\mu_{L}^{0}$, $\mu_{H}^{0}$ are the original, uncorrected bag model quantities (magnetic or M1 transition moments) for the 
light ($u$, $d$, or $s$) and heavy ($c$ or $b$) quarks, respectively.

\section{State mixing}  \label{sec_mix}

As was mentioned earlier the color-magnetic interaction mixes the
states with different intermediate spins $\left|B[q_{1}q_{2}]q_{3}\right\rangle $ and
$\left|B\{q_{1}q_{2}\}q_{3}\right\rangle $. The physical states $\left|B\right\rangle $
and $\left|B'\right\rangle $ are the linear combinations of these
states

\begin{equation}
\left.\begin{array}{c}
\left|B\right\rangle =C_{1}\left|B_{[q_{1}q_{2}]q_{3}}\right\rangle +C_{2}\left|B_{\{q_{1}q_{2}\}q_{3}}\right\rangle ,\\
\left|B^{\prime}\right\rangle =C_{1}\left|B_{\{q_{1}q_{2}\}q_{3}}\right\rangle -C_{2}\left|B_{[q_{1}q_{2}]q3}\right\rangle .
\end{array}\right.\label{mix 01}
\end{equation}

The magnetic observables for such baryons are given by

\begin{eqnarray}
\mu(B) & = & C_{1}^{2}\,\mu(B_{[q_{1}q_{2}]q_{3}})+C_{2}^{2}\,\mu(B_{\{q_{1}q_{2}\}q_{3}})\nonumber \\
 & +  &  2C_{1}\,C_{2}\,\mu(B_{\{q_{1}q_{2}\}q_{3}}\rightarrow B_{[q_{1}q_{2}]q_{3}}),\label{mix 02}
\end{eqnarray}

\begin{eqnarray}
\mu(B^{\prime}) & = & C_{1}^{2}\,\mu(B_{\{q_{1}q_{2}\}q_{3}})+C_{2}^{2}\,\mu(B_{[q_{1}q_{2}]q_{3}})\nonumber \\
 & -  &  2C_{1}\,C_{2}\,\mu(B_{\{q_{1}q_{2}\}q_{3}}\rightarrow B_{[q_{1}q_{2}]q_{3}}),\label{mix 03}
\end{eqnarray}

\begin{eqnarray}
\mu(B^{\prime}\rightarrow B) & = & (C_{1}^{2}\,-C_{2}^{2}\,)\,\mu(B_{\{q_{1}q_{2}\}q_{3}}\rightarrow B_{[q_{1}q_{2}]q_{3}})\nonumber \\
 &+& \! C_{1}\,C_{2}\,[\mu(B_{\{q_{1}q_{2}\}q_{3}})-\mu(B_{[q_{1}q_{2}]q_{3}})], \label{mix 04} 
\end{eqnarray}

\begin{eqnarray}
\mu(B^{*}\rightarrow B) & = & C_{1}\,\mu(B^{*}\rightarrow
 B_{[q_{1}q_{2}]q_{3}})\nonumber \\
 & + &   C_{2}\,\mu(B^{*}\rightarrow B_{\{q_{1}q_{2}\}q_{3}}),  \label{mix 05}
\end{eqnarray}

\begin{eqnarray}
\mu(B^{*}\rightarrow B^{\prime}) & = & C_{1}\,\mu(B^{*}\rightarrow B_{\{q_{1}q_{2}\}q_{3}}) \nonumber \\
& - &   C_{2}\,\mu(B^{*}\rightarrow B_{[q_{1}q_{2}]q_{3}}).   \label{mix 06}
\end{eqnarray}

In Eqs. (\ref{mix 02}) and (\ref{mix 03}) the static values of the
transition moments must be used. On the other hand, in Eqs. (\ref{mix 04})--(\ref{mix 06})
all the entries are $k$-dependent moments.

\begin{table*}
\caption{\label{t3.1} Expansion coefficients $C_{1},C_{2}$  in the expression
 $\left|B\right\rangle =C_{1}\left|B_{[q_{1}q_{2}]q_{3}}\right\rangle +C_{2}\left|B_{\{q_{1}q_{2}\}q_{3}}\right\rangle$.}

\begin{tabular}{ccccccc}
\hline\hline  
Baryons  & Quark ordering  & Our  & NR \cite{FLNC81}  & PM \cite{AHN10b}\;  & PM \cite{RP08}  & QCDSR \cite{AOZ11,AAS12b}\\
\hline 
$\Xi_{c},\Xi_{c}^{\prime}$  & $(qs)c$  & $0.9975,\;$$0.0703$  & $0.9978,\;$$0.0660$  & $\cdot\cdot\cdot$  & $0.9919,\;$$0.0438$  & $0.996,\;$$0.086$\\
``  & $(qc)s$  & $-0.5597,\;$$0.8287$  & $\cdot\cdot\cdot$  & $\cdot\cdot\cdot$  & $\cdot\cdot\cdot$  & $\cdot\cdot\cdot$\\
``  & $(sc)q$  & $0.4379,\;$$0.8990$  & $\cdot\cdot\cdot$  & $\cdot\cdot\cdot$  & $\cdot\cdot\cdot$  & $\cdot\cdot\cdot$\\
\hline 
$\Xi_{b},\Xi_{b}^{\prime}$  & $(qs)b$  & $0.9998,\;$$0.0175$  & $0.9999,\;$$0.0170$  & $\cdot\cdot\cdot$  & $0.9913,\;$$0.0330$  & $0.995,\;$$0.100$\\
``  & $(qb)s$  & $-0.5151,\;$$0.8571$  & $\cdot\cdot\cdot$  & $\cdot\cdot\cdot$  & $\cdot\cdot\cdot$  & $\cdot\cdot\cdot$\\
``  & $(sb)q$  & $0.4848,\;$$0.8746$  & $\cdot\cdot\cdot$  & $\cdot\cdot\cdot$  & $\cdot\cdot\cdot$  & $\cdot\cdot\cdot$\\
\hline 
$\Xi_{cb},\Xi_{cb}^{\prime}$  & $(qc)b$  & $0.9918,\;$$0.1278$  & $0.9916,\;$$0.1296$  & $\cdot\cdot\cdot$  & $\cdot\cdot\cdot$  & $\cdot\cdot\cdot$\\
``  & $(qb)c$  & $-0.6066,\;$$0.7950$  & $\cdot\cdot\cdot$  & $\cdot\cdot\cdot$  & $\cdot\cdot\cdot$  & $\cdot\cdot\cdot$\\
``  & $(cb)q$  & $-0.3852,\;$$-0.9228$  & $\cdot\cdot\cdot$  & $0.431,\;$$0.902\quad$  & $0.3839,\;$$0.8976$  & $0.249,\;$$0.969$\\
\hline 
$\Omega_{cb},\Omega_{cb}^{\prime}$  & $(sc)b$  & $0.9937,\;$$0.1120$  & $0.9928,\;$$0.1197$  & $\cdot\cdot\cdot$  & $\cdot\cdot\cdot$  & $\cdot\cdot\cdot$\\
``  & $(sb)c$  & $-0.5939,\;$$0.8046$  & $\cdot\cdot\cdot$  & $\cdot\cdot\cdot$  & $\cdot\cdot\cdot$  & $\cdot\cdot\cdot$\\
``  & $(cb)s$  & $-0.3998,\;$$-0.9166$  & $\cdot\cdot\cdot$  & $0.437,\;$$0.899\quad$  & $0.4149,\;$$0.8947$  & $0.279,\;$$0.960$\\
\hline\hline  
\end{tabular}
\end{table*}

We have performed the detailed analysis of the state mixing effect using all possible
quark ordering schemes. In Table~\ref{t3.1} we present the expansion
coefficients obtained for all possible quark orderings. A symbol $q$
is used for the lightest ($u$ or $d$) quarks. Of course, if the
state mixing is taken into account, the final results for the
calculated physical quantities do not depend on the basis. If the
state mixing is ignored, the results, in general, become dependent
on the quark ordering scheme. 

As pointed out in Ref.~\cite{FLNC81},
there is a favored ordering scheme (the optimal basis), which to some
extent minimizes the state mixing effect. This is the basis with the
specific quark ordering scheme, in which the heaviest quark $q_{3}$
is chosen as the third one in the wave function $\left|(q_{1}q_{2})^{S}q_{3}\right\rangle $.
In this basis the effect of the hyperfine mixing on the baryon masses
is generally negligible. This is because in the case of weak mixing
the mass shifts are second order in the mixing angles. Thus, if
this optimal basis is chosen for the calculation of baryon masses,
one could not bother very much about the state mixing problem. On the other hand,
the shifts due the hyperfine state mixing for the magnetic moments
are first order in the mixing angle, and in this case the
effect can be important. We postpone the detailed analysis of this
effect on the magnetic observables of heavy baryons to subsequent
sections. 

In Table~\ref{t3.1} we also compare our results for the
expansion coefficients with the results obtained using some other
approaches. These are the nonrelativistic quark model (NR),
two different nonrelativistic potential models (PM), and the
approach based on the QCD sum rules (QCDSR).
We see that our results are very close to the values obtained in the
simple nonrelativistic model (NR). Presumably the reason is the same
algebraic structure of the hyperfine interaction in both models. The
results obtained in other, slightly different  version of the bag model \cite{BS08a} 
are also practically indistinguishable from the present ones.

The results obtained using potential models and QCDSR method demonstrate
some model dependence, however, the main pattern remains similar in
all cases. For instance, in the case of $\Xi_{c}$ type baryons all
methods give very similar predictions. In the case of $\Xi_{b}$ type
baryons the potential model \cite{RP08} as well as QCDSR method predict
stronger mixing than our model. For the baryons of $\Xi_{cb}$ and
$\Omega_{cb}$ type our results are similar to the results given by
both potential models, but QCDSR method predicts weaker mixing.
In general, our predictions seem to be quite reasonable, with possible
exception of the ($\Xi_{b}$, $\Xi_{b}^{\prime}$) baryons, where the real state mixing
presumably should be somewhat stronger.

\section{Hyperfine splittings} \label{sec_hyp}

If one seeks reliable predictions for the decay widths, it is 
important to have accurate values of the photon momentum $k$, because
the decay widths are third order in $k$. For this reason, we prefer
to use in the calculations of decay widths the experimental values
of the hadron masses. Unfortunately, not all masses of heavy ground-state
hadrons are measured. Such are the $B_{c}^{*}$ meson, singly heavy
baryon $\Omega_{b}^{*}$, and all (now, except for the $\Xi_{cc}$)
doubly and triply heavy baryons. In these cases we need as accurate
as possible theoretical estimate. For heavy hadrons the momentum of
the emitted photon is approximately equal to the corresponding mass
difference. In what follows we will use the semiempirical approach
to estimate these mass splittings.
The mass spectra of doubly heavy baryons were studied using various approaches,
such as lattice QCD \cite{NeA13,BDMO14,AK17,AeA00a,MLW02a,LW09,FMT03,ACCDGP12,ADJKK14,B15b,PCB15},
QCD sum rules \cite{ZH08b,AN10b,W10d,W10e,AAS12a,AAS13b}, nonrelativistic
phenomenological quark model \cite{KR14}, semiempirical sum rules
\cite{RLP95,LRP96}, NRQCD based effective theory \cite{BVR05}, potential
models \cite{S96e,RP08,ER12,AHN10a,IMMW00,KLPS02,L09,VGV08,YHHOS15},
covariant quark model \cite{CDR98}, model based on Bethe-Salpeter
equations \cite{MMMP06}, relativistic quark model with the hyperspherical
expansion \cite{M08b}, relativistic quark-diquark model
 \cite{EFGM02}, Salpeter model with AdS/QCD inspired potential
\cite{G09}. The spectra of triply heavy baryons were calculated using
lattice QCD \cite{BDMO14}, QCD sum rules \cite{AAS13a,AAS14a}, potential
models \cite{RP08,FHN12}, relativistic quark model \cite{M08b}, and
the model with AdS/QCD inspired potential \cite{G09}.

\begin{table*}
\caption{\label{t4.1} Hyperfine splittings (in MeV) of ground-state singly heavy baryons
obtained using various approaches.}

\begin{tabular}{lcccccc}
\hline\hline 
Baryons  & Exp \cite{PDG17}  & Lattice$^{a}$ \cite{BDMO14}  & PM(AL1) \cite{AAHN04}  & PM(RP) \cite{RP08}  & PM(VGV) \cite{VGV08}  & Bag \\
\hline 
$\Sigma_{c}^{*}-\Lambda_{c}$  & $232$  & $297(33)(43)$  & $253$  & $251$  & $217$  & $195$ \\
$\Sigma_{c}^{*}-\Sigma_{c}$  & $65$  & $78(7)(11)$  & $79$  & $64$  & $67$  & $92$ \\
$\Sigma_{c}-\Lambda_{c}$  & $167$  & $219(36)(43)$  & $174$  & $187$  & $150$  & $103$ \\
$\varXi_{c}^{*}-\Xi_{c}$  & $177$  & $214(16)(39)$  & $181$  & $183$  & $171$  & $161$ \\
$\Xi_{c}^{*}-\Xi_{c}^{\prime}$  & $68$  & $73.7(5.0)8.7)$  & $77$  & $55$  & $68$  & $86$ \\
$\Xi_{c}^{\prime}-\Xi_{c}$  & $109$  & $140(16)(38)$  & $104$  & $128$  & $103$  & $75$ \\
$\Omega_{c}^{*}-\Omega_{c}$  & $71(1)$  & $75.3(1.9)(7.6)$  & $74$  & $58$  & $68$  & $80$ \\
$\Sigma_{b}^{*}-\Lambda_{b}$  & $214$  & $251(46)(40)$  & $239$  & $246$  & $209$  & $162$ \\
$\Sigma_{b}^{*}-\Sigma_{b}$  & $21(2)$  & $21.2(4.9)(7.3)$  & $31$  & $25$  & $22$  & $31$ \\
$\Sigma_{b}-\Lambda_{b}$  & $193$  & $230(47)(40)$  & $208$  & $221$  & $187$  & $131$ \\
$\varXi_{b}^{*}-\Xi_{b}$  & $160$  & $189(29)(33)$  & $167$  & $174$  & $160$  & $131$ \\
$\Xi_{b}^{*}-\Xi_{b}^{\prime}$  & $20$  & $27.0(3.2)(8.6)$  & $29$  & $10$  & $22$  & $29$ \\
$\Xi_{b}^{\prime}-\Xi_{b}$  & $140$  & $162(29)(33)$  & $138$  & $164$  & $138$  & $102$ \\
$\Omega_{b}^{*}-\Omega_{b}$  & $\cdot\cdot\cdot$  & $28.4(2.2)(7.7)$  & $30$  & $21$  & $23$  & $27$ \\
\hline\hline 
\multicolumn{7}{l}{$^{a}$The first uncertainty is statistical, the second one is systematic.}\\
\end{tabular}
\end{table*}

\begin{table*}
\caption{\label{t4.2} Hyperfine splittings (in MeV) of ground-state doubly and triply heavy
baryons obtained using various approaches.}

\begin{tabular}{lccccc}
\hline\hline 
Baryons  & Lattice$^{a}$ \cite{BDMO14}  & PM(AL1) \cite{AHN10a,FHN12}  & PM(RP) \cite{RP08}  & PM(VGV) \cite{VGV08}  & Bag \\
\hline 
$\Xi_{cc}^{*}-\Xi_{cc}$  & $82.8(7.2)(5.8)$  & $94$  & $77$  & $77$  & $95$ \\
$\Omega_{cc}^{*}-\Omega_{cc}$  & $83.8(1.4)(5.3)$  & $83$  & $61$  & $72$  & $82$\\
$\varXi_{cb}^{*}-\Xi_{cb}$  & $43(19)(38)$  & $77$  & $63$  & $\cdot\cdot\cdot$  & $70$ \\
$\Xi_{cb}^{*}-\Xi_{cb}^{\prime}$  & $26.7(3.3)(8.4)$  & $29$  & $27$  & $\cdot\cdot\cdot$  & $25$ \\
$\Xi_{cb}^{\prime}-\Xi_{cb}$  & $16(18)(38)$  & $48$  & $36$  & $\cdot\cdot\cdot$  & $45$ \\
$\Omega_{cb}^{*}-\Omega_{cb}$  & $62(9)25()$  & $70$  & $51$  & $\cdot\cdot\cdot$  & $60$ \\
$\Omega_{cb}^{*}-\Omega_{cb}^{\prime}$  & $27.4(2.0)(6.7)$  & $29$  & $22$  & $\cdot\cdot\cdot$  & $23$ \\
$\varOmega_{cb}^{\prime}-\Omega_{cb}$  & $35(9)(25)$  & $41$  & $29$  & $\cdot\cdot\cdot$  & $37$ \\
$\Xi_{bb}^{*}-\Xi_{bb}$  & $34.6(2.5)(7.4)$  & $39$  & $27$  & $29$  & $36$ \\
$\Omega_{bb}^{*}-\Omega_{bb}$  & $35.7(1.3)(5.5)$  & $38$  & $32$  & $28$  & $32$ \\
$\Omega_{ccb}^{*}-\Omega_{ccb}$  & $29.6(0.7)(4.2)$  & $28$  & $20$  & $\cdot\cdot\cdot$  & $18$ \\
$\Omega_{cbb}^{*}-\Omega_{cbb}$  & $33.5(0.6)(4.1)$  & $31$  & $19$  & $\cdot\cdot\cdot$  & $21$ \\
\hline\hline 
\multicolumn{6}{l}{$^{a}$ The first uncertainty is statistical, the second one is systematic.}\\
\end{tabular}
\end{table*}

In Tables~\ref{t4.1} and \ref{t4.2} we have collected some predictions
for the hyperfine splittings of heavy baryons obtained using lattice
QCD, three variants of the potential model (labeled as AL1, RP, VGV), and our
current variant of the bag model, accompanied with available experimental
data (Exp). We see from Table~\ref{t4.2} that all these approaches
give more or less similar estimates. Seemingly, one can use any of
them, but it is not quite clear to what extent these results are trustworthy.
To test the quality of the predictions it could be useful to compare
the estimates given by these approaches with the available experimental data (see
Table~\ref{t4.1}). The results of a comparison are a little bit disappointing.
We see that the predictions given by various approaches differ considerably.
The best agreement with experiment provides the VGV potential model with quark-quark
interaction mediated via one-gluon plus scalar and pseudoscalar boson
exchanges \cite{VGV08}. However, it is not very helpful for us, because
in this case there are no predictions for $\Xi_{cb}$, $\Omega_{bc}$,
and triply heavy baryons. The lattice QCD predictions, as a rule,
exceed experimental data (sometimes by $\sim50\%$), but remain compatible
with them within errors.

Our strategy to obtain the improved baryon mass splittings is based
on the assumption that these mass differences are defined by mutual
quark-quark interaction, as, e.g. in the phenomenological approach
used in Ref.~\cite{KR14}. We propose to rescale the corresponding
quark-quark interaction strengths to match the model predictions for the singly heavy baryons with experimental data. Then, suggesting the dynamics of singly, doubly, and triply
heavy baryons to be interdependent, we can rescale the corresponding
interaction parameters for doubly (and triply) heavy baryons and obtain
new (plausibly improved) predictions. For this purpose we invoke quark
model relations, which can be obtained, for example, using the parameters
presented in Table~\ref{t2.1}. The use of the quark model relations
can be partially justified by the observation from the lattice QCD
calculations \cite{PEMP15} that heavy baryon spectra are similar
to the expectations from nonrelativistic quark models. For the baryons
containing three quarks of different flavors in order to avoid the
state mixing problem the optimal basis is used. For the charmed baryons
of $B_{Q}$ and $B_{QQ}$ type we have

\begin{align}
M(\Sigma_{c}^{*})-M(\Sigma_{c}) &  =6K_{qc},\label{hyp 001}\\
M(\Sigma_{c}^{*})-M(\Lambda_{c}) &  =4K_{qq}+2K_{qc},\label{hyp 002}\\
M(\Sigma_{c})-M(\Lambda_{c}) &  =4K_{qq}-4K_{qc},\label{hyp 003}\\
M(\Xi_{c}^{*})-M(\Xi_{c}^{\prime}) &  =3K_{qc}+3K_{sc},\label{hyp 004}\\
M(\Xi_{c}^{*})-M(\Xi_{c}) &  =4K_{qs}+K_{qc}+K_{sc},\label{hyp 005}\\
M(\Xi_{c}^{\prime})-M(\Xi_{c}) &  =4K_{qs}-2K_{qc}-2K_{sc},\label{hyp 006}\\
M(\Omega_{c}^{*})-M(\Omega_{c}) &  =6K_{sc},\label{hyp 007}\\
M(\Xi_{cc}^{*})-M(\Xi_{cc}) &  =6K_{qc},\label{hyp 008}\\
M(\Omega_{cc}^{*})-M(\Omega_{cc}) &  =6K_{sc}.\label{hyp 009}
\end{align}

The corresponding relations for bottom baryons can be obtained
by the substitution $c\rightarrow b$, for example,

\begin{align}
M(\Xi_{bb}^{*})-M(\Xi_{bb})  & =6K_{qb},\label{hyp 008a}\\
M(\Omega_{bb}^{*})-M(\Omega_{bb})  & =6K_{sb},\label{hyp 009b}
\end{align}
etc. The expressions for doubly
heavy baryons of $\Xi_{cb}$ and $\Omega_{cb}$ type are

\begin{align}
M(\Xi_{cb}^{*})-M(\Xi_{cb}^{\prime})  & =3K_{qb}+3K_{cb},\label{hyp 010}\\
M(\Xi_{cb}^{*})-M(\Xi_{cb})  & =4K_{qc}+K_{qb}+K_{cb},\label{hyp 011}\\
M(\Xi_{cb}^{\prime})-M(\Xi_{cb})  & =4K_{qc}-2K_{qb}-2K_{cb},\label{hyp 012}\\
M(\Omega_{cb}^{*})-M(\Omega_{cb}^{\prime})  & =3K_{sb}+3K_{cb},\label{hyp 013}\\
M(\Omega_{cb}^{*})-M(\Omega_{cb})  & =4K_{sc}+K_{sb}+K_{cb},\label{hyp 014}\\
M(\Omega_{cb}^{\prime})-M(\Omega_{cb})  & =4K_{sc}-2K_{sb}-2K_{cb}.\label{hyp 015}
\end{align}

For the triply heavy baryons one has

\begin{align}
M(\Omega_{ccb}^{*})-M(\Omega_{ccb})  & =6K_{cb},\label{hyp 016}\\
M(\Omega_{cbb}^{*})-M(\Omega_{cbb})  & =6K_{cb}.\label{hyp 017}
\end{align}

In the expressions above the index $q$ denotes the lightest ($u$ or $d$) quarks. In general, parameters $K_{ij}$ depend on the hadron.
To proceed with we make an assumption that in baryons with the same
heavy quark content the interaction strengths between quarks do not
differ substantially, i.e. $K_{ij}(qqQ)\approx K_{ij}(qsQ),$ $K_{ij}(qsQ)\approx K_{ij}(ssQ),$
and $K_{ij}(qQ_{1}Q_{2})\approx K_{ij}(sQ_{1}Q_{2})$. Now we can invert
the equations above in order to estimate the parameters of the quark-quark
interaction. For singly charmed baryons we obtain

\begin{align}
K_{qq}  & =\frac{1}{12}\left\{2\left[M(\Sigma_{c}^{*})-M(\Lambda_{c})\right]+\left[M(\Sigma_{c})-M(\Lambda_{c})\right]\right\},\label{hyp 018a}\\
K_{qs}  & =\frac{1}{12}\left\{2\left[M(\Xi_{c}^{*})-M(\Xi_{c})\right]+\left[M(\Xi_{c}^{\prime})-M(\Xi_{c})\right]\right\},\label{hyp 018b}\\
K_{qc}  & =\frac{1}{6}\left\{2\left[M(\Xi_{c}^{*})-M(\Xi_{c}^{\prime})\right]-\left[M(\Omega_{c}^{*})-M(\Omega_{c})\right]\right\},\label{hyp 019a}\\
K_{sc}  & =\frac{1}{6}\left\{2\left[M(\Xi_{c}^{*})-M(\Xi_{c}^{\prime})\right]-\left[M(\Sigma_{c}^{*})-M(\Sigma_{c})\right]\right\},\label{hyp 019b}\\
K_{qc}  & =\frac{1}{6}\left[M(\Sigma_{c}^{*})-M(\Sigma_{c})\right],\label{hyp 020}\\
K_{sc}  & =\frac{1}{6}\left[M(\Omega_{c}^{*})-M(\Omega_{c})\right].\label{hyp 021}
\end{align}

In analogy with the above relations for the bottom baryons we have

\begin{align}
K_{qq}  & =\frac{1}{12}\left\{2\left[M(\Sigma_{b}^{*})-M(\Lambda_{b})\right]+\left[M(\Sigma_{b})-M(\Lambda_{b})\right]\right\},\label{hyp 022a}\\
K_{qs}  & =\frac{1}{12}\left\{2\left[M(\Xi_{b}^{*})-M(\Xi_{b})\right]+\left[M(\Xi_{b}^{\prime})-M(\Xi_{b})\right]\right\},\label{hyp 022b}\\
K_{qb}  & =\frac{1}{6}\left\{2\left[M(\Xi_{b}^{*})-M(\Xi_{b}^{\prime})\right]-\left[M(\Omega_{b}^{*})-M(\Omega_{b})\right]\right\},\label{hyp 023a}\\
K_{sb}  & =\frac{1}{6}\left\{2\left[M(\Xi_{b}^{*})-M(\Xi_{b}^{\prime})\right]-\left[M(\Sigma_{b}^{*})-M(\Sigma_{b})\right]\right\},\label{hyp 023b}\\
K_{qb}  & =\frac{1}{6}\left[M(\Sigma_{b}^{*})-M(\Sigma_{b})\right],\label{hyp 024}\\
K_{sb}  & =\frac{1}{6}\left[M(\Omega_{b}^{*})-M(\Omega_{b})\right].\label{hyp 025}
\end{align}

For the doubly heavy baryons $\Xi_{QQ}$ and $\Omega_{QQ}$ the corresponding expressions are

\begin{align}
K_{qc}  & =\frac{1}{6}\left[M(\Xi_{cc}^{*})-M(\Xi_{cc})\right],\label{hyp 026}\\
K_{sc}  & =\frac{1}{6}\left[M(\Omega_{cc}^{*})-M(\Omega_{cc})\right],\label{hyp 027}\\
K_{qb}  & =\frac{1}{6}\left[M(\Xi_{bb}^{*})-M(\Xi_{bb})\right],\label{hyp 028}\\
K_{sb}  & =\frac{1}{6}\left[M(\Omega_{bb}^{*})-M(\Omega_{bb})\right],\label{hyp 029}
\end{align}
and similarly for the triply heavy baryons

\begin{align}
K_{cb}  & =\frac{1}{6}\left[M(\Omega_{ccb}^{*})-M(\Omega_{ccb})\right],\label{hyp 030}\\
K_{cb}  & =\frac{1}{6}\left[M(\Omega_{cbb}^{*})-M(\Omega_{cbb})\right].\label{hyp 031}
\end{align}

The case of the $B_{cb}$ type baryons is more complicated. We can
write down

\begin{align}
K_{qc}  & =\frac{1}{6}\left\{2\left[M(\Xi_{cb}^{*})-M(\Xi_{cb})\right]+\left[M(\Xi_{cb}^{\prime})-M(\Xi_{cb})\right]\right\},\label{hyp 032}\\
K_{sc}  & =\frac{1}{6}\left\{2\left[M(\Omega_{cb}^{*})-M(\Omega_{cb})\right]+\left[M(\Omega_{cb}^{\prime})-M(\Omega_{cb})\right]\right\},\label{hyp 034}
\end{align}

\begin{align}
K_{qb}+K_{cb}  & =\frac{1}{3}\left[M(\Xi_{cb}^{*})-M(\Xi_{cb}^{\prime})\right],\label{hyp 033}\\
K_{sb}+K_{cb} & =\frac{1}{3}\left[M(\Omega_{cb}^{*})-M(\Omega_{cb}^{\prime})\right],\label{hyp 035}
\end{align}
but the problem is that the information we can get from the mass splittings
is insufficient to disentangle the three interaction parameters $K_{qb}$,
$K_{sb}$, and $K_{cb}$. We will use some interpolation to relate
parameters $K_{qb}$ and $K_{sb}$ to the corresponding parameters
obtained from the Eqs. (\ref{hyp 028}) and (\ref{hyp 029}).

\begin{table}
\caption{\label{t4.3} Interaction parameters (in MeV) for singly heavy baryons obtained
using various input data.}

\begin{tabular}{lcccccc}
\hline\hline 
 & Baryons  & Eqs.  & Exp  & PM(AL1)  & PM(VGV)  & Bag \\
\hline 
$6K_{qq}(qqc)$  & $\Sigma_{c}^{*},\:\Sigma_{c},\:\Lambda_{c}$  &  (\ref{hyp 018a})  & $314$  & $340$  & $292$  & $246$ \\
$6K_{qc}(qqc)$  & $\Sigma_{c}^{*},\:\Sigma_{c},\:\Lambda_{c}$  &  (\ref{hyp 020})  & $65$  & $79$  & $67$  & $92$\\
$6K_{qs}(qsc)$  & $\varXi_{c}^{*},\:\Xi_{c}^{\prime},\:\Xi_{c}$  &  (\ref{hyp 018b})  & $230$  & $233$  & $223$  & $198$ \\
$6K_{qc}(qsc)$  & $\varXi_{c}^{*},\:\Xi_{c}^{\prime},\:\Xi_{c}$  &  (\ref{hyp 019a})  & $65$  & $80$  & $68$  & $92$ \\
$6K_{sc}(qsc)$  & $\varXi_{c}^{*},\:\Xi_{c}^{\prime},\:\Xi_{c}$  &  (\ref{hyp 019b})  & $71$  & $75$  & $69$  & $80$ \\
$6K_{sc}(ssc)$  & $\Omega_{c}^{*},\:\Omega_{c}$  &  (\ref{hyp 021})  & $71$  & $74$  & $68$  & $80$ \\
\hline 
$6K_{qq}(qqb)$  & $\Sigma_{b}^{*},\:\Sigma_{b},\:\Lambda_{b}$  &  (\ref{hyp 022a})  & $311$  & $343$  & $303$  & $227$ \\
$6K_{qb}(qqb)$  & $\Sigma_{b}^{*},\:\Sigma_{b},\:\Lambda_{b}$  &  (\ref{hyp 024})  & $20$  & $31$  & $22$  & $31$ \\
$6K_{qs}(qsb)$  & $\varXi_{b}^{*},\:\Xi_{b}^{\prime},\:\Xi_{b}$  &  (\ref{hyp 022b})  & $229$  & $236$  & $229$  & $182$ \\
$6K_{qb}(qsb)$  & $\varXi_{b}^{*},\:\Xi_{b}^{\prime},\:\Xi_{b}$  & (\ref{hyp 023a})  & $\cdot\cdot\cdot$  & $28$  & $21$  & $31$ \\
$6K_{sb}(qsb)$  & $\varXi_{b}^{*},\:\Xi_{b}^{\prime},\:\Xi_{b}$  &  (\ref{hyp 023b})  & $20$  & $27$  & $22$  & $27$ \\
$6K_{sb}(ssb)$  & $\Omega_{b}^{*},\:\Omega_{b}$  &  (\ref{hyp 025})  & $\cdot\cdot\cdot$  & $30$  & $23$  & $27$ \\
\hline\hline 
\end{tabular}
\end{table}

In Table~\ref{t4.3} we present interaction parameters $K_{ij}$ obtained
using as an input the mass splittings given in Table~\ref{t4.1}.
In the calculations of the experimental interaction parameters 
the value $M(\Sigma_{b}^{*})-M(\Sigma_{b})=20$ MeV
was used instead of 21 MeV. With the original value
one would have $6K_{qb}=21$ MeV and $6K_{sb}=19$ MeV.
For singly charmed baryons $K_{qc}<K_{sc}$, and therefore for bottom
baryons we expect $K_{qb}\lesssim K_{sb}$. The new choice, while
staying within the error bars, is consistent with this expectation.
We have not included the results of Roberts and Pervin \cite{RP08}
into our analysis because this model, due to its complexity, is not completely
compatible with the simple picture we have used to
describe the quark-quark interactions.

From Table~\ref{t4.3} we see that some regularities exist. For instance,
the approximate relations $K_{qq}(qqc)\approx K_{qq}(qqb)$ and $K_{qs}(qsc)\approx K_{qs}(qsb)$ are valid.
This is the manifestation of the heavy quark symmetry (HQS) \cite{LT88,IW91,FI92,EQ17,M97a,MW00},
which states that the light degrees of freedom are to some extent
independent of heavy flavors. As we see, it works perfectly in the
systems with one heavy quark. Indeed, the coincidences for the experimental
values $314\approx311$, and $230\approx229$ are really impressive.
On the other hand, for the heavy baryons ($\Xi_{cb}$, $\Xi_{cb}^{\prime}$)
and ($\Omega_{cb}$, $\Omega_{cb}^{\prime}$) the utility of HQS
is limited. For example, the HQS suggestion that the spin of heavy
diquark is conserved for these baryons could be reliable only if the
quark-quark hyperfine interaction was ignored.

Next, we see that our assumptions $K_{ij}(qqQ)\approx K_{ij}(qsQ)$ and
$K_{ij}(qsQ)\approx K_{ij}(ssQ)$ work very well. In fact, for the
experimental values we have $K_{qc}(qqc)=K_{qc}(qsc)$ and $K_{sc}(qsc)=K_{sc}(ssc)$.
Therefore, we expect $K_{sb}(qsb)=K_{sb}(ssb)$, and, as a consequence, 
\begin{equation}
M(\Omega_{b}^{*})-M(\Omega_{b})=6K_{sb}(ssb)=20\,\mathrm{MeV}.\label{hyp 036}
\end{equation}

This result is similar to the prediction $M(\Omega_{b}^{*})-M(\Omega_{b})=23$ MeV,
obtained using mass sum rules \cite{LRP96}, and the prediction $19.8\pm3.1$ MeV,
obtained using $1/N_{c}$ expansion \cite{J08}. The predictions obtained
using various potential models are: 21 MeV \cite{RP08},
23 MeV \cite{VGV08}, and 20.4 MeV \cite{WQGW17}.
The relativistic quark model \cite{EFG11b} gives $M(\Omega_{b}^{*})-M(\Omega_{b})=24$ MeV.

Another symmetry that is worthwhile to discuss is the approximate
independence of the hyperfine interaction on light quark flavor ($u$,
$d$, or $s$) observed in the meson spectra

\begin{align}
M(D^{*})-M(D)  & \approx M(D_{s}^{*})-M(D_{s}),\label{hyp 037}\\
M(B^{*})-M(B) & \approx M(B_{s}^{*})-M(B_{s}),\label{hyp 038}
\end{align}
which looks like some reminiscence of the flavor SU(3) symmetry. The interpretation
of this phenomenon is not clear: is it a kind of an accidental
symmetry \cite{RY94}, or a consequence of relativistic kinematics \cite{ACOV04}).
In any case, it could be useful to check if such symmetry also survives in the heavy baryon sector.
For the singly charmed baryons this symmetry implies $K_{qc}(qqc)\approx K_{sc}(qsc)$.
However, from Table~\ref{t4.3} we see that $K_{sc}(qsc)$ obtained using
as an input experimental data is $\approx10\%$ larger than $K_{qc}(qqc)$.
Neither AL1 type potential model, nor the bag model can reproduce
such behavior. Moreover, the lattice QCD predictions \cite{BDMO14} seem
also to suffer from this drawback. On the other hand, the potential model
used in Ref.~\cite{VGV08} has succeeded, but at the expense of the
additional chiral interaction. For the singly bottom baryons this symmetry
seems to be restored $M(\Sigma_{b}^{*})-M(\Sigma_{b})\approx M(\Xi_{b}^{*})-M(\Xi_{b}^{\prime})$,
and consequently $K_{qb}(qqb)\approx K_{sb}(qsb)$. We expect similar
behavior in the case of doubly heavy baryons, too.

In order to obtain the improved predictions for the hyperfine splittings
of doubly and triply heavy baryons we have chosen as an input the
predictions of AL1 potential model. Alternative choice could be the
bag model predictions. In the case of doubly heavy baryons both approaches
give similar results, however, for singly heavy baryons AL1 predictions
are evidently more reliable. This is a serious indication that internal
quark dynamics is better described in this approach. Using Eqs. (\ref{hyp 026})--(\ref{hyp 034}) we immediately obtain 

\begin{align}
6K_{qc}(qcc)  & =94\,\mathrm{MeV},\\
6K_{sc}(scc)  & =83\,\mathrm{MeV},\\
6K_{qb}(qbb)  & =39\,\mathrm{MeV},\\
6K_{sb}(sbb)  & =38\,\mathrm{MeV},\\
6K_{cb}(ccb)  & =28\,\mathrm{MeV},\\
6K_{cb}(cbb)  & =31\,\mathrm{MeV},\\
6K_{qc}(qcb)  & =101\,\mathrm{MeV},\\
6K_{sc}(scb)  & =90.5\,\mathrm{MeV}.
\end{align}

In order to get an estimation
for the interaction parameters $K_{qb}(qcb)$ and $K_{sb}(scb)$ some
additional assumption is necessary. After the inspection of the listed
above parameters we note that the substitution of the bottom quark
instead the charmed one increases the strengths of the interaction
by $\approx10\%$. Assuming 
\begin{equation}
\frac{K_{qb}(qcb)}{K_{qb}(qbb)}=\frac{K_{sb}(scb)}{K_{sb}(sbb)}=\frac{K_{cb}(ccb)}{K_{cb}(cbb)}\,,
\label{hyp 039}
\end{equation}
we obtain 

\begin{align}
6K_{qb}(qcb)  & =35.2\,\mathrm{MeV},\\
6K_{sb}(scb)  & =34.3\,\mathrm{MeV}.
\end{align}

Now, using Eqs. (\ref{hyp 033}) and (\ref{hyp 035}), we get 

\begin{align}
6K_{cb}(qcb)  & =22.8\,\mathrm{MeV},\\
6K_{cb}(scb) & =23.7\,\mathrm{MeV}.
\end{align}

The next step is to rescale the interaction parameters obtained above.
We define 
\begin{equation}
K_{ij}^{\prime}(q_{a}Q_{1}Q_{2})=K_{ij}(q_{a}Q_{1}Q_{2})\frac{K_{ij}(q_{a}q_{b}Q)\mathrm{^{Exp}}}{K_{ij}(q_{a}q_{b}Q)\mathrm{^{AL1}}}\,,\label{hyp 040}
\end{equation}
where $(i,j)=(q,c)$, $(s,c)$, $(q,b)$, and $(s,b)$. Parameters $K_{ij}^{\mathrm{Exp}}$ and
$K_{ij}^{\mathrm{AL1}}$  belong to the singly heavy baryon
sector. In order to rescale the interaction parameters $K_{cb}$ we
need some reliable theoretical estimate, because the experimental
data are absent. In this case we turn to the lattice
QCD result $M(\Omega_{cbb}^{*})-M(\Omega_{cbb})=33.5(0.6)(4.1)$ MeV
\cite{BDMO14}. From Table~\ref{t4.1} we see that for the singly
heavy baryons lattice QCD predictions always overestimate the experimental
data. We expect similar tendencies to hold also in the sectors of
doubly and triply heavy baryons. Having this in mind we take the smallest
still compatible with lattice QCD value and define $K_{cb}^{\prime}(cbb)=29$ MeV.
Now we can rescale the remaining parameters using relations

\begin{equation}
K_{cb}^{\prime}(\cdot\cdot\cdot)=K_{cb}(\cdot\cdot\cdot)\frac{K_{cb}^{\prime}(cbb)}{K_{cb}(cbb)}\,.\label{hyp 041}
\end{equation}

The new (improved) parameters are 

\begin{align}
6K_{qc}^{\prime}(qcc)  & =77\,\mathrm{MeV},\\
6K_{sc}^{\prime}(scc)  & =79\,\mathrm{MeV},\\
6K_{qc}^{\prime}(qcb)  & =83.1\,\mathrm{MeV},\\
6K_{qb}^{\prime}(qcb)  & =22.7\,\mathrm{MeV},\\
6K_{cb}^{\prime}(qcb)  & =21.3\,\mathrm{MeV},\\
6K_{sc}^{\prime}(scb)  & =86.9\,\mathrm{MeV}, \\
6K_{sb}^{\prime}(scb)  & =22.9\,\mathrm{MeV},\\
6K_{cb}^{\prime}(scb)  & =22\,\mathrm{MeV},\\
6K_{qb}^{\prime}(qbb)  & =25\,\mathrm{MeV},\\
6K_{sb}^{\prime}(sbb)  & =25\,\mathrm{MeV},\\ 
6K_{cb}^{\prime}(ccb)  & =26\,\mathrm{MeV},\\
6K_{cb}^{\prime}(cbb)  & =29\,\mathrm{MeV}. 
\end{align}

We see that the interaction
parameter $K_{sc}^{\prime}(scc)$ now became larger than $K_{qc}^{\prime}(qcc)$
in agreement with the expectations from the singly heavy baryon sector. Moreover,
the approximate independence of quark-quark interaction strength on light quark
flavor holds in the bottom sector, i.e. $K_{qb}^{\prime}(qbb)\approx K_{sb}^{\prime}(sbb)$,
$K_{qb}^{\prime}(qcb)\approx K_{sb}^{\prime}(scb)$, and is slightly
broken in the case of doubly charmed baryons.

In addition one new interesting feature emerges. We see that for the
baryons $\Xi_{cb}$ and $\Omega_{cb}$ the interaction strength
between light and bottom quarks is very similar to the interaction
strength between charmed and bottom quarks. This could be an indication
that in the systems containing one bottom quark the charmed quark
behaves in some ways more like the light quark than the heavy one.
Such tendency was also observed in the full lattice QCD calculations
\cite{MDFHL12}.

\begin{table}
\caption{\label{t4.4} Improved predictions for the hyperfine splittings of doubly and triply
heavy baryons (in MeV).}

\begin{tabular}{lcccc}
\hline\hline 
Baryons  & PM(AL1)  & PM(VGV)  & PM(YHHOS)  & Bag \\
\hline 
$\Xi_{cc}^{*}-\Xi_{cc}$  & $77$  & $75$  & $71$  & $67$ \\
$\Omega_{cc}^{*}-\Omega_{cc}$  & $79$  & $75$  & $75$  & $73$\\
$\varXi_{cb}^{*}-\Xi_{cb}$  & $63$  & $\cdot\cdot\cdot$  & $\cdot\cdot\cdot$  & $52$ \\
$\Xi_{cb}^{*}-\Xi_{cb}^{\prime}$  & $22$  & $\cdot\cdot\cdot$  & $\cdot\cdot\cdot$  & $24$ \\
$\Xi_{cb}^{\prime}-\Xi_{cb}$  & $41$  & $\cdot\cdot\cdot$  & $\cdot\cdot\cdot$  & $28$ \\
$\Omega_{cb}^{*}-\varOmega_{cb}$  & $65$  & $\cdot\cdot\cdot$  & $\cdot\cdot\cdot$  & $55$ \\
$\Omega_{cb}^{*}-\varOmega_{cb}^{\prime}$  & $22$  & $\cdot\cdot\cdot$  & $\cdot\cdot\cdot$  & $23$ \\
$\varOmega_{cb}^{\prime}-\Omega_{cb}$  & $43$  & $\cdot\cdot\cdot$  & $\cdot\cdot\cdot$  & $32$ \\
$\Xi_{bb}^{*}-\Xi_{bb}$  & $25$  & $26$  & $23$  & $23$ \\
$\Omega_{bb}^{*}-\Omega_{bb}$  & $25$  & $24$  & $22$  & $24$ \\
$\Omega_{ccb}^{*}-\Omega_{ccb}$  & $26$  & $\cdot\cdot\cdot$  & $\cdot\cdot\cdot$  & $25$ \\
$\Omega_{cbb}^{*}-\Omega_{cbb}$  & $29$  & $\cdot\cdot\cdot$  & $\cdot\cdot\cdot$  & $29$ \\
\hline\hline 
\end{tabular}
\end{table}

Putting new (primed) interaction parameters into Eqs.~(\ref{hyp 008})--(\ref{hyp 017}) we obtain the new
(improved) predictions for the hyperfine splittings of doubly and triply
heavy baryons. The results are presented in Table~\ref{t4.4}. We
also performed analogous analysis taking as input the bag model
results and the predictions given by two potential models. These are
the model used in Ref.~\cite{VGV08} (labeled as VGV) and the model
used in Ref.~\cite{YHHOS15} (labeled as YHHOS). For the sake of comparison
we have included in Table~\ref{t4.4} these results, too. We see that
new hyperfine splittings are almost always smaller than
the original ones. At this point we can make one (more or less reasonable)
prediction. Taking the average of the potential model based predictions
given in Table~\ref{t4.4} we can write down $M(\Xi_{cc}^{*})-M(\Xi_{cc})=74\pm4$ MeV.
Next, adding this value to the experimental mass of $\Xi_{cc}^{*}$ we
obtain an estimate $M(\Xi_{cc}^{*})=3695\pm5$ MeV.

We now have at our disposal the values for the hyperfine splittings
(and corresponding photon momenta) of all ground-state heavy baryons. 
However, in the meson sector one
mass splitting needed for the further analysis is still missing. This
is the mass difference $M(B_{c}^{*})-M(B_{c})$. There exist a number
of theoretical predictions for this quantity obtained using various
approaches, such as nonrelativistic QCD \cite{PPSS04a}, QCD sum rules
\cite{W13a}, lattice QCD \cite{GeA10,DDHH12,WLW15}, semiempirical
mass formulae \cite{RDLP95}, nonrelativistic potential models \cite{GKLT95a,CO96,CIKM97,MZ98a,PV09},
various variants of relativistic or semirelativistic models \cite{Cv88,IMMW92,ZVR95,GJ96,BBD10a,EFG11a},
Bethe-Salpeter model \cite{ALV99,BP00,IS05b,GHK16}, and so on. More
references can be found in Ref.~\cite{GHK16}. The latest predictions
for this hyperfine splitting \cite{KR14,GeA10,DDHH12,WLW15,EFG11a}
are more or less similar. Let us take the typical nonrelativistic
potential model prediction $M(B_{c}^{*})-M(B_{c})=68\pm8$ MeV \cite{KR14}
and lattice QCD result $53\pm7$ MeV \cite{GeA10}. The
only estimate compatible with these two predictions is $M(B_{c}^{*})-M(B_{c})=60$ MeV.
It is this value we will use in the further calculations. Some other
approaches also provide results compatible with our choice. For instance,
another potential model based estimate can be obtained using the semiempirical
relationship \cite{CIKM97} 
\begin{equation}
\triangle B_{c}=(0.7)[M(J/\psi)-M(\eta_{c})]^{0.65}[M(\Upsilon)-M(\eta_{b})]^{0.35},\nonumber
\end{equation}
which is expected to hold at about $10\%$ level. Using the experimental
masses of $J/\psi$, $\eta_{c}$, $\Upsilon$, and $\eta_{b}$ mesons
one gets $\triangle B_{c}=64\pm6$ MeV, the result similar
to the one obtained in Ref.~\cite{KR14}. The full lattice QCD 
including the charm quarks in the see \cite{DDHH12} gives very accurate
prediction $\triangle B_{c}/\triangle B_{s}=1.166(79)$.
Taking PDG average value $\triangle B_{s}=48.6(1.8)$ MeV
one obtains $\triangle B_{c}=56.7(5.8)$ MeV, in good agreement
with our choice. Their original prediction is somewhat lower ($54\pm3$ MeV)
because of the different choice of $\triangle B_{s}$. The recent
free-form smearing lattice QCD \cite{WLW15} prediction $\triangle B_{c}=57.5(3)$ MeV
is slightly lower than our choice. On the other hand, the relativistic
quark model \cite{EFG11a}, which gives good fits to the meson spectrum,
predicts $\triangle B_{c}=61$ MeV. Note that these two
predictions are respectively $\approx0.5$ MeV and $\approx1$ MeV
lower than the  estimates for the $\Upsilon-\eta_{b}$
hyperfine splittings obtained using these approaches. If one defines
this splitting as the difference between averaged PDG masses $M(\Upsilon)=9460.30(26)$ MeV
and $M(\eta_{b})=9399.0(2.3)$ MeV, one obtains $M(\Upsilon)-M(\eta_{b})=61(2)$ MeV.
Subtracting 1 MeV from this value one again obtains $M(B_{c}^{*})-M(B_{c})\approx60$ MeV.

\begin{table}
\caption{\label{t4.5} Mass splittings and momenta of emitted photons (in MeV) for the ground-state
mesons.}

\begin{tabular}{lccc}
\hline\hline  
Transition  & $\triangle E$  & \quad & $k$\\
\hline 
$\rho^{0}\rightarrow\pi^{0}$  & $640$  & & $376$\\
$\rho^{+}\rightarrow\pi^{+}$  & $635$  & & $375$ \\
$\omega^{0}\rightarrow\pi^{0}$  & $647$  & & $379$ \\
$K^{*0}\rightarrow K^{0}$  & $398$  & & $310$ \\
$K^{*+}\rightarrow K^{+}$  & $399$  & & $310$ \\
$D^{*0}\rightarrow D^{0}$  & $142$  & & $137$ \\
$D^{*+}\rightarrow D^{+}$  & $140$  & & $135$ \\
$D_{s}^{*+}\rightarrow D_{s}^{+}$  & $144$  & & $139$ \\
$J/\psi\rightarrow\eta_{c}$  & $114$ &  & $112$ \\
$B^{*0}\rightarrow B^{0}$  & $46$  & & $46$ \\
$B^{*+}\rightarrow B^{+}$  & $46$  & & $46$ \\
$B_{s}^{*0}\rightarrow B_{s}^{0}$ & $48$  & & $48$ \\
$B_{c}^{*+}\rightarrow B_{c}^{+}$  & $60$  & & $60$ \\
$\Upsilon\rightarrow\eta_{b}$  & $61$  & & $61$ \\
\hline\hline  
\end{tabular}
\end{table}

\begin{table}
\caption{Mass splittings and momenta of emitted photons (in MeV) for the singly
heavy baryons.\label{t4.6}}
\begin{tabular}{lccclcc}
\hline\hline  
Transition  & $\triangle E$  & $k$ & $\quad$ & Transition  & $\triangle E$  & $k$\\
\hline 
$\Sigma_{c}^{*0}\rightarrow\Sigma_{c}^{0}$  & $65$  & $64$  &  & $\Sigma_{b}^{*-}\rightarrow\Sigma_{b}^{-}$  & $20$ & $20$\\
$\Sigma_{c}^{*+}\rightarrow\Sigma_{c}^{+}$  & $65$  & $64$  &  & $\Sigma_{b}^{*0}\rightarrow\Sigma_{b}^{0}$  & $20$ & $20$\\
$\Sigma_{c}^{*+}\rightarrow\Lambda_{c}^{+}$  & $231$  & $220$  &  & $\Sigma_{b}^{*0}\rightarrow\Lambda_{b}^{0}$  & $214$ & $210$\\
$\Sigma_{c}^{+}\rightarrow\Lambda_{c}^{+}$  & $166$  & $160$  &  & $\Sigma_{b}^{0}\rightarrow\Lambda_{b}^{0}$  & $193$ & $190$\\
$\Sigma_{c}^{*++}\rightarrow\Sigma_{c}^{++}$  & $65$  & $64$  &  & $\Sigma_{b}^{*+}\rightarrow\Sigma_{b}^{+}$  & $20$ & $20$\\
$\Xi_{c}^{*0}\rightarrow\Xi_{c}^{0}$  & $175$  & $169$ &  & $\Xi_{b}^{*-}\rightarrow\Xi_{b}^{-}$  & $160$ & $158$\\
$\Xi_{c}^{*0}\rightarrow\Xi_{c}^{\prime\,0}$  & $68$  & $67$ &  & $\Xi_{b}^{*-}\rightarrow\Xi_{b}^{\prime\,-}$ & $20$ & $20$\\
$\Xi_{c}^{\prime\,0}\rightarrow\Xi_{c}^{0}$  & $107$  & $105$ &  & $\Xi_{b}^{\prime\,-}\rightarrow\Xi_{b}^{-}$  & $140$ & $138$\\
$\Xi_{c}^{*+}\rightarrow\Xi_{c}^{+}$  & $178$  & $172$  &  & $\Xi_{b}^{*0}\rightarrow\Xi_{b}^{0}$  & $157$ & $155$\\
$\Xi_{c}^{*+}\rightarrow\Xi_{c}^{\prime\,+}$  & $70$  & $69$  &  & $\Xi_{b}^{*0}\rightarrow\Xi_{b}^{\prime\,0}$  & $20$ & $20$\\
$\Xi_{c}^{\prime\,+}\rightarrow\Xi_{c}^{+}$  & $108$  & $106$ &  & $\Xi_{b}^{\prime\,0}\rightarrow\Xi_{b}^{0}$  & $137$ & $135$\\
$\Omega_{c}^{*0}\rightarrow\Omega_{c}^{0}$  & $71$  & $70$ &  & $\Omega_{b}^{*-}\rightarrow\Omega_{b}^{-}$  & $20$ & $20$\\
\hline\hline  
\end{tabular}
\end{table}

Now we have all ingredients to estimate the momenta of the outgoing
photons necessary for the calculation of the M1 decay widths.
The results for the mesons and singly heavy baryons are presented
in Tables~\ref{t4.5}, \ref{t4.6}. Where available, the experimental
data of the hadron masses were used. When the experimental data for
the masses of the isospin multiplet members are absent, the isospin
symmetry is assumed. For the mass splittings $\triangle\Omega_{b}$
and $\triangle B_{c}$ the estimates obtained above are given. In the
case of doubly and triply heavy baryons the momenta of the emitted
photons and the hyperfine splittings practically coincide.
The corresponding results are given in Table~\ref{t4.4} (labeled
as AL1).

\section{Parameter setting and the magnetic properties of mesons}  \label{sec_mes}

In the previous paper \cite{S16a} we have proposed the improved version
of bag model suitable for the description of the magnetic properties
of light and heavy mesons. Now we want to extend our analysis to the
heavy baryon sector. However, before switching to this new field,
it could be useful in this context to discuss the magnetic properties of the mesons
once more. The utility of the meson sector for our investigation is
twofold. First, the existing experimental data may be used to determine
the free parameters of the model. Next, the comparison of the model
predictions with other data and with some most reliable estimates
obtained using other approaches may serve as a preliminary test of
the capability of the model to describe the magnetic properties of
other hadrons.

We think it could be useful to change slightly the procedure used
to fit the model parameters $C_{L}$ and $C_{H}$ necessary to set
the scale of the magnetic observables of the light ($u,$ $d$, or
$s$) and heavy ($c$ or $b$) quarks. We expect this new choice to
be more appropriate for the description of the magnetic properties
of heavy hadrons. In Ref.~\cite{S16a} parameter $C_{L}$ was adjusted
to reproduce exactly the value of the proton magnetic moment. Our
new choice is to set the scale factor $C_{L}$ to reproduce the decay
widths of the light mesons $\rho\rightarrow\pi\gamma$ and $\omega\rightarrow\pi\gamma$
simultaneously. From the 2017 PDG update of Review of Particle Physics
\cite{PDG17} we have

\begin{align}
\Gamma(\rho^{+}\rightarrow\pi^{+}\gamma)  & =68\pm7\,\mathrm{keV},\\
\Gamma(\rho^{0}\rightarrow\pi^{0}\gamma)  & =70\pm9\,\mathrm{keV},\\
\Gamma(\omega^{0}\rightarrow\pi^{0}\gamma)  & =713\pm26\,\mathrm{keV}.
\end{align}

Using relation inverse to Eq. (\ref{mbag 034}) the experimental values
of the corresponding M1 transition moments can be deduced:

\begin{align}
\mu(\rho^{+}\rightarrow\pi^{+}\gamma)  & =0.68\pm0.04\,\mathrm{\mu_{N}},\\
\mu(\rho^{0}\rightarrow\pi^{0}\gamma)  & =0.69\pm0.05\,\mathrm{\mu_{N}},\\
\mu(\omega^{0}\rightarrow\pi^{0}\gamma)  & =2.18\pm0.04\,\mathrm{\mu_{N}}.
\end{align}

Setting $C_{L}=1.52$ we obtain $\mu(\rho\rightarrow\pi\gamma)=0.72\,\mathrm{\mu_{N}}$,
and $\mu(\omega\rightarrow\pi\gamma)=2.15\,\mathrm{\mu_{N}}$, the
predictions within error bars compatible with the data. The scale
factor $C_{H}$ for the heavy quarks is now adjusted to reproduce
the PDG average of the partial decay width $\Gamma(J/\psi\rightarrow\eta_{c})=1.59\pm0.42$ keV.
The experimental M1 transition moment in this case is $\mu(J/\psi\rightarrow\eta_{c}\gamma)=0.64\pm0.09\,\mathrm{\mu_{N}}$.
We have set $C_{H}=0.94$. This choice leads to $\mu(J/\psi\rightarrow\eta_{c}\gamma)=0.634\,\mathrm{\mu_{N}}$.

\begin{table*}
\caption{\label{t5.1a} Magnetic moments of light mesons (in nuclear magnetons $\mu_{N}$).}

\begin{tabular}{lcccccccccc}
\hline\hline  
Particle  & Our  & NR  & RH  & DSE  & NJL  & NJL  & QCDSR  & Latt  & Latt  & Latt \\
 &  &  & \cite{BS13d}  & \cite{BM08}  & \cite{CBCT15}  & \cite{LCD15}  & \cite{AOS09}  & \cite{LMW08}  & \cite{OKLMM15a}  & \cite{SDE15} \\
\hline 
$\rho^{+}$  & $2.65$  & $2.79$  & $2.37$  & $2.54$  & $3.14$  & $2.54$  & $2.9(5)$  & $3.25(3)$  & $2.61(10)$  & $2.00(9)$ \\
$K^{*0}$  & $-0.229$  & $-0.32$  & $-0.183$  & $-0.26$  & $\cdot\cdot\cdot$  & $\cdot\cdot\cdot$  & $0.29(4)$  & $\cdot\cdot\cdot$  & $\cdot\cdot\cdot$  & $\cdot\cdot\cdot$ \\
$K^{*+}$  & $2.35$  & $2.47$  & $2.19$  & $2.23$  & $\cdot\cdot\cdot$  & $2.26$  & $2.1(4)$  & $2.81(1)$  & $\cdot\cdot\cdot$  & $\cdot\cdot\cdot$ \\
\hline\hline  
\end{tabular}
\end{table*}

\begin{table*}
\caption{\label{t5.2a} Magnetic moment of $\rho^{+}$ meson (in nuclear magnetons $\mu_{N}$).}

\begin{tabular}{lccccccccccc}
\hline\hline  
Our  & Exp  & EFT  & QCDSR  & DSE  & RQM  & RQM  & LFM  & LFM  & LFM  & LFM  & LFM \\
 & \cite{GS15a}  & \cite{DEGM14}  & \cite{S03a}  & \cite{RBGRW11}  & \cite{BS14}  & \cite{KPT18}  & \cite{MS02}  & \cite{J03}  & \cite{CJ04}  & \cite{MddF15}  & \cite{SD17a} \\
\hline 
$2.65$  & $2.6(6)$  & $2.71$  & $2.4(4)$  & $2.13$  & $2.68$  & $2.63$  & $2.86$  & $2.23$  & $2.34$  & $2.69$  & $2.51$ \\
\hline\hline  
\end{tabular}
\end{table*}

\begin{table*}
\caption{\label{t5.1b} Magnetic moments of light mesons in natural magnetons $(1/(2M_{i})$).}

\begin{tabular}{lcccccccccc}
\hline\hline  
Particle  & Our  & NR  & DSE  & NJL  & NJL  & QCDSR  & Latt  & Latt  & Latt  & Latt \\
 &  &  & \cite{BM08}  & \cite{CBCT15}  & \cite{LCD15}  & \cite{AOS09}  & \cite{LMW08}  & \cite{LST17}  & \cite{OKLMM15a}  & \cite{SDE15} \\
\hline 
$\rho^{+}$  & $2.17$  & $2.31$  & $2.01$  & $2.59$  & $2.09$  & $2.4(4)$  & $2.39(1)$  & $2.11(1)$  & $2.21(8)$  & $2.17(10)$ \\
$K^{*0}$  & $-0.218$  & $-0.31$  & $-0.26$  & $\cdot\cdot\cdot$  & $\cdot\cdot\cdot$  & $0.28(4)$  & $\cdot\cdot\cdot$  & $\cdot\cdot\cdot$  & $\cdot\cdot\cdot$  & $\cdot\cdot\cdot$ \\
$K^{*+}$  & $2.24$  & $2.35$  & $2.23$  & $\cdot\cdot\cdot$  & $2.21$  & $2.0(4)$  & $2.38(1)$  & $2.29(19)$  & $\cdot\cdot\cdot$  & $\cdot\cdot\cdot$ \\
\hline\hline  
\end{tabular}
\end{table*}

\begin{table*}
\caption{\label{t5.2b} Magnetic moment of $\rho^{+}$ meson in natural magnetons $(1/(2M_{i})$).}

\begin{tabular}{lccccccccccc}
\hline\hline  
Our  & Exp  & EFT  & QCDSR  & DSE  & RQM  & RQM  & LFM  & LFM  & LFM  & LFM  & LFM \\
 & \cite{GS15a}  & \cite{DEGM14}  & \cite{S03a}  & \cite{RBGRW11}  & \cite{BS14}  & \cite{KPT18}  & \cite{MS02}  & \cite{J03}  & \cite{CJ04}  & \cite{MddF15}  & \cite{SD17a} \\
\hline 
$2.17$  & $2.1(5)$  & $2.24$  & $2.0(3)$  & $2.11$  & $2.20$  & $2.16$  & $2.35$  & $1.83$  & $1.92$  & $2.21$  & $2.06$ \\
\hline\hline  
\end{tabular}
\end{table*}

\begin{table}
\caption{\label{t5.3} Magnetic moments of heavy mesons (in nuclear magnetons $\mu_{N}$).}

\begin{tabular}{lrrrr}
\hline\hline  
Particle  & Our  & NR  & BSLT  & NJL \\
 &  &  & \cite{L03b}  & \cite{LCD15} \\
\hline 
$D^{*0}$  & $-1.28$  & $-1.47$  & $\cdot\cdot\cdot$  & $\cdot\cdot\cdot$ \\
$D^{*+}$  & $1.13$  & $1.32$  & $\cdot\cdot\cdot$  & $1.16$ \\
$D_{s}^{*+}$  & $0.927$  & $1.00$  & $\cdot\cdot\cdot$  & $0.98$ \\
$B^{*0}$  & $-0.693$  & $-0.87$  & $\cdot\cdot\cdot$  & $\cdot\cdot\cdot$ \\
$B^{*+}$  & $1.56$  & $1.92$  & $\cdot\cdot\cdot$  & $1.47$ \\
$B_{s}^{*0}$  & $-0.513$  & $-0.55$  & $\cdot\cdot\cdot$  & $\cdot\cdot\cdot$ \\
$B_{c}^{*+}$  & $0.380$  & $0.45$  & $0.426$  & $\cdot\cdot\cdot$ \\
\hline\hline  
\end{tabular}
\end{table}

Now, the calculation of the magnetic observables is rather 
simple task. We only need to use the expressions presented in 
Sec.~\ref{sec_bag}. Let us begin with the magnetic moments of the mesons. The
results are presented in Tables~\ref{t5.1a}--\ref{t5.3}, where
they are compared with the estimates obtained using various other
approaches, such as the model based on relativistic Hamiltonian (RH),
various models based on the Dyson-Schwinger equation (DSE), the effective
field theory (EFT), standard and  light cone QCD sum rules (QCDSR), 
the relativistic quark model (RQM), 
various light front quark models (LFM), lattice QCD simulations
(Latt), two variants of the Nambu-Jona-Lasinio model (NJL), and the
formalism based on the Blankenbecler-Sugar equation (BSLT). The label
NR stands for the simple nonrelativistic quark model. The 
input values for the quark magnetic moments in this model  $\mu_{u}=1.86~\mu_{N}$,
$\mu_{d}=-0.93~\mu_{N}$, $\mu_{s}=-0.61~\mu_{N}$, $\mu_{c}=0.39~\mu_{N}$,
and $\mu_{b}=-0.06~\mu_{N}$ were taken from Ref.~\cite{FLNC81}. Note that
in this approach, as in almost all nonrelativistic quark models \cite{MNSMMT93},
the values of the magnetic moments of the light quarks are adjusted to
reproduce the magnetic moments of light baryons. The adopted in Ref.~\cite{FLNC81}
values $\mu_{c}$ and $\mu_{b}$ are about $10\%$ smaller than the
corresponding values deduced in Ref.~\cite{KL08} using nonrelativistic framework from the hyperfine splittings of $\Sigma_{c}$ and $\Sigma_{b}$ baryons.

The experimental data for the magnetic moments of the mesons are so far absent
(the possibilities to measure the magnetic moments of another short living particles, 
i.e. heavy baryons, are discussed in the Refs.~\cite{B16,FeA17}). 
Nevertheless, there exist indirect, to some extent model dependent,
estimate of the $\rho^{+}$ magnetic moment based on the analysis
of preliminary experimental data from the BaBar Collaboration. We have
included this result in Tables~\ref{t5.2a}, \ref{t5.2b} under the label Exp.

In this paper we prefer to express magnetic moments in the units of
nuclear magneton ($\mu_{N}$). Nevertheless, it is sometimes convenient
to express them in so-called natural magnetons $1/(2M_{i})$, where
$M$$_{i}$ is the mass of the particle under consideration. For instance,
it is a frequent practice in the analysis of the magnetic
moments of light mesons. For the discussion about the utility of natural magnetons
in the lattice QCD calculations see Ref.~\cite{PSTWCDO17}. We present,
the predictions for the magnetic moments of light mesons expressed
in nuclear magnetons in Tables~\ref{t5.1a}, \ref{t5.2a} and expressed
in natural magnetons in Tables~\ref{t5.1b}, \ref{t5.2b}. The differences
between the two approaches is visible, for example, in the comparison
of our results for the $\rho^{+}$ meson with the corresponding prediction
obtained using lattice QCD with three degenerate flavors of dynamical
quark tuned to approximately the physical strange quark mass \cite{SDE15}. 
We see that for the quantities expressed in natural magnetons the
coincidence is excellent, but this is not the case when the nuclear
magnetons are used. The similar behavior we observe for the predictions
obtained in Ref.~\cite{RBGRW11} using Dyson-Schwinger formalism.
On the other hand, the comparison of our predictions for the $\rho^{+}$
meson with the results obtained using full lattice QCD near physical
masses \cite{OKLMM15a} shows good agreement, no matter in which units
the results are expressed. With a few exceptions the agreement with
the results obtained using other approaches is also good. Note the
excellent agreement between our results and the predictions obtained
using instant-form relativistic quark models \cite{BS14,KPT18}. The
agreement with the predictions obtained in Ref.~\cite{MddF15} in
the framework of the light-front quark model is also very good. Our
estimates for $K^{*+}$ meson are lower than quenched lattice QCD
predictions \cite{LMW08}, larger than the predictions obtained using
approach based on a relativistic Hamiltonian \cite{BS13d}, and approximately
$5\%$ larger than the results obtained using Dyson-Schwinger \cite{BM08}
and NJL \cite{LCD15} frameworks. In the case of $K^{*0}$ meson almost
all approaches predict small and negative ($\sim-0.2\mu_{N}$) magnetic
moment value. In the heavy meson sector there are only few other predictions
to compare our results with. The overall agreement is good, and this
could be an indication that all these approaches provide more or less
reasonable predictions.

The inclusion of the results obtained in the framework
of simple nonrelativistic quark model (NR) needs some comment. In
all cases these estimates are larger in magnitude than ours. However,
we see, that these naive predictions in all likelihood are not so
bad as would be expected. More specifically, the whole pattern provided
by NR approach resembles the predictions obtained using other approaches.
Of course, in the case of light mesons there is little benefit of
it. In the cases when various subtle effects come into the game the NR
approach plausibly also becomes useless. But sometimes for heavy
hadrons the NR estimates could serve as a useful guide if one wants
to gain some preliminary insight how the things look like. For example,
the NR results for the magnetic moments of neutral heavy mesons $D^{*0}$,
$B^{*0}$, and $B_{s}^{*0}$, or magnetic transition moments of triply heavy baryons, are so far the only estimates to compare
our predictions with.

\begin{table}
\caption{\label{t5.4a} Transition moments of light vector mesons (in nuclear
magnetons). Only absolute values $|\mu|$ are presented.}
\begin{tabular}{ccrrrrcc}
\hline\hline  
Transition  & Exp  & Our  & EQM  & SRPM  & RPM  & RPM  & NR \\
 & \cite{PDG17}  &  & \cite{HI82}  & \cite{GI85}  & \cite{BD94}  & \cite{JPT02}  & \\
\hline 
$\rho^{0}\rightarrow\pi^{0}$  & $0.69(5)$  & $0.720$  & $0.72$  & $0.69$  & $0.70$  & $0.68$  & $0.93$ \\
$\rho^{+}\rightarrow\pi^{+}$  & $0.68(4)$  & $0.720$  & $0.72$  & $0.69$  & $0.70$  & $0.68$  & $0.93$ \\
$\omega^{0}\rightarrow\pi^{0}$  & $2.18(4$)  & $2.15$  & $2.14$  & $2.07$  & $2.07$  & $2.02$  & $2.79$ \\
$K^{*0}\rightarrow K^{0}$  & $1.19(5)$  & $1.27$  & $1.25$  & $1.20$  & $1.35$  & $1.17$  & $1.54$ \\
$K^{*+}\rightarrow K^{+}$  & $0.78(4)$  & $0.905$  & $0.90$  & $0.91$  & $0.79$  & $0.84$  & $1.25$ \\
\hline \hline 
\end{tabular}
\end{table}

\begin{table}
\caption{\label{t5.4b} Transition moments of heavy vector mesons (in nuclear
magnetons). Only absolute values $|\mu|$ are presented.}
\begin{tabular}{lcrrrcc}
\hline\hline  
Transition  & Exp  & Our  & SRPM  & RPM  & RPM  & NR \\
 & \cite{PDG17}  &  & \cite{GI85}  & \cite{BD94}  & \cite{JPT02}  &  \\
\hline  
$D^{*0}\rightarrow D^{0}$  & $\cdot\cdot\cdot$  & $1.80$  & $1.78$  & $1.98$  & $1.64$  & $2.25$ \\
$D^{*+}\rightarrow D^{+}$  & $0.44(6)$  & $0.418$  & $0.35$  & $0.39$  & $0.49$  & $0.54$ \\
$D_{s}^{*+}\rightarrow D_{s}^{+}$  & $\cdot\cdot\cdot$  & $0.240$  & $0.13$  & $0.23$  & $0.25$  & $0.22$ \\
$J/\psi\rightarrow\eta_{c}$  & $0.64(9)$  & $0.634$  & $0.69$  & $\cdot\cdot\cdot$  & $\cdot\cdot\cdot$  & $0.78$ \\
$B^{*0}\rightarrow B^{0}$  & $\cdot\cdot\cdot$  & $0.783$  & $0.78$  & $0.89$  & $0.72$  & $0.99$ \\
$B^{*+}\rightarrow B^{+}$  & $\cdot\cdot\cdot$  & $1.39$  & $1.37$  & $1.59$  & $1.39$  & $1.80$ \\
$B_{s}^{*0}\rightarrow B_{s}^{0}$  & $\cdot\cdot\cdot$  & $0.614$  & $0.55$  & $0.68$  & $0.47$  & $0.67$ \\
$B_{c}^{*+}\rightarrow B_{c}^{+}$  & $\cdot\cdot\cdot$  & $0.257$  & $\cdot\cdot\cdot$  & $0.32$  & $0.33$  & $0.33$ \\
$\Upsilon\rightarrow\eta_{b}$  & $\cdot\cdot\cdot$  & $0.118$  & $0.13$  & $\cdot\cdot\cdot$  & $\cdot\cdot\cdot$  & $0.12$ \\
\hline \hline 
\end{tabular}
\end{table}

\begin{table*}
\caption{\label{t5.5} M1 decay widths of light vector mesons (in keV).}
\begin{tabular}{ccccccccccccc}
\hline\hline  
Transition  & Exp  & Our  &  & QM  & PM(AL1)  & PM(AP1)  & RPM  & RPM   & LFM  & LFM  & \\
 & \cite{PDG17}  &  &  & \cite{KX92}  & \cite{BSG02}  & \cite{BSG02}  & \cite{BD94}  & \cite{JPT02}   & \cite{J99}  & \cite{CJ99}  & \\
\hline 
$\rho^{0}\rightarrow\pi^{0}$  & $70(9)$  & $76$  &  & $74.6$  & $48.7$  & $60.6$  & $69.0$  & $65.45$   & $81.3$  & $69$  & \\
$\rho^{+}\rightarrow\pi^{+}$  & $68(7)$  & $76$  &  & $74.6$  & $48.5$  & $60.4$  & $68.3$  & $64.8$   & $81.3$  & $69$  & \\
$\omega^{0}\rightarrow\pi^{0}$  & $713(26)$  & $694$  &  & $716$  & $459$  & $572$  & $645$  & $613$   & $\cdot\cdot\cdot$  & $667$  & \\
$K^{*0}\rightarrow K^{0}$  & $116(10)$  & $134$  &  & $114$  & $98.3$  & $116$  & $150$  & $112$    & $124$  & $117$  & \\
$K^{*+}\rightarrow K^{+}$  & $50(5)$  & $68$  &  & $82.3$  & $79.1$  & $104$  & $51.1$  & $58.1$   & $54.4$  & $71.4$  & \\
\hline\hline  
\end{tabular}
\end{table*}

\begin{table*}
\caption{\label{t5.6a} M1 decay widths of heavy vector mesons (in keV).}

\begin{tabular}{lccccccccccc}
\hline\hline  
Transition  & Exp  & Our  & QM  & PM(AL1)  & PM(AP1)  & PM & PM & HQET  & $\chi$PT  & Latt  & Latt \\

 & \cite{PDG17}  &  & \cite{KX92}  & \cite{BSG02}  & \cite{BSG02}  & \cite{BEH03}  & \cite{CS05}  & \cite{CH14}  & \cite{HM06b}  & \cite{BH11}  & \cite{DDKL14} \\
\hline
$D^{*0}\rightarrow D^{0}$  & $\cdot\cdot\cdot$  & $22.9$ & $21.7$ & $33.6$ & $44.7$ & $33.5$ & $32$ & $22.7(2.2)$ & $26(6)$  & $27(14)$  & $\cdot\cdot\cdot$ \\
$D^{*+}\rightarrow D^{+}$  & $1.34(36)$  & $1.19$  & $1.42$  & $2.48$  & $3.58$  & $1.63$ & $1.8$ & $0.9(3)$  & $1.54(35)$  & $0.8(7)$  & $\cdot\cdot\cdot$ \\
$D_{s}^{*+}\rightarrow D_{s}^{+}$  & $\cdot\cdot\cdot$  & $0.430$  & $0.21$  & $0.23$  & $0.31$  & \emph{$0.43$} & $0.2$ & $\cdot\cdot\cdot$  & $\cdot\cdot\cdot$  & $\cdot\cdot\cdot$  & $0.066(26)$ \\
$J/\psi\rightarrow\eta_{c}$  & $1.59(42)$  & $1.56$  & $1.27$  & $1.85$  & $1.87$  & $\cdot\cdot\cdot$  & $\cdot\cdot\cdot$  & $\cdot\cdot\cdot$  & $\cdot\cdot\cdot$  & $\cdot\cdot\cdot$  & $\cdot\cdot\cdot$ \\
$B^{*0}\rightarrow B^{0}$  & $\cdot\cdot\cdot$  & $0.165$  & $\cdot\cdot\cdot$  & $0.28$  & $0.36$  & $0.24$ & $\cdot\cdot\cdot$  & $0.148(20)$  & $\cdot\cdot\cdot$  & $\cdot\cdot\cdot$  & $\cdot\cdot\cdot$ \\
$B^{*+}\rightarrow B^{+}$  & $\cdot\cdot\cdot$  & $0.520$  & $\cdot\cdot\cdot$  & $0.97$  & $1.20$  & $0.78$ & $\cdot\cdot\cdot$  & $0.468(75)$  & $\cdot\cdot\cdot$  & $\cdot\cdot\cdot$  & $\cdot\cdot\cdot$ \\
$B_{s}^{*0}\rightarrow B_{s}^{0}$  & $\cdot\cdot\cdot$  & $0.115$  & $\cdot\cdot\cdot$  & $\cdot\cdot\cdot$  & $\cdot\cdot\cdot$  & $0.15$ & $\cdot\cdot\cdot$  & $\cdot\cdot\cdot$  & $\cdot\cdot\cdot$  & $\cdot\cdot\cdot$  & $\cdot\cdot\cdot$ \\
$B_{c}^{*+}\rightarrow B_{c}^{+}$  & $\cdot\cdot\cdot$  & $0.039$  & $\cdot\cdot\cdot$  & $\cdot\cdot\cdot$  & $\cdot\cdot\cdot$  & $\cdot\cdot\cdot$  & $\cdot\cdot\cdot$  & $\cdot\cdot\cdot$  & $\cdot\cdot\cdot$  & $\cdot\cdot\cdot$  & $\cdot\cdot\cdot$ \\
\hline \hline 
\end{tabular}
\end{table*}

\begin{table*}
\caption{\label{t5.6b} M1 decay widths of heavy vector mesons (in keV).}

\begin{tabular}{lccccccccccc}
\hline\hline  
Transition  & Exp & Our  & RPM & RPM & RPM  & RQM  & RQM  & BSLT  & QCDSR  & LFM & LFM\\
 & \cite{PDG17}  &  & \cite{BD94}  & \cite{JPT02}  & \cite{PDKPB16,PDKPB17}  & \cite{IV95}  & \cite{EFG02b,EFG03a}  & \cite{LNR00,L03b}  & \cite{ADIP96}  & \cite{J96b}  & \cite{C07,CJ09} \\
\hline 
$D^{*0}\rightarrow D^{0}$  & $\cdot\cdot\cdot$  & $22.9$  & $28.4$  & $19.5$ & $26.5$  & $23.6$  & $11.5$  & $1.25$  & $14.4$  & $21.7$ & $20.0$\\
$D^{*+}\rightarrow D^{+}$  & $1.34(36)$  & $1.19$  & $1.08$ & $1.63$ & $0.932$  & $0.950$  & $1.04$  & $1.10$  & $1.5$  & $0.56$ & $0.90$\\
$D_{s}^{*+}\rightarrow D_{s}^{+}$  & $\cdot\cdot\cdot$  & $0.430$  & $0.38$ & $0.44$ & $0.213$  & $\cdot\cdot\cdot$  & $0.19$  & $0.337$  & $\cdot\cdot\cdot$  & $\cdot\cdot\cdot$  & $0.18$\\
$J/\psi\rightarrow\eta_{c}$  & $1.59(42)$  & $1.56$  & $\cdot\cdot\cdot$  & $\cdot\cdot\cdot$  & $2.32$  & $\cdot\cdot\cdot$  & $1.05$  & $1.25$  & $\cdot\cdot\cdot$  & $\cdot\cdot\cdot$  & $1.69$\\
$B^{*0}\rightarrow B^{0}$  & $\cdot\cdot\cdot$  & $0.165$  & $0.21$ & $0.14$ & $0.181$  & $0.131$  & $0.070$  & $0.0096$  & $0.16$  & $0142$ & $0.13$\\
$B^{*+}\rightarrow B^{+}$  & $\cdot\cdot\cdot$  & $0.520$  & $0.67$ & $0.52$ & $0.577$  & $0.401$  & $0.19$  & $0.0674$  & $0.63$  & $0.43$ & $0.40$\\
$B_{s}^{*0}\rightarrow B_{s}^{0}$  & $\cdot\cdot\cdot$  & $0.115$  & $0.13$ & $0.06$ & $0.119$  & $\cdot\cdot\cdot$  & $0.054$  & $0.148$  & $\cdot\cdot\cdot$  & $\cdot\cdot\cdot$  & $0.068$\\
$B_{c}^{*+}\rightarrow B_{c}^{+}$  & $\cdot\cdot\cdot$  & $0.039$  & $0.020$ & $0.030$ & $0.023$  & $\cdot\cdot\cdot$  & $0.033$  & $0.034$  & $\cdot\cdot\cdot$  & $\cdot\cdot\cdot$  & $0.0224$\\
$\Upsilon\rightarrow\eta_{b}$  & $\cdot\cdot\cdot$  & $0.0087$  & $\cdot\cdot\cdot$  & $\cdot\cdot\cdot$  & $0.011$  & $\cdot\cdot\cdot$  & $0.0058$  & $0.0077$  & $\cdot\cdot\cdot$  & $\cdot\cdot\cdot$  & $0.045$\\
\hline\hline  
\end{tabular}
\end{table*}

\begin{table*}
\caption{\label{t5.7} $J/\psi\rightarrow\eta_{c}\,\gamma$ and $\Upsilon\rightarrow\eta_{b}\,\gamma$
decay widths (in keV).}

\begin{tabular}{ccccccccccccc}
\hline\hline  
 Transition & Exp & Our  & QM  & pNRQCD  & PM  & PM  & PM  & PM  & SRPM  & SRPM  & RPM  & SIE \\
 & \cite{PDG17} &  & \cite{BeA04b}  & \cite{PS13}  & \cite{ZSG91}  & \cite{SOEF16}  & \cite{BVM11}  & \cite{DLGZ17a,DLGZ17b}  & \cite{RR07}  & \cite{GM15}  & \cite{BGS17}  & \cite{ADMNS07} \\
\hline 
$J/\psi\rightarrow\eta_{c}$ &  $1.59(42)$  & $1.56$  & $1.96$  & $2.12(4)$  & $1.8$  & $\cdot\cdot\cdot$  & $3.28$  & $2.44$  & $1.8$  & $\cdot\cdot\cdot$  & $\cdot\cdot\cdot$  & $\cdot\cdot\cdot$ \\
$\Upsilon\rightarrow\eta_{b}$ & $\cdot\cdot\cdot$ & $0.0087$  & $0.00895$  & $0.0152(5)$  & $0.004$  & $0.0093$  & $0.0154$  & $0.01$  & $0.001$  & $0.01$  & $0.0031$  & $0.010$\\
\hline\hline  
\end{tabular}
\end{table*}

\begin{table*}
\caption{\label{t5.8}  $J/\psi\rightarrow\eta_{c}\,\gamma$ decay widths (in keV).}

\begin{tabular}{cccccccccccccc}
\hline\hline  
Exp & Our  & PM  & PM  & SRPM  & RQM  & QCDSR  & QCDSR  & QCDSR  & QCDSR  & Latt  & Latt  & Latt  & Latt\\
\cite{PDG17} & & \cite{EGMR08}  & \cite{CYC12}  & \cite{BGS05}  & \cite{DS87}  & \cite{A84}  & \cite{S80d}  & \cite{BR85b}  & \cite{V08}  & \cite{DET09}  & \cite{CeA11b}  & \cite{DeA12a}  & \cite{BS13c} \\
\hline 
 $1.59(42)$ & $1.56$  & $2.85$  & $2.2$  & $2.4$  & $1.05$  & $2.1(4)$  & $2.7(5)$  & $2.6(5)$  & $2.9$  & $2.51(8)$  & $2.84(6)$  & $2.49(19)$  & $2.64(14)$\\
\hline\hline  
\end{tabular}
\end{table*}

\begin{table}
\caption{\label{t5.9}  $B_{c}^{*+}\rightarrow B_{c}^{+}\gamma$ decay widths (in keV).}
\begin{tabular}{ccccccccc}
\hline\hline  
Our  & PM  & PM  & PM  & PM  & SRPM  & RQM  & RQM  & BSE\\
 & \cite{EQ94}  & \cite{GKLT95a}  & \cite{F99d}  & \cite{MBV17a}  & \cite{G04}  & \cite{NW00}  & \cite{MBV17b}  & \cite{ASV05} \\
\hline 
$0.039$  & $0.135$  & $0.060$  & $0.059$  & $0.058$  & $0.080$  & $0.050$  & $0.0185$  & $0.019$\\
\hline\hline  
\end{tabular}
\end{table}

\begin{table}
\caption{\label{t5.10} Rescaled results for decay widths $\Gamma(\Upsilon\rightarrow\eta_{b}\,\gamma)$
(in keV).}
\begin{tabular}{ccccccc}
\hline\hline  
 & LFM  & PM  & PM  & BSLT  & SRPM  & pNRQCD \\
 & \cite{C07}  & \cite{ZSG91}  & \cite{BVM11}  & \cite{L03b}  & \cite{RR07}  & \cite{PS13} \\
\hline 
Old  & 0.045  & 0.004  & 0.0154  & 0.0077  & 0.001  & 0.0152\\
New  & 0.014  & 0.008  & 0.0106  & 0.0094  & 0.004  & 0.0101\\
\hline\hline  
\end{tabular}
\end{table}

Now we proceed with the analysis of the magnetic dipole transitions.
The calculated M1 transition moments of ground-state
vector mesons are presented in Tables~\ref{t5.4a} and \ref{t5.4b}. These
results were used as an input to calculate M1 decay widths (presented
in Tables~\ref{t5.5}--\ref{t5.9}). We compare
our predictions with the estimates obtained using nonrelativistic
quark model (QM), extended quark model (EQM), potential nonrelativistic QCD (pNRQCD), 
various potential models (PM), light front quark
models (LFM), relativistic quark models (RQM), semirelativistic potential
model (SRPM), relativistic potential models (RPM), formalism
based on the Bethe-Salpeter equation (BSE), Blankenbecler-Sugar equation
(BSLT), heavy quark effective theory (HQET), chiral perturbation theory
($\chi$PT), usual and light cone QCD sum rules (QCDSR), method based
on the spectral integral equations (SIE), and lattice QCD (Latt).
The experimental data are from the 1917 update of Review of Particle
Physics \cite{PDG17}, or are derived from.

From Table~\ref{t5.4a} we see that our results for the light mesons
are practically indistinguishable from the predictions obtained in
the extended quark model of Hayne and Isgur \cite{HI82}, probably
the best predictions possible in the quark model framework. For the
$\rho\rightarrow\pi\gamma$ and $\omega\rightarrow\pi\gamma$ transitions
this is the only way to obtain satisfactory fit simultaneously. In
the case of $K$ mesons our predictions are somewhat larger than experimental
values, and the predictions obtained in the extended quark model also
suffer from this drawback. More detailed comparison shows that in
the light meson sector the results obtained in our model are
of similar quality as the predictions obtained using other approaches. The lattice QCD for the time being
likely has some trouble in the describing the magnetic
transition moments of light mesons. The full lattice QCD prediction for
the $\rho\rightarrow\pi\gamma$ transition moment obtained near physical
value of $m_{\pi}=156$ MeV \cite{OKLMM15b} is lower by
about $33\%$ than the experimental estimate. This is a serious contradiction
between experimental data and the theoretical prediction obtained using
method expected to provide sufficiently accurate results. The situation
looks somewhat strange because similar calculation of $\rho^{+}$
magnetic moment \cite{OKLMM15a} seems to provide quite reasonable
prediction consistent with other recent estimates, including ours.

The comparison of our results for the transition moments of heavy mesons
(Table~\ref{t5.4b}) with the predictions obtained using semirelativistic
and relativistic potential models shows a good agreement in almost
all cases. The exception is the decay $B_{c}^{*+}\rightarrow B_{c}^{+}\gamma$.
In this case other predictions are about $25\%$
larger than ours.

Somewhat more complicated picture emerges when we compare our predictions
for the decay widths. The agreement with the results obtained
within the framework of RPM \cite{BD94,JPT02,PDKPB16,PDKPB17} remains
good. But the predictions obtained using other approaches are varied.
The lattice QCD predictions for the decay widths of heavy-light
mesons (see Table~\ref{t5.6a}) suffer from large uncertainties. Moreover,
the prediction for the decay $D_{s}^{*+}\rightarrow D_{s}^{+}$ is
order of magnitude smaller than other estimates.

Often the reason of differences in the calculated decay widths is
the differences in photon energies used in calculations. The model
dependence of the transition moments sometimes also plays his role.
In order to illustrate the dependence of the potential model results
on the particular potential we included in Tables~\ref{t5.5}, \ref{t5.6a}
the predictions obtained in Ref.~\cite{BSG02} using two different
potentials (AL1 and AP1). In the light meson sector both these variants
give the decay widths lower than ours, the $K^{*+}\rightarrow K^{+}$transition
being an exception. For heavy mesons the tendency becomes opposite,
and these predictions are, as a rule, larger than ours. On the other
hand, the predictions obtained using relativistic quark model \cite{EFG02b,EFG03a}
are substantially lower than our results. BSLT approach \cite{LNR00,L03b}
also gives lower predictions, however, in the case of doubly heavy
mesons the estimates given by this approach become close to our results.
Our prediction for the transition $\Upsilon\rightarrow\eta_{b}\gamma$
practically coincide with the typical nonrelativistic result 
$\Gamma(\Upsilon\rightarrow\eta_{b}\gamma)=0.00895$ keV
\cite{BeA04b}. This result seems to differ from the 
pNRQCD prediction $\Gamma(\Upsilon\rightarrow\eta_{b}\gamma)=0.0152$ keV.
However, the latter estimate was obtained using old experimental data
$\Delta M=70$ MeV for the $\Upsilon\rightarrow\eta_{b}$
hyperfine splitting. After simple rescaling one obtains 
$\Gamma(\Upsilon\rightarrow\eta_{b}\gamma)=0.0101$ keV,
much closer to the usual nonrelativistic result. The comparison of the
rescaled pNRQCD prediction with our result enables us to estimate
the accuracy of the decay width obtained in our model.
We conclude that our model predicts bottonium decay widths
with the accuracy $\approx16\%$, and, consequently, the possible
uncertainty for the magnetic moment of the bottom quark is approximately 8\%. 

The pNRQCD result \cite{PS13} for the bottonium decay width is not
the only one that requires rescaling. For example, such are also predictions
obtained in Refs.~\cite{C07,ZSG91,BVM11,RR07,L03b}. For illustration,
we have listed the corresponding rescaled values in Table~\ref{t5.10}.
From this table we see that new, rescaled predictions are essentially
closer to our result. In addition, our prediction for the $\Upsilon\rightarrow\eta_{b}\gamma$
decay is close to the estimates obtained in Refs.~\cite{SOEF16,DLGZ17b}
using potential models, semirelativistic potential model (Ref.~\cite{GM15}),
and the approach based on the spectral integral equations (Ref.~\cite{ADMNS07}).

Our result for the decay width $\Gamma(J/\psi\rightarrow\eta_{c}\gamma)$
is not the prediction. It is the fit to the PDG average of the partial
decay width $\Gamma(J/\psi\rightarrow\eta_{c}\gamma)=1.59\pm0.42$ keV.
Some approaches (such as the quark model used in Ref.~\cite{KX92},
relativistic quark models \cite{EFG02b,DS87}, and the model based on BSLT framework
\cite{L03b}) give lower predictions, other (light front quark model
\cite{C07}, potential models \cite{ZSG91,BSG02}, and one variant
of semirelativistic potential model \cite{RR07}) give similar or
slightly larger results. On the other hand, many variants of potential
model \cite{EGMR08,BVM11,CYC12,DLGZ17a}, semirelativistic potential
model \cite{BGS05}, relativistic potential model \cite{PDKPB16},
and lattice QCD \cite{DET09,CeA11b,DeA12a,BS13c} predict substantially
larger decay widths. In this context our choice looks like some kind of compromise.

In summary, we have chosen the free parameter $C_{L}$ to reproduce
the experimental values of the decay widths $\rho\rightarrow\pi\gamma$
and $\omega\rightarrow\pi\gamma$ simultaneously. Another parameter
$C_{H}$ was chosen to reproduce the PDG average of the decay width
$\Gamma(J/\psi\rightarrow\eta_{c}\gamma)$. Predictions for the static
magnetic moments seem to be more or less reliable and are similar
to almost all other predictions obtained using various approaches
including recent lattice QCD results. 
Since the transition $J/\psi\rightarrow\eta_{c}\gamma$ was chosen
for the fit of the model parameter $C_{H}$, the only experimental
value we can use to test our approach in the heavy meson sector remains
the transition moment of the $D^{*+}\rightarrow D^{+}\gamma$ decay.
We see from Tables~\ref{t5.4b}, \ref{t5.6a} that the agreement is
quite good. Thus the predictions of our model for M1
transitions of the ground-state heavy mesons seem to be consistent with
available data. The problem is that only two decay widths are measured,
and the uncertainties in data are large (15--25\%). One more field
in which the quality of the predictions can be tested is the nonrelativistic
limit. For this purpose the decay $\Upsilon\rightarrow\eta_{b}\gamma$
can be used. Strictly speaking, this is the exclusive quark-antiquark
system in which the nonrelativistic approximation should be valid. Our
prediction for the transition moment $\mu(\Upsilon\rightarrow\eta_{b})$
is very close to the NR result ($0.12~\mu_{N}$). 
Prediction for the decay width $\Gamma(\Upsilon\rightarrow\eta_{b}\gamma)$
approximately coincides with the typical nonrelativistic estimate,
indicating the proper behavior of the model in the nonrelativistic
limit.

\section{Magnetic properties of heavy baryons}  \label{sec_bar-1}

\subsection{Magnetic moments}

We begin our analysis of the magnetic properties of
heavy baryons with the calculation of magnetic moments. The results
are presented in Tables~\ref{t6.1}--\ref{t6.11}. The mixing of
the baryons ($\Xi_{c}$, $\Xi_{c}^{\prime}$), ($\Xi_{b}$, $\Xi_{b}^{\prime}$), 
($\Xi_{cb}$, $\Xi_{cb}^{\prime}$), and ($\Omega_{cb}$, $\Omega_{cb}^{\prime}$) 
is taken into account. For the transition moments the static ($k=0)$
values are given.

\begin{table*} [p]
\caption{\label{t6.1} Magnetic moments of $\Sigma_{c}$ and $\Omega_{c}$ baryons (in nuclear
magnetons $\mu_{N}$).}
\begin{tabular}{ccccccccccccc}
\hline\hline 
 & Our  & NR  & PM(BD)  & PM(AL1)  & PM  & Hyp  & EM\&C  & $\chi$QM  & RPM  & RQM  & QCDSR  & Latt\\
 &  &  & \cite{S96d}  & \cite{S96d}  & \cite{MPV08}  & \cite{PRV08a}  & \cite{KDV05}  & \cite{SDCG10}  & \cite{BD83a}  & \cite{FGIKLNP06}  & \cite{AOS02b,ABS15a,AOS02a}  & \cite{CEIOT14b,CEOT15}\\
\hline 
$\Sigma_{c}^{0}$  & $-1.31$  & $-1.37$  & $-1.35$  & $-1.44$  & $-1.16$  & $-1.01$  & $-1.17$  & $-1.60$  & $-1.39$  & $-1.04$  & $-1.50(35)$  & $-1.12(20)$\\
$\Sigma_{c}^{+}$  & $0.487$  & $0.49$  & $0.507$  & $0.548$  & $0.392$  & $0.500$  & $0.63$  & $0.30$  & $0.525$  & $0.36$  & $0.50(15)$  & $\cdot\cdot\cdot$ \\
$\Lambda_{c}^{+}$  & $0.335$  & $0.39$  & $0.335$  & $0.341$  & $0.408$  & $0.384$  & $0.37$  & $0.392$  & $0.341$  & $0.42$  & $0.40(5)$  & $\cdot\cdot\cdot$ \\
$\Sigma_{c}^{+}\rightarrow\Lambda_{c}^{+}$  & $-1.56$  & $-1.61$  & $\cdot\cdot\cdot$  & $\cdot\cdot\cdot$  & $\cdot\cdot\cdot$  & $\cdot\cdot\cdot$  & $-1.51$  & $1.56$  & $\cdot\cdot\cdot$  & $\cdot\cdot\cdot$  & $-1.5(4)$  & $\cdot\cdot\cdot$ \\
$\Sigma_{c}^{++}$  & $2.28$  & $2.35$  & $2.36$  & $2.53$  & $1.95$  & $2.27$  & $2.18$  & $2.20$  & $2.44$  & $1.76$  & $2.4(5)$  & $2.03(39)$\\
$\Omega_{c}^{0}$  & $-0.950$  & $-0.94$  & $-0.806$  & $-0.835$  & $-0.950$  & $-0.958$  & $-0.92$  & $-0.90$  & $-0.85$  & $-0.85$  & $-0.9(2)$  & $-0.668(31)$\\
\hline\hline  
\end{tabular}
\end{table*}

\begin{table*} [p]
\caption{\label{t6.2} Magnetic moments of $\Sigma_{b}$ and $\Omega_{b}$ baryons (in nuclear
magnetons $\mu_{N}$).}
\begin{tabular}{cccccccccccc}
\hline\hline 
 & Our  & NR  & PM(BD)  & PM(AL1)  & PM  & PM  & Hyp  & EM\&C  & RPM  & RQM  & QCDSR\\
 &  &  & \cite{S96d}  & \cite{S96d}  & \cite{MTV16}  & \cite{MPV08}  & \cite{PRV08a}  & \cite{DKV13}  & \cite{BD83a}  & \cite{FGIKLNP06}  & \cite{AOS02b,ABS15a,AOS02a}\\
\hline 
$\Sigma_{b}^{-}$  & $-1.15$  & $-1.22$  & $-1.22$  & $-1.31$  & $-1.16$  & $-1.03$  & $-1.05$  & $-1.11$  & $-1.26$  & $-1.01$  & $-1.3(3)$ \\
$\Sigma_{b}^{0}$  & $0.603$  & $0.64$  & $0.639$  & $0.682$  & $0.609$  & $0.547$  & $0.591$  & $0.53$  & $0.659$  & $0.53$  & $0.6(2)$ \\
$\Lambda_{b}^{0}$  & $-0.060$  & $-0.06$  & $-0.059$  & $-0.060$  & $-0.063$  & $-0.064$  & $-0.064$  & $-0.060$  & $\cdot\cdot\cdot$  & $-0.06$  & $-0.18(5)$\\
$\Sigma_{b}^{0}\rightarrow\Lambda_{b}^{0}$  & $-1.53$  & $-1.61$  & $\cdot\cdot\cdot$  & $\cdot\cdot\cdot$  & $-1.55$  & $\cdot\cdot\cdot$  & $\cdot\cdot\cdot$  & $-1.54$  & $\cdot\cdot\cdot$  & $\cdot\cdot\cdot$  & $-1.6(4)$ \\
$\Sigma_{b}^{+}$  & $2.25$  & $2.50$  & $2.50$  & $2.67$  & $2.38$  & $2.12$  & $2.23$  & $2.17$  & $2.58$  & $2.07$  & $2.4(5)$ \\
$\Omega_{b}^{-}$  & $-0.806$  & $-0.79$  & $-0.676$  & $-0.703$  & $\cdot\cdot\cdot$  & $-0.805$  & $-0.958$  & $-0.863$  & $-0.714$  & $-0.82$  & $-0.8(2)$ \\
\hline\hline 
\end{tabular}
\end{table*}

\begin{table*} [p]
\caption{\label{t6.3} Magnetic moments of $\Xi_{c}$ and $\Xi_{c}^{\prime}$ baryons
(in nuclear magnetons $\mu_{N}$).}
\begin{tabular}{ccccccccccccc}
\hline\hline 
 & Our  & NR  & PM(BD)  & PM(AL1)  & Our  & PM   & EM\&C  & $\chi$QM  & RPM  & RQM  & QCDSR  & Latt\\
 & Mixed  & Mixed  & \cite{S96d}  & \cite{S96d}  & Unmix  & \cite{MPV08}  & \cite{KDV05}  & \cite{SDCG10}  & \cite{BD83a}  & \cite{FGIKLNP06}  & \cite{AAO08,ABS15a,AAS14b}  & \cite{BCEOT17}\\
\hline 
$\Xi_{c}^{\prime\,0}$  & $-1.13$  & $-1.18$  & $\cdot\cdot\cdot$  & $\cdot\cdot\cdot$  & $-1.12$  & $-0.987$   & $-0.93$  & $-1.32$  & $-1.12$  & $-0.95$  & $-1.2(3)$  & $-0.599(71)$ \\
$\Xi_{c}^{0}$  & $0.346$  & $0.41$  & $0.357$  & $0.360$  & $0.334$   & $\cdot\cdot\cdot$  & $0.366$  & $0.28$  & $0.341$  & $0.39$  & $0.35(5)$  & $0.192(17)$ \\
$\Xi_{c}^{\prime\,0}\rightarrow\Xi_{c}^{0}$  & $0.035$  & $0.08$  & $\cdot\cdot\cdot$  & $\cdot\cdot\cdot$  & $0.138$  & $\cdot\cdot\cdot$   & $0.13$  & $-0.31$  & $\cdot\cdot\cdot$  & $\cdot\cdot\cdot$  & $0.18(2)$  & $0.009(13)$\\
$\Xi_{c}^{\prime\,+}$  & $0.825$  & $0.89$  & $\cdot\cdot\cdot$  & $\cdot\cdot\cdot$  & $0.633$  & $0.509$   & $0.76$  & $0.76$  & $0.796$  & $0.47$  & $0.8(2)$  & $0.315(144)$ \\
$\Xi_{c}^{+}$  & $0.142$  & $0.20$  & $0.166$  & $0.211$  & $0.334$   & $\cdot\cdot\cdot$  & $0.37$  & $0.40$  & $0.341$  & $0.37$  & $0.50(5)$  & $0.235(25)$\\
$\Xi_{c}^{\prime\,+}\rightarrow\Xi_{c}^{+}$  & $-1.35$  & $-1.40$  & $\cdot\cdot\cdot$  & $\cdot\cdot\cdot$  & $-1.38$  & $\cdot\cdot\cdot$  & $-1.39$  & $1.30$  & $\cdot\cdot\cdot$  & $\cdot\cdot\cdot$  & $1.3(1)$  & $0.729(103)$ \\
\hline\hline 
\end{tabular}
\end{table*}

\begin{table*} [p]
\caption{\label{t6.4} Magnetic moments of $\Xi_{b}$ and $\Xi_{b}^{\prime}$ baryons
(in nuclear magnetons $\mu_{N}$).}
\begin{tabular}{cccccccccccc}
\hline\hline  
 & Our  & NR  & PM(BD)  & PM(AL1)  & Our  & PM  & Hyp  & EM\&C  & RPM  & RQM  & QCDSR\\
 & Mixed  & Mixed  & \cite{S96d}  & \cite{S96d}  & Unmix  & \cite{MPV08}  & \cite{PRV08a}  & \cite{DKV13}  & \cite{BD83a}  & \cite{FGIKLNP06}  & \cite{AAO08,ABS15a,AAS14b}\\
\hline 
$\Xi_{b}^{\prime\,-}$  & $-0.968$  & $-1.02$  & $\cdot\cdot\cdot$  & $\cdot\cdot\cdot$  & $-0.966$  & $-0.941$  & $-0.902$  & $-0.996$  & $-0.985$  & $-0.91$  & $-1.2(3)$\\
$\Xi_{b}^{-}$  & $-0.0555$  & $-0.05$  & $-0.052$  & $-0.055$  & $-0.0596$  & $\cdot\cdot\cdot$  & $\cdot\cdot\cdot$  & $-0.066$  & $\cdot\cdot\cdot$  & $-0.06$  & $-0.08(2)$\\
$\Xi_{b}^{\prime\,-}\rightarrow\Xi_{b}^{-}$  & $0.113$  & $0.16$  & $\cdot\cdot\cdot$  & $\cdot\cdot\cdot$  & $0.128$  & $\cdot\cdot\cdot$  & $\cdot\cdot\cdot$  & $0.142$  & $\cdot\cdot\cdot$  & $\cdot\cdot\cdot$  & $0.21(1)$\\
$\Xi_{b}^{\prime\,0}$  & $0.782$  & $0.90$  & $\cdot\cdot\cdot$  & $\cdot\cdot\cdot$  & $0.737$  & $0.658$  & $0.766$  & $0.676$  & $0.893$  & $0.66$  & $0.7(2)$\\
$\Xi_{b}^{0}$  & $-0.106$  & $-0.11$  & $-0.106$  & $-0.086$  & $-0.0596$  & $\cdot\cdot\cdot$  & $\cdot\cdot\cdot$  & $-0.060$  & $\cdot\cdot\cdot$  & $-0.06$  & $-0.045(5)$\\
$\Xi_{b}^{\prime\,0}\rightarrow\Xi_{b}^{0}$  & $-1.33$  & $-1.41$  & $\cdot\cdot\cdot$  & $\cdot\cdot\cdot$  & $-1.34$  & $\cdot\cdot\cdot$  & $\cdot\cdot\cdot$  & $1.35$  & $\cdot\cdot\cdot$  & $\cdot\cdot\cdot$  & $1.4(1)$\\
\hline\hline  
\end{tabular}
\end{table*}

\begin{table*} [p]
\caption{\label{t6.5} Magnetic moments of $J=\frac{3}{2}$ singly charmed baryons (in nuclear
magnetons $\mu_{N}$).}
\begin{tabular}{lcccccccccc}
\hline\hline  
Baryon  & Our  & NR  & PM  & Hyp  & Hyp  & Hyp  & EM\&C  & $\chi$QM  & QCDSR  & Latt\\
 &  &  & \cite{MPV08}  & \cite{PRV08a}  & \cite{GRQR14}  & \cite{STRV16b}  & \cite{DV09}  & \cite{SDCG10}  & \cite{AAO09a}  & \cite{CEOT15}\\
\hline 
$\Sigma_{c}^{*0}$  & $-1.49$  & $-1.47$  & $-1.15$  & $-0.848$  & $-1.44$  & $-1.02$  & $-1.18$  & $-1.99$  & $-0.81(20)$  & $\cdot\cdot\cdot$\\
$\Sigma_{c}^{*+}$  & $1.25$  & $1.32$  & $1.13$  & $1.25$  & $1.32$  & $\cdot\cdot\cdot$  & $1.18$  & $0.97$  & $2.00(46)$  & $\cdot\cdot\cdot$\\
$\Sigma_{c}^{*++}$  & $3.98$  & $4.11$  & $3.41$  & $3.84$  & $4.10$  & $\cdot\cdot\cdot$  & $3.63$  & $3.92$  & $4.81(1.22)$  & $\cdot\cdot\cdot$\\
$\Xi_{c}^{*0}$  & $-1.20$  & $-1.15$  & $-0.987$  & $-0.688$  & $-1.18$  & $-0.825$  & $-1.02$  & $-1.43$  & $-0.68(18)$  & $\cdot\cdot\cdot$\\
$\Xi_{c}^{*+}$  & $1.47$  & $1.64$  & $1.26$  & $1.51$  & $1.04$  & $\cdot\cdot\cdot$  & $1.39$  & $1.59$  & $1.68(42)$  & $\cdot\cdot\cdot$\\
$\Omega_{c}^{*0}$  & $-0.936$  & $-0.83$  & $-0.834$  & $-0.865$  & $-0.92$  & $-0.625$  & $-0.84$  & $-0.86$  & $-0.62(18)$  & $-0.730(23)$\\
\hline\hline  
\end{tabular}
\end{table*}

\begin{table*} [p]
\caption{\label{t6.6} Magnetic moments of $J=\frac{3}{2}$ singly bottom baryons (in nuclear
magnetons $\mu_{N}$).}
\begin{tabular}{lcccccccc}
\hline\hline  
Baryon  & Our  & NR  & PM  & Hyp  & Hyp  & EM\&C  & $\chi$QM  & QCDSR\\
 &  &  & \cite{MTV16}  & \cite{PRV08a}  & \cite{GRQR14}  & \cite{DKV13}  & \cite{SDCG10}  & \cite{AAO09a}\\
\hline 
$\Sigma_{b}^{*-}$  & $-1.82$  & $-1.92$  & $-1.82$  & $-1.66$  & $-1.91$  & $-1.75$  & $-1.63$  & $-1.50(36)$ \\
$\Sigma_{b}^{*0}$  & $0.820$  & $0.87$  & $0.819$  & $0.791$  & $0.89$  & $0.705$  & $0.724$  & $0.50(15)$ \\
$\Sigma_{b}^{*+}$  & $3.46$  & $3.56$  & $3.46$  & $3.23$  & $3.69$  & $3.10$  & $3.08$  & $2.52(50)$ \\
$\Xi_{b}^{*-}$  & $-1.55$  & $-1.60$  & $\cdot\cdot\cdot$  & $-1.10$  & $-1.65$  & $-1.59$  & $-1.48$  & $-1.42(35)$ \\
$\Xi_{b}^{*0}$  & $1.03$  & $1.19$  & $\cdot\cdot\cdot$  & $1.04$  & $1.16$  & $0.915$  & $0.875$  & $0.50(15)$ \\
$\Omega_{b}^{*-}$  & $-1.31$  & $-1.28$  & $\cdot\cdot\cdot$  & $-1.20$  & $-1.38$  & $-1.39$  & $-1.29$  & $-1.40(35)$ \\
\hline\hline  
\end{tabular}
\end{table*}

\begin{table*} [p]
\caption{\label{t6.7} Magnetic moments of $B_{cb}$ and $B_{cb}^{\prime}$  baryons (in nuclear magnetons $\mu_{N}$)
with and without state mixing.}

\begin{tabular}{ccccccc}
\hline\hline  
 & Our  & NR  & PM(BD)  & PM(AL1)  &  & Our\\
 & Mixed  & Mixed  & \cite{S96d}  & \cite{S96d}  &  & Unmix\\
\hline 
$\Xi_{cb}^{\prime\,0}$  & $-0.452$  & $-0.53$  & $\cdot\cdot\cdot$  & $\cdot\cdot\cdot$  & $\Xi_{\{dc\}b}^{0}$  & $-0.291$\\
$\Xi_{cb}^{0}$  & $0.102$  & $0.13$  & $0.117$  & $0.058$  & $\Xi_{[dc]b}^{0}$  & $-0.0596$\\
$\Xi_{cb}^{\prime\,0}\rightarrow\Xi_{cb}^{0}$  & $0.600$  & $0.70$  & $\cdot\cdot\cdot$  & $\cdot\cdot\cdot$  & $\Xi_{\{dc\}b}^{0}\rightarrow\Xi_{[dc]b}^{0}$  & $0.651$\\
$\Xi_{cb}^{\prime\,+}$  & $1.46$  & $1.71$  & $\cdot\cdot\cdot$  & $\cdot\cdot\cdot$  & $\Xi_{\{uc\}b}^{+}$  & $1.30$\\
$\Xi_{cb}^{+}$  & $-0.222$  & $-0.25$  & $-0.254$  & $-0.198$  & $\Xi_{[uc]b}^{+}$  & $-0.0596$\\
$\Xi_{cb}^{\prime\,+}-\Xi_{cb}^{+}$  & $-0.532$  & $-0.62$  & $\cdot\cdot\cdot$  & $\cdot\cdot\cdot$  & $\Xi_{\{uc\}b}^{+}\rightarrow\Xi_{[uc]b}^{+}$  & $-0.729$\\
$\Omega_{cb}^{\prime\,0}$  & $-0.275$  & $-0.27$  & $\cdot\cdot\cdot$  & $\cdot\cdot\cdot$  & $\Omega_{\{sc\}b}^{0}$  & $-0.157$\\
$\Omega_{cb}^{0}$  & $0.058$  & $0.08$  & $0.047$  & $0.009$  & $\Omega_{[sc]b}^{0}$  & $-0.0596$\\
$\Omega_{cb}^{\prime\,0}\rightarrow\Omega_{cb}^{0}$  & $0.510$  & $0.56$  & $\cdot\cdot\cdot$  & $\cdot\cdot\cdot$  & $\Omega_{\{sc\}b}^{0}\rightarrow\Omega_{[sc]b}^{0}$  & $0.534$\\
\hline\hline  
\end{tabular}
\end{table*}

\begin{table*} [p]
\caption{\label{t6.8} Magnetic moments of $B_{cb}$ and $B_{cb}^{\prime}$ baryons (in nuclear magnetons $\mu_{N}$)
without state mixing.}

\begin{tabular}{ccccccccr}
\hline\hline  
 & Our  & PM(AL1)  & Hyp  & Hyp  & EM\&C  & RQM  & RQM  & $\chi$PT\\
 & Unmix  & \cite{AHNV07b}  & \cite{PRV08a}  & \cite{STR16,SR17a}  & \cite{DKV13}  & \cite{FGIKLNP06}  & \cite{GSP16}  & \cite{LMLZ17a}\\
\hline 
$\Xi_{[cb]d}^{0}$  & $-0.796$  & $-0.993$  & $\cdot\cdot\cdot$  & $\cdot\cdot\cdot$  & $-0.814$  & $-0.76$  & $\cdot\cdot\cdot$  & $-0.59$\\
$\Xi_{\{cb\}d}^{0}$  & $0.446$  & $0.518$  & $0.477$  & $0.354$  & $0.480$  & $0.42$  & $0.63$  & $0.56$\\
$\Xi_{[cb]d}^{0}\rightarrow\Xi_{\{cb\}d}^{0}$  & $-0.225$  & $\cdot\cdot\cdot$  & $\cdot\cdot\cdot$  & $\cdot\cdot\cdot$  & $0.242$  & $\cdot\cdot\cdot$  & $\cdot\cdot\cdot$  & $\cdot\cdot\cdot$ \\
$\Xi_{[cb]u}^{+}$  & $1.59$  & $1.99$  & $\cdot\cdot\cdot$  & $\cdot\cdot\cdot$  & $1.72$  & $1.52$  & $\cdot\cdot\cdot$  & $0.69$\\
$\Xi_{\{cb\}u}^{+}$  & $-0.350$  & $-0.475$  & $-0.400$  & $-0.204$  & $-0.369$  & $-0.12$  & $-0.52$  & $-0.54$\\
$\Xi_{[cb]u}^{+}\rightarrow\Xi_{\{cb\}u}^{+}$  & $-0.225$  & $\cdot\cdot\cdot$  & $\cdot\cdot\cdot$  & $\cdot\cdot\cdot$  & $0.250$  & $\cdot\cdot\cdot$  & $\cdot\cdot\cdot$  & $\cdot\cdot\cdot$ \\
$\Omega_{[cb]s}^{0}$  & $-0.595$  & $-0.542$  & $\cdot\cdot\cdot$  & $\cdot\cdot\cdot$  & $-0.624$  & $-0.61$  & $\cdot\cdot\cdot$  & $0.24$\\
$\Omega_{\{cb\}s}^{0}$  & $0.378$  & $0.368$  & $0.395$  & $0.439$  & $0.407$  & $0.45$  & $0.49$  & $0.49$\\
$\Omega_{\{cb\}s}^{0}\rightarrow\Omega_{[cb]s}^{0}$  & $-0.225$  & $\cdot\cdot\cdot$  & $\cdot\cdot\cdot$  & $\cdot\cdot\cdot$  & $0.243$  & $\cdot\cdot\cdot$  & $\cdot\cdot\cdot$  & $\cdot\cdot\cdot$ \\
\hline\hline  
\end{tabular}
\end{table*}

\begin{table*} [p]
\caption{\label{t6.9} Magnetic moments of $J=\frac{1}{2}$ doubly heavy baryons (in nuclear
magnetons $\mu_{N}$).}

\begin{tabular}{lcccccccccccc}
\hline\hline  
Baryon  & Our  & NR  & PM(BD)  & PM(AL1)  & Hyp  & Hyp  & EM\&C  & RPM  & RQM  & RQM  & $\chi$PT  & Latt\\
 &  &  & \cite{S96d}  & \cite{S96d}  & \cite{PRV08a}  & \cite{STR16,SR17a}  & \cite{KDV05,DKV13}  & \cite{BD83a}  & \cite{FGIKLNP06}  & \cite{GSP16}  & \cite{LMLZ17a}  & \cite{CEIOT13,CEOT15} \\
\hline 
$\Xi_{cc}^{+}$  & $0.719$  & $0.83$  & $0.775$  & $0.784$  & $0.859$  & $0.784$  & $0.77$  & $0.774$  & $0.72$  & $0.853$  & $0.85$  & $0.43(3)$\\
$\Xi_{cc}^{++}$  & $-0.110$  & $-0.10$  & $-0.172$  & $-0.206$  & $-0.137$  & $-0.031$  & $-0.11$  & $-0.184$  & $0.13$  & $-0.169$  & $-0.25$  & $\cdot\cdot\cdot$ \\
$\Omega_{cc}^{+}$  & $0.645$  & $0.72$  & $0.620$  & $0.635$  & $0.783$  & $0.692$  & $0.70$  & $0.639$  & $0.67$  & $0.74$  & $0.78$  & $0.407(7)$ \\
$\Xi_{bb}^{-}$  & $0.171$  & $0.23$  & $0.230$  & $0.251$  & $0.190$  & $0.196$  & $0.218$  & $0.236$  & $0.18$  & $0.32$  & $0.26$  & $\cdot\cdot\cdot$ \\
$\Xi_{bb}^{0}$  & $-0.581$  & $-0.70$  & $-0.698$  & $-0.742$  & $-0.656$  & $-0.663$  & $-0.630$  & $-0.722$  & $-0.53$  & $-0.89$  & $-0.84$  & $\cdot\cdot\cdot$ \\
$\Omega_{bb}^{-}$  & $0.112$  & $0.12$  & $0.095$  & $0.101$  & $0.109$  & $0.108$  & $0.139$  & $0.10$  & $0.04$  & $0.16$  & $0.19$  & $\cdot\cdot\cdot$ \\
\hline\hline  
\end{tabular}
\end{table*}


\begin{table*} [p]
\caption{\label{t6.10} Magnetic moments of $J=\frac{3}{2}$ doubly heavy baryons (in nuclear
magnetons $\mu_{N}$).}

\begin{tabular}{lcccccccr}
\hline\hline  
Baryon  & Our  & NR  & PM (AL1)  & Hyp  & Hyp  & EM\&C  & RQM  & $\chi$PT\\
 &  &  & \cite{AHNV07b}  & \cite{PRV08a}  & \cite{STR16,SR17a}  & \cite{KDV05,DKV13}  & \cite{GSP16}  & \cite{MLLZ17} \\
\hline 
$\Xi_{cc}^{*+}$  & $-0.178$  & $-0.15$  & $-0.311$  & $-0.168$  & $0.068$  & $0.035$  & $-0.23$  & $-0.18$\\
$\Xi_{cc}^{*++}$  & $2.35$  & $2.64$  & $2.67$  & $2.75$  & $2.22$  & $2.52$  & $2.72$  & $2.61$\\
$\Omega_{cc}^{*+}$  & $0.0475$  & $0.139$  & $0.139$  & $0.121$  & $0.285$  & $0.21$  & $0.16$  & $0.17$\\
$\Xi_{bb}^{*-}$  & $-0.880$  & $-1.05$  & $-1.11$  & $-0.951$  & $-1.74$  & $-1.052$  & $-1.32$  & $-1.33$\\
$\Xi_{bb}^{*0}$  & $1.40$  & $1.74$  & $1.87$  & $1.58$  & $1.61$  & $1.51$  & $2.30$  & $2.83$ \\
$\Omega_{bb}^{*-}$  & $-0.697$  & $-0.73$  & $-0.662$  & $-0.711$  & $-1.24$  & $-0.805$  & $-0.86$  & $-1.54$ \\
$\Xi_{cb}^{*0}$  & $-0.534$  & $-0.60$  & $-0.712$  & $-0.567$  & $-0.372$  & $-0.508$  & $-0.76$  & $-0.84$ \\
$\Xi_{cb}^{*+}$  & $1.88$  & $2.19$  & $2.27$  & $2.05$  & $1.56$  & $2.02$  & $2.68$  & $3.72$ \\
$\Omega_{cb}^{*0}$  & $-0.329$  & $-0.280$  & $-0.261$  & $-0.316$  & $-0.181$  & $-0.309$  & $-0.32$  & $-1.09$\\
\hline\hline  
\end{tabular}
\end{table*}

\begin{table*}  [p]
\caption{\label{t6.11} Magnetic moments of triply heavy baryons (in nuclear magnetons $\mu_{N}$).}

\begin{tabular}{lcccccccccc}
\hline\hline  
Baryon  & Our  & NR  & PM(BD)  & PM(AL1)  & PM  & Hyp  & EM\&C  & RPM  & RQM  & Latt\\
 &  &  & \cite{S96d}  & \cite{S96d}  & \cite{TMV16}  & \cite{PMV09}  & \cite{DV09,DKV13}  & \cite{BD83a}  & \cite{FGIKLNP06}  & \cite{CEOT15} \\
\hline 
$\Omega_{ccc}^{*++}$  & $0.989$  & $1.17$  & $1.00$  & $1.02$  & $1.18$  & $1.19$  & $1.16$  & $\cdot\cdot\cdot$  & $\cdot\cdot\cdot$  & $0.676(5)$\\
$\Omega_{ccb}^{+}$  & $0.455$  & $0.54$  & $0.466$  & $0.475$  & $0.565$  & $0.502$  & $0.522$  & $0.476$  & $0.53$  & $\cdot\cdot\cdot$ \\
$\Omega_{ccb}^{*+}$  & $0.594$  & $0.72$  & $\cdot\cdot\cdot$  & $\cdot\cdot\cdot$  & $0.751$  & $0.651$  & $0.703$  & $\cdot\cdot\cdot$  & $\cdot\cdot\cdot$  & $\cdot\cdot\cdot$ \\
$\Omega_{cbb}^{0}$  & $-0.187$  & $-0.21$  & $-0.191$  & $-0.193$  & $-0.223$  & $-0.203$  & $-0.200$  & $-0.197$  & $-0.20$  & $\cdot\cdot\cdot$ \\
$\Omega_{cbb}^{*0}$  & $0.204$  & $0.27$  & $\cdot\cdot\cdot$  & $\cdot\cdot\cdot$  & $0.285$  & $0.216$  & $0.225$  & $\cdot\cdot\cdot$  & $\cdot\cdot\cdot$  & $\cdot\cdot\cdot$ \\
$\Omega_{bbb}^{*-}$  & $-0.178$  & $-0.18$  & $-0.178$  & $-0.180$  & $-0.196$  & $-0.195$  & $-0.198$  & $\cdot\cdot\cdot$  & $\cdot\cdot\cdot$  & $\cdot\cdot\cdot$ \\
\hline\hline  
\end{tabular}
\end{table*}

We compare our predictions with the results obtained using other approaches,
such as simple nonrelativistic quark model (NR), various potential
models (PM), hypercentral approach (Hyp), effective mass and charge
scheme (EM\&C), chiral quark model ($\chi$QM), relativistic quark
models (RQM), relativistic potential models (RPM), QCD sum rules (QCDSR),
lattice QCD (Latt), and chiral perturbation theory ($\chi$PT). From
Ref.~\cite{S96d} we have taken predictions corresponding to two
potentials (BD and AL1) in order to illustrate the sensitivity of
the results to the form of the potential. For the baryons ($\Xi_{c}$, $\Xi_{c}^{\prime}$),
($\Xi_{b}$, $\Xi_{b}^{\prime}$), ($\Xi_{cb}$, $\Xi_{cb}^{\prime}$),
and ($\Omega_{cb}$, $\Omega_{cb}^{\prime}$) we have included our
results with and without state mixing. This is done for two reasons.
First, the comparison of the mixed predictions with the corresponding
unmixed results allows us to estimate the effect of this mixing on
the values of the magnetic (transition) moments. Second, it is not very
meaningful to compare the mixed estimates obtained in one framework
with the unmixed results obtained using other approaches. For some
baryons ($\Xi_{c}^{0}$, $\Xi_{c}^{+}$, $\Xi_{b}^{-}$, $\Xi_{b}^{0}$,
$\Xi_{cb}^{0}$, $\Xi_{cb}^{+}$, and $\Omega_{cb}^{0}$) there exist
the predictions of the magnetic moments with the state mixing
taken into account. These calculations (Refs.~\cite{S96d,S96e})
are based on the solution of the Fadeev equations in the framework
of nonrelativistic potential model, and the state mixing effect is
present in this method by construction. In all other cases it is more
reasonable to compare presented estimates with our unmixed predictions.
We include also the mixed NR results \cite{FLNC81} in order to have
one more prediction to compare our results with. From Tables~\ref{t6.1}--\ref{t6.4},
\ref{t6.7}, \ref{t6.9}, and \ref{t6.11} we see that our predictions
agree well with the results obtained using Fadeev formalism, and the
best agreement is achieved in the case of BD potential. For the $\Lambda_{Q}$,
$\Sigma_{Q}$, $\Xi_{Q,}$ and $\Xi_{Q}^{\prime}$ baryons the agreement
is almost excellent, however, for the baryons $\Omega_{Q}$ PM(BD)
predictions are  15--20$\%$ smaller than ours. For doubly heavy baryons the agreement is good,
but not so impressive, and for triply heavy baryons the agreement
is sufficiently good again. Note the excellent agreement between our
result, the PM (BD and AL1) predictions, and the NR estimate in the
obviously nonrelativistic case of $\Omega_{bbb}^{*-}$ baryon. This
is an indication that in the baryon sector our model also has the proper nonrelativistic limit, the feature desirable for any more or less reliable phenomenological
model. Next, we see that almost all predictions of various potential
models are similar. Moreover, these predictions in almost all cases
are close to the predictions obtained using simple nonrelativistic
approach (NR). As was noted in Ref.~\cite{S96d} the reason may be similar 
values of the quark masses
used in the otherwise different potential models. Our approach for
the singly heavy baryons gives predictions similar to the NR estimates.
In the case of doubly (and triply) heavy baryons our predictions are
lower than NR results, while the predictions obtained using potential
models remain, as a rule, similar to the NR estimates. For example,
for the baryons $\Xi_{cc}^{+}$ and $\Xi_{bb}^{-}$ our results are
closer to the predictions obtained in RQM \cite{FGIKLNP06} than to
the results obtained in PM or NR models. The predictions obtained using various other approaches differ. For instance, relativistic potential model \cite{BD83a} give
predictions similar to the ones obtained using potential models or
NR approach. On the other hand, the predictions obtained in the relativistic
quark model (Ref.~\cite{FGIKLNP06}) are lower by $\approx30\%$.
The predictions obtained using light cone QCD sum rules (QCDSR) are
similar to ours. Lattice QCD results for $\Sigma_{c}^{0}$ and $\Sigma_{c}^{++}$
baryons within the error bars agree with other predictions, but for
$\Omega_{c}^{0}$ and $\Omega_{c}^{*0}$ are by $\approx30\%$ lower.
In the case of $\Xi_{c}$, $\Xi_{c}^{\prime}$, doubly heavy ($\Xi_{cc}^{+}$,
$\Omega_{cc}^{+}$), and triply heavy ($\Omega_{ccc}^{*++}$) baryons
the lattice QCD results are again substantially lower, and the disagreement
between lattice QCD results and the predictions obtained using other
approaches is evident.

In order to estimate the effect of the state mixing one can compare
the results for the mixed and unmixed cases presented in Tables~\ref{t6.3},
\ref{t6.4}, \ref{t6.7}, and \ref{t6.8}. We see that for the magnetic
moments of $\Xi_{c}^{0}$, $\Xi_{c}^{\prime\,0}$, $\Xi_{b}^{-}$,
 $\Xi_{b}^{\prime\,-}$ baryons the effect is not large.
The transition moment $\mu\left(\Xi_{c}^{\prime\,0}\rightarrow\Xi_{c}^{0}\right)$
in the physical basis is smaller by $\approx4$ times. For the ($\Xi_{b}^{-}$, 
$\Xi_{b}^{\prime\,-}$)
baryons the transition moment is lower by $\approx10\%$. 
For ($\Xi_{c}^{+}$, $\Xi_{c}^{\prime\,+}$) and ($\Xi_{b}^{0}$, $\Xi_{b}^{\prime\,0}$) baryons the tendency is opposite, i.e. the state mixing has practically no effect
on the transition moments, but the effect on the values of the corresponding
magnetic moments is significant. The magnetic moment of $\Xi_{c}^{\prime\,+}$
baryon becomes larger by $\approx25\%$, and magnetic moment of $\Xi_{c}^{+}$
smaller by $\approx2.4$ times. For ($\Xi_{b}^{0}$, $\Xi_{b}^{\prime\,0}$)
baryons the effect is slightly weaker. The magnetic moment of $\Xi_{b}^{\prime\,0}$
baryon is larger by $\approx10\%$, and magnetic moment of $\Xi_{b}^{0}$
becomes larger twice. Note that we compare the predictions obtained
in the physical (mixed) basis with the results obtained in the optimal
$(q_{1}q_{2})Q$ basis. No one uses the spin coupling scheme such
as $(q_{1}Q)q_{2}$. It can be readily checked that the effect of
the state mixing in this basis would be very large. In the case of
doubly heavy baryons ($\Xi_{cb}$, $\Xi_{cb}^{\prime}$), and ($\Omega_{cb}$, $\Omega_{cb}^{\prime}$)
the mixing effect is always important, no matter which basis is used.
The effect is not so severe for the optimal $(qc)b$ basis, however,
traditionally another basis $(cb)q$ is hooked up. We have compared our
unmixed predictions obtained in this basis with the corresponding
results obtained using various other approaches in Table~\ref{t6.8}.
As expected, when calculated in the same basis, the results obtained
using various approaches are similar. 

As we have seen, the state mixing
effect  for the singly heavy ($\Xi_{c}$, $\Xi_{c}^{\prime}$),
($\Xi_{b}$, $\Xi_{b}^{\prime}$), as well as for doubly heavy ($\Xi_{cb}$, $\Xi_{cb}^{\prime}$),
 ($\Omega_{cb}$, $\Omega_{cb}^{\prime}$) baryons is important
and in the realistic calculations must be taken into account.

\subsection{M1 transition moments and decay widths}

In this section we present our results for the M1 decay
characteristics of heavy baryons. In Tables~\ref{t7.1}--\ref{t7.5}
the magnetic dipole transitions moments are given. Alongside we present
the static moments in order to demonstrate the dependence of the transition
moment on the photon momentum $k$. Static values also may be useful
for the comparison of our predictions with the estimates obtained
using other approaches. We see from Tables~\ref{t7.1}--\ref{t7.5}
that the static moments are slightly ($\lesssim10\%$) larger. So,
the effect of setting $k=0$ is not large, but it is not negligible.

\begin{table*} [p]
\caption{\label{t7.1} Transition moments of $\Sigma_{c}$ and $\Omega_{c}$ baryons (in
nuclear magnetons $\mu_{N}$).}
\begin{tabular}{ccccccccr}
\hline\hline  
Transition  & Our  & Our  & NR  & PM  & Hyp  & Hyp  & EM\&C  & $\chi$QM\\
 &  & Static  &  & \cite{MPV09}  & \cite{MPV09}  & \cite{STRV16b}  & \cite{KDV05,DV09}  & \cite{SDCG10}\\
\hline 
$\Sigma_{c}^{*0}\rightarrow\Sigma_{c}^{0}$  & $-1.14$  & $-1.17$  & $-1.24$  & $-1.06$  & $-1.12$  & $-1.04$  & $1.07$  & $1.48$\\
$\Sigma_{c}^{*+}\rightarrow\Sigma_{c}^{+}$  & $0.102$  & $0.111$  & $0.07$  & $0.008$  & $0.100$  & $\cdot\cdot\cdot$  & $0.08$  & $-0.003$\\
$\Sigma_{c}^{*+}\rightarrow\Lambda_{c}^{+}$  & $2.07$  & $2.23$  & $2.2$  & $1.86$  & $2.12$  & $1.84$  & $2.15$  & $2.40$\\
$\Sigma_{c}^{+}\rightarrow\Lambda_{c}^{+}$  & $-1.48$  & $-1.56$  & $-1.61$  & $-1.35$  & $-1.54$  & $\cdot\cdot\cdot$  & $-1.54$  & $1.56$\\
$\Sigma_{c}^{*++}\rightarrow\Sigma_{c}^{++}$  & $1.34$  & $1.39$  & $1.39$  & $1.08$  & $1.32$  & $\cdot\cdot\cdot$  & $1.23$  & $-1.37$\\
$\Omega_{c}^{*0}\rightarrow\Omega_{c}^{0}$  & $-0.892$  & $-0.911$  & $-0.94$  & $-0.908$  & $-0.916$  & $-0.876$  & $0.90$  & $1.24$\\
\hline\hline  
\end{tabular}
\end{table*}

\begin{table*} [p]
\caption{\label{t7.2} Transition moments of $\Sigma_{b}$ and $\Omega_{b}$ baryons (in
nuclear magnetons $\mu_{N}$).}
\begin{tabular}{lrrrrrr}
\hline\hline  
Transition  & Our  & Our  & NR  & PM  & PM  & Hyp\\
 &  & Static  &  & \cite{MTV16}  & \cite{MPV09}  & \cite{MPV09}\\
\hline 
$\Sigma_{b}^{*-}\rightarrow\Sigma_{b}^{-}$  & $-0.760$  & $-0.768$  & $-0.82$  & $-0.774$  & $-0.683$  & $-0.692$\\
$\Sigma_{b}^{*0}\rightarrow\Sigma_{b}^{0}$  & $0.464$  & $0.468$  & $0.49$  & $0.473$  & $0.429$  & $0.459$\\
$\Sigma_{b}^{*0}\rightarrow\Lambda_{b}^{0}$  & $2.02$  & $2.16$  & $2.28$  & $2.19$  & $1.93$  & $2.00$\\
$\Sigma_{b}^{0}\rightarrow\Lambda_{b}^{0}$  & $-1.43$  & $-1.53$  & $-1.61$  & $-1.55$  & $-1.37$  & $-1.42$\\
$\Sigma_{b}^{*+}\rightarrow\Sigma_{b}^{+}$  & $1.69$  & $1.70$  & $1.81$  & $1.72$  & $2.22$  & $2.30$\\
$\Omega_{b}^{*-}\rightarrow\Omega_{b}^{-}$  & $-0.523$  & $-0.528$  & $-0.52$  & $\cdot\cdot\cdot$  & $-0.523$  & $-0.476$\\
\hline\hline  
\end{tabular}
\end{table*}

\begin{table*} [p]
\caption{\label{t7.3} Transition moments of $\Xi_{c}$  baryons (in nuclear magnetons
$\mu_{N}$).}
\begin{tabular}{lcccccccr}
\hline\hline  
Transition  & Our  & Our  & NR  & Our  & PM  & Hyp  & EM\&C  & $\chi$QM\\
 & Mixed  & Static  & Mixed  & Unmix  & \cite{MPV09}  & \cite{MPV09}  & \cite{KDV05,DV09}  & \cite{SDCG10}\\
\hline 
$\Xi_{c}^{*0}\rightarrow\Xi_{c}^{\prime\,0}$  & $-0.994$  & $-1.01$  & $-1.07$  & $-1.03$  & $\cdot\cdot\cdot$  & $\cdot\cdot\cdot$  & $0.99$  & $1.24$\\
$\Xi_{c}^{*0}\rightarrow\Xi_{c}^{0}$  & $-0.249$  & $-0.268$  & $-0.33$  & $-0.193$  & $-0.120$  & $-0.208$  & $0.18$  & $-0.50$\\
$\Xi_{c}^{\prime\,0}\rightarrow\Xi_{c}^{0}$  & $0.0339$  & $0.035$  & $0.08$  & $0.139$  & $\cdot\cdot\cdot$  & $\cdot\cdot\cdot$  & $0.13$  & $-0.31$\\
$\Xi_{c}^{*+}\rightarrow\Xi_{c}^{\prime\,+}$  & $0.0664$  & $0.0738$  & $0.09$  & $0.216$  & $\cdot\cdot\cdot$  & $\cdot\cdot\cdot$  & $0.17$  & $-0.23$\\
$\Xi_{c}^{*+}\rightarrow\Xi_{c}^{+}$  & $1.86$  & $1.96$  & $2.03$  & $1.97$  & $0.991$  & $1.11$  & $1.94$  & $2.08$\\
$\Xi_{c}^{\prime\,+}\rightarrow\Xi_{c}^{+}$  & $-1.33$  & $-1.35$  & $-1.40$  & $-1.38$  & $\cdot\cdot\cdot$  & $\cdot\cdot\cdot$  & $-1.39$  & $1.30$\\
\hline\hline  
\end{tabular}
\end{table*}

\begin{table*} [p]
\caption{\label{t7.4} Transition moments of $\Xi_{b}$  baryons (in nuclear magnetons
$\mu_{N}$).}
\begin{tabular}{ccccccc}
\hline\hline  
Transition  & Our  & Our  & NR  & Our  & PM  & Hyp\\
 & Mixed  & Static  & Mixed  & Unmix  & \cite{MPV09}  & \cite{MPV09}\\
\hline 
$\Xi_{b}^{*-}\rightarrow\Xi_{b}^{\prime\,-}$  & $-0.6229$  & $-0.636$  & $-0.66$  & $-0.641$  & $\cdot\cdot\cdot$  & $\cdot\cdot\cdot$ \\
$\Xi_{b}^{*-}\rightarrow\Xi_{b}^{-}$  & $-0.182$  & $-0.193$  & $-0.26$  & $-0.181$  & $-0.124$  & $-0.196$\\
$\Xi_{b}^{\prime\,-}\rightarrow\Xi_{b}^{-}$  & $0.109$  & $0.113$  & $0.16$  & $0.128$  & $\cdot\cdot\cdot$  & $\cdot\cdot\cdot$ \\
$\Xi_{b}^{*0}\rightarrow\Xi_{b}^{\prime\,0}$  & $0.521$  & $0.529$  & $0.61$  & $0.563$  & $\cdot\cdot\cdot$  & $\cdot\cdot\cdot$ \\
$\Xi_{b}^{*0}\rightarrow\Xi_{b}^{0}$  & $1.83$  & $1.91$  & $2.03$  & $1.89$  & $1.04$  & $1.05$ \\
$\Xi_{b}^{\prime\,0}\rightarrow\Xi_{b}^{0}$  & $-1.30$  & $-1.39$  & $-1.41$  & $-1.34$  & $\cdot\cdot\cdot$  & $\cdot\cdot\cdot$ \\
\hline\hline  
\end{tabular}
\end{table*}

\begin{table*} [p]
\caption{\label{t7.6} Transition moments of $\Xi_{cb}$ and $\Omega_{cb}$ baryons (in nuclear
magnetons $\mu_{N}$).}
\begin{tabular}{lccclclcr}
\hline\hline  
Transition  & Our  & Our  & NR  & Transition  & Our  & Transition  & Our  & $\chi$PT\\
 & Mixed  & Static  & Mixed  &  & Unmix  &  & Unmix  & \cite{LMLZ18}\\
\hline 
$\Xi_{cb}^{*0}\rightarrow\Xi_{cb}^{\prime\,0}$  & $-0.0416$  & $-0.0445$  & $-0.06$  & $\Xi_{cb}^{*0}\rightarrow\Xi_{\{dc\}b}^{0}$  & $-0.162$  & $\Xi_{cb}^{*0}\rightarrow\Xi_{[cb]d}^{0}$  & $0.319$  & $-0.36$\\
$\Xi_{cb}^{*0}\rightarrow\Xi_{cb}^{0}$  & $-0.919$  & $-0.934$  & $-1.09$  & $\Xi_{cb}^{*0}\rightarrow\Xi_{[dc]b}^{0}$  & $-0.899$  & $\Xi_{cb}^{*0}\rightarrow\Xi_{\{cb\}d}^{0}$  & $0.879$  & $1.34$\\
$\Xi_{cb}^{\prime\,0}\rightarrow\Xi_{cb}^{0}$  & $0.598$  & $0.600$  & $0.70$  & $\Xi_{\{dc\}b}^{0}\rightarrow\Xi_{[dc]b}^{0}$  & $0.649$  & $\Xi_{[cb]d}^{0}\rightarrow\Xi_{\{cb\}d}^{0}$  & $-0.225$  & $\cdot\cdot\cdot$ \\
$\Xi_{cb}^{*+}\rightarrow\Xi_{cb}^{\prime\,+}$  & $0.814$  & $0.823$  & $0.95$  & $\Xi_{cb}^{*+}\rightarrow\Xi_{\{uc\}b}^{+}$  & $0.958$  & $\Xi_{cb}^{*+}\rightarrow\Xi_{[cb]u}^{+}$  & $0.319$  & $-0.36$\\
$\Xi_{cb}^{*+}\rightarrow\Xi_{cb}^{+}$  & $1.12$  & $1.15$  & $1.33$  & $\Xi_{cb}^{*+}\rightarrow\Xi_{[uc]b}^{+}$  & $0.998$  & $\Xi_{cb}^{*+}\rightarrow\Xi_{\{cb\}u}^{+}$  & $-1.37$  & $-2.56$\\
$\Xi_{cb}^{\prime\,+}\rightarrow\Xi_{cb}^{+}$  & $-0.531$  & $-0.532$  & $-0.62$  & $\Xi_{\{uc\}b}^{+}\rightarrow\Xi_{[uc]b}^{+}$  & $-0.727$  & $\Xi_{[cb]u}^{+}\rightarrow\Xi_{\{cb\}u}^{+}$  & $-0.225$  & $\cdot\cdot\cdot$ \\
$\Omega_{cb}^{*0}\rightarrow\Omega_{cb}^{\prime\,0}$  & $0.0173$  & $0.0161$  & $0.05$  & $\Omega_{cb}^{*0}\rightarrow\Omega_{\{sc\}b}^{0}$  & $-0.0684$  & $\Omega_{cb}^{*0}\rightarrow\Omega_{[cb]s}^{0}$  & $0.318$  & $-0.36$\\
$\Omega_{cb}^{*0}\rightarrow\Omega_{cb}^{0}$  & $-0.748$  & $-0.758$  & $-0.82$  & $\Omega_{cb}^{*0}\rightarrow\Omega_{[sc]b}^{0}$  & $-0.741$  & $\Omega_{cb}^{*0}\rightarrow\Omega_{\{cb\}s}^{0}$  & $0.688$  & $1.33$\\
$\Omega_{cb}^{\prime\,0}\rightarrow\Omega_{cb}^{0}$  & $0.508$  & $0.510$  & $0.56$  & $\Omega_{\{sc\}b}^{0}\rightarrow\Omega_{[sc]b}^{0}$  & $0.532$  & $\Omega_{[cb]s}^{0}\rightarrow\Omega_{\{cb\}s}^{0}$  & $-0.225$  & $\cdot\cdot\cdot$ \\
\hline\hline  
\end{tabular}
\end{table*}

\begin{table*} [p]
\caption{\label{t7.5} Transition moments of doubly and triply heavy baryons (in nuclear magnetons
$\mu_{N}$).}
\begin{tabular}{ccccccc}
\hline\hline  
Decay  & Our  & Our  & NR  & EM\&C  & $\chi$QM  & $\chi$PT\\
 &  & Static  &  & \cite{DV09}  & \cite{SDCG10}  & \cite{LMLZ18}\\
\hline 
$\Xi_{cc}^{*+}\rightarrow\Xi_{cc}^{+}$  & $1.07$  & $1.10$  & $1.24$  & $1.06$  & $-1.41$  & $1.55$\\
$\Xi_{cc}^{*++}\rightarrow\Xi_{cc}^{++}$  & $-1.21$  & $-1.27$  & $-1.39$  & $1.35$  & $1.33$  & $-2.35$\\
$\Omega_{cc}^{*+}\rightarrow\Omega_{cc}^{+}$  & $0.869$  & $0.891$  & $0.94$  & $0.88$  & $-0.89$  & $1.54$\\
$\Xi_{bb}^{*-}\rightarrow\Xi_{bb}^{-}$  & $0.643$  & $0.653$  & $0.82$ & $\cdot\cdot\cdot$ & $\cdot\cdot\cdot$   & $\cdot\cdot\cdot$\\
$\Xi_{bb}^{*0}\rightarrow\Xi_{bb}^{0}$  & $-1.45$  & $-1.48$  & $-1.81$ & $\cdot\cdot\cdot$ & $\cdot\cdot\cdot$ & $\cdot\cdot\cdot$  \\
$\Omega_{bb}^{*-}\rightarrow\Omega_{bb}^{-}$  & $0.478$  & $0.484$  & $0.52$ & $\cdot\cdot\cdot$ & $\cdot\cdot\cdot$ & $\cdot\cdot\cdot$  \\
$\Omega_{ccb}^{*+}\rightarrow\Omega_{ccb}^{+}$  & $0.362$  & $0.364$  & $0.42$ & $\cdot\cdot\cdot$ & $\cdot\cdot\cdot$ & $\cdot\cdot\cdot$  \\
$\Omega_{cbb}^{*0}\rightarrow\Omega_{cbb}^{0}$  & $-0.352$  & $-0.360$  & $-0.42$ & $\cdot\cdot\cdot$  & $\cdot\cdot\cdot$ & $\cdot\cdot\cdot$\\
\hline\hline  
\end{tabular}
\end{table*}

\begin{table*} [p]
\caption{M1 decay widths (in keV) of $\Sigma_{c}$ and $\Omega_{c}$ baryons.\label{t7.9}}
\begin{tabular}{lcccccccccc}
\hline\hline  
Transition  & Our  &  PM  & PM  & Hyp  & QM & RQM  & $\chi$PT & $\chi$PT & QCDSR & Latt\\
 &    & \cite{DDSV94} & \cite{MPV09}  & \cite{MPV09}  & \cite{WYZZ17} & \cite{IKLR99b}  & \cite{C97e}  & \cite{JCZ15}  & \cite{AAO09b,AAS15b,ABS16}  & \cite{BCEO15} \\
\hline 
$\Sigma_{c}^{*0}\rightarrow\Sigma_{c}^{0}$  & $1.41$   & $1.2$  & $1.12$  & $1.44$  & $3.43$ & $\cdot\cdot\cdot$  & $1.2$  & $2.52$  & $0.08(3)$  & $\cdot\cdot\cdot$ \\
$\Sigma_{c}^{*+}\rightarrow\Sigma_{c}^{+}$  & $0.011$   & $1\cdot10^{-3}$  & $<10^{-4}$  & $0.01$  & $0.004$ & $0.14$  & $0.002$  & $0.85$  & $0.40(16)$  & $\cdot\cdot\cdot$ \\
$\Sigma_{c}^{*+}\rightarrow\Lambda_{c}^{+}$  & $190$  & $230$  & $155$  & $244$  & $373$ & $151$  & $147$  & $893$ & $130(45)$ & $\cdot\cdot\cdot$ \\
$\Sigma_{c}^{+}\rightarrow\Lambda_{c}^{+}$  & $74.1$   & $100$  & $60.6$  & $98.0$  & $80.6$  & $60.7$  & $88$  & $164$  & $50(17)$  & $\cdot\cdot\cdot$ \\
$\Sigma_{c}^{*++}\rightarrow\Sigma_{c}^{++}$  & $1.96$    & $1.60$  & $1.15$  & $1.98$  & $3.94$  & $\cdot\cdot\cdot$  & $1.4$  & $11.6$  & $2.65(1.20)$  & $\cdot\cdot\cdot$ \\
$\Omega_{c}^{*0}\rightarrow\Omega_{c}^{0}$  & $1.13$   & $0.69$  & $2.02$  & $0.82$  & $0.89$ & $\cdot\cdot\cdot$  & $\cdot\cdot\cdot$  & $4.82$  & $0.932$ & $0.074(8)$ \\
\hline\hline  
\end{tabular}
\end{table*}

\begin{table*} [p]
\caption{M1 decay widths (in keV) of $\Sigma_{b}$ and $\Omega_{b}$ baryons.\label{t7.10}}
\begin{tabular}{lccccccc}
\hline\hline  
Transition  & Our    & PM  & PM  & Hyp & QM & $\chi$PT & QCDSR\\
 &    & \cite{MTV16}  & \cite{MPV09}  & \cite{MPV09} & \cite{WYZZ17} & \cite{JCZ15}  & \cite{AAO09b,AAS15b,ABS16} \\
\hline 
$\Sigma_{b}^{*-}\rightarrow\Sigma_{b}^{-}$  & $0.0192$   & $0.023$  & $0.01$  & $0.02$ & $0.06$  & $0.08$  & $0.11(6)$ \\
$\Sigma_{b}^{*0}\rightarrow\Sigma_{b}^{0}$  & $0.0083$    & $0.0086$  & $<10^{-3}$  & $0.01$ & $0.02$  & $0.05$  & $0.028(16)$ \\
$\Sigma_{b}^{*0}\rightarrow\Lambda_{b}^{0}$  & $158$    & $114$  & $129$  & $142$  & $335$ & $435$  & $114(45)$ \\
$\Sigma_{b}^{0}\rightarrow\Lambda_{b}^{0}$  & $116$    & $78$  & $94.8$  & $100$  & $130$  & $288$  & $152(60)$ \\
$\Sigma_{b}^{*+}\rightarrow\Sigma_{b}^{+}$  & $0.110$   & $0.13$  & $0.08$  & $0.11$  & $0.25$  & $0.60$  & $0.46(22)$ \\
$\Omega_{b}^{*-}\rightarrow\Omega_{b}^{-}$  & $0.0091$    & $\cdot\cdot\cdot$ & $0.03$  & $0.20$  & $0.10$  & $\cdot\cdot\cdot$  & $0.092$ \\
\hline\hline  
\end{tabular}
\end{table*}

\begin{table*} [p]
\caption{M1 decay widths (in keV) of $\Xi_{c}$ baryons.\label{t7.11}}
\begin{tabular}{lccccccccccc}
\hline\hline  
Transition  & Our    & Our  & PM & PM  & Hyp  & QM  & RQM & $\chi$PT & QCDSR & Latt\\
 & Mixed    & Unmix  & \cite{DDSV94} & \cite{MPV09}  & \cite{MPV09}  & \cite{WYZZ17} & \cite{IKLR99b} & \cite{JCZ15}  & \cite{AAO09b,AAS15b,ABS16}  & \cite{BCEOT17} \\
\hline 
$\Xi_{c}^{*0}\rightarrow\Xi_{c}^{\prime\,0}$  & $1.23$    & $1.33$  & $1.1$  & $\cdot\cdot\cdot$  & $\cdot\cdot\cdot$  & $3.03$  & $\cdot\cdot\cdot$  & $3.83$ & $2.14$  & $\cdot\cdot\cdot$ \\
$\Xi_{c}^{*0}\rightarrow\Xi_{c}^{0}$  & $1.24$   & $0.745$  & $1.22$  & $0.30$  & $1.15$  & $0.0$  & $0.68$  & $0.36$  & $0.66(32)$  & $\cdot\cdot\cdot$ \\
$\Xi_{c}^{\prime\,0}\rightarrow\Xi_{c}^{0}$  & $0.011$   & $0.185$  & $0.23$  & $\cdot\cdot\cdot$  & $\cdot\cdot\cdot$  & $0.0$  & $0.17$  & $0.02$  & $0.27(6)$  & $0.002(4)$\\
$\Xi_{c}^{*+}\rightarrow\Xi_{c}^{\prime\,+}$  & $0.006$    & $0.063$  & $0.03$  & $\cdot\cdot\cdot$  & $\cdot\cdot\cdot$  & $0.004$  & $\cdot\cdot\cdot$  & $1.10$  & $0.274$  & $\cdot\cdot\cdot$ \\
$\Xi_{c}^{*+}\rightarrow\Xi_{c}^{+}$  & $72.7$   & $81.6$  & $16$  & $63.3$  & $99.9$  & $139$  & $54$  & $502$  & $52.(25)$  & $\cdot\cdot\cdot$ \\
$\Xi_{c}^{\prime\,+}\rightarrow\Xi_{c}^{+}$  & $17.3$  & $18.6$  & $5.7$  & $\cdot\cdot\cdot$  & $\cdot\cdot\cdot$  & $42.3$  & $12.7$  & $54.3$  & $8.5(2.5)$  & $5.5(1.5)$ \\
\hline\hline  
\end{tabular}
\end{table*}

\begin{table*} [p]
\caption{M1 decay widths (in keV) of $\Xi_{b}$ baryons.\label{t7.12}}
\begin{tabular}{lcccccccc}
\hline\hline  
Transition  & Our   & Our  & PM  & Hyp & QM & $\chi$PT & QCDSR\\
 & Mixed   & Unmix  & \cite{MPV09}  & \cite{MPV09} & \cite{WYZZ17} & \cite{JCZ15}  & \cite{AAO09b,AAS15b,ABS16}\\
\hline 
$\Xi_{b}^{*-}\rightarrow\Xi_{b}^{\prime\,-}$  & $0.0131$    & $0.0136$  & $\cdot\cdot\cdot$  & $\cdot\cdot\cdot$  & $15.0$  & $\cdot\cdot\cdot$  & $0.303$ \\
$\Xi_{b}^{*-}\rightarrow\Xi_{b}^{-}$  & $0.542$    & $0.536$  & $0.69$  & $0.03$  & $0.0$  & $1.87$  & $1.50(75)$ \\
$\Xi_{b}^{\prime\,-}\rightarrow\Xi_{b}^{-}$  & $0.259$    & $0.357$  & $\cdot\cdot\cdot$  & $\cdot\cdot\cdot$  & $0.0$  & $\cdot\cdot\cdot$  & $3.3(1.3)$ \\
$\Xi_{b}^{*0}\rightarrow\Xi_{b}^{\prime\,0}$  & $0.009$   & $0.0105$  & $\cdot\cdot\cdot$  & $\cdot\cdot\cdot$  & $5.19$  & $\cdot\cdot\cdot$  & $0.131$ \\
$\Xi_{b}^{*0}\rightarrow\Xi_{b}^{0}$  & $51.8$   & $55.3$  & $18.8$  & $3.60$  & $104$ & $136$  & $135(65)$ \\
$\Xi_{b}^{\prime\,0}\rightarrow\Xi_{b}^{0}$  & $34.3$   & $36.4$  & $\cdot\cdot\cdot$  & $\cdot\cdot\cdot$  & $84.6$  & $\cdot\cdot\cdot$  & $47(21)$ \\
\hline\hline  
\end{tabular}
\end{table*}

\begin{table*} [p]
\caption{M1 decay widths (in keV) of $\Xi_{cb}$ and $\Omega_{cb}$ baryons.\label{t7.13}}
\begin{tabular}{lccclcc}
\hline\hline  
Transition  & Our    & PM \cite{AHN10b} & RQM \cite{BFGIKLO10} & Transition  & PM \cite{AHN10b}  & $\chi$PT \cite{LMLZ18}\\
 & Mixed    & Mixed  & Mixed  &  & Unmix & Unmix\\
\hline 
$\Xi_{cb}^{*0}\rightarrow\Xi_{cb}^{\prime\,0}$  & $7.6\cdot10^{-5}$  & $0.0012$  & $2\cdot10^{-6}$ & $\Xi_{cb}^{*0}\rightarrow\Xi_{[cb]d}^{0}$  & $0.0404$ & $0.52$\\
$\Xi_{cb}^{*0}\rightarrow\Xi_{cb}^{0}$  & $0.876$   & $1.03$  & $0.51$  & $\Xi_{cb}^{*0}\rightarrow\Xi_{\{cb\}d}^{0}$  & $0.505$ & $7.19$\\
$\Xi_{cb}^{\prime\,0}\rightarrow\Xi_{cb}^{0}$  & $0.204$   & $0.209$  & $0.31$  & $\Xi_{[cb]d}^{0}\rightarrow\Xi_{\{cb\}d}^{0}$  & $0.00992$ & $\cdot\cdot\cdot$ \\
$\Xi_{cb}^{*+}\rightarrow\Xi_{cb}^{\prime\,+}$  & $0.0293$   & $0.0605$  & $0.0015$  & $\Xi_{cb}^{*+}\rightarrow\Xi_{[cb]u}^{+}$  & $0.0404$ & $0.52$\\
$\Xi_{cb}^{*+}\rightarrow\Xi_{cb}^{+}$  & $1.31$   & $0.739$  & $0.46$  & $\Xi_{cb}^{*+}\rightarrow\Xi_{\{cb\}u}^{+}$  & $1.05$ & $26.2$\\
$\Xi_{cb}^{\prime\,+}\rightarrow\Xi_{cb}^{+}$  & $0.161$   & $0.124$  & $0.14$  & $\Xi_{[cb]u}^{+}\rightarrow\Xi_{\{cb\}u}^{+}$  & $0.00992$ & $\cdot\cdot\cdot$ \\
$\Omega_{cb}^{*0}\rightarrow\Omega_{cb}^{\prime\,0}$  & $1.3\cdot10^{-5}$  & $0.0031$  & $1\cdot10^{-6}$ & $\Omega_{cb}^{*0}\rightarrow\Omega_{[cb]s}^{0}$  & $0.0369$ & $0.52$\\
$\Omega_{cb}^{*0}\rightarrow\Omega_{cb}^{0}$  & $0.637$   & $0.502$  & $0.29$  & $\Omega_{cb}^{*0}\rightarrow\Omega_{\{cb\}s}^{0}$  & $0.209$ & $7.08$\\
$\Omega_{cb}^{\prime\,0}\rightarrow\Omega_{cb}^{0}$  & $0.170$  & $0.0852$  & $0.21$  & $\Omega_{[cb]s}^{0}\rightarrow\Omega_{\{cb\}s}^{0}$  & $0.00568$ & $\cdot\cdot\cdot$ \\
\hline\hline  
\end{tabular}
\end{table*}

\begin{table*} [p]
\centering\protect\protect\caption{M1 decay widths (in keV) of doubly and triply heavy baryons.\label{t7.14}}
\begin{tabular}{lccccccc}
\hline\hline  
Decay  & Our    & QM & SRPM  & RQM & RQM  & $\chi$PT\\
 &   & \cite{XWLZZ17}  & \cite{LWXZ17}  & \cite{BFGIKLO10} & Resc  & \cite{LMLZ18}\\
\hline 
$\Xi_{cc}^{*+}\rightarrow\Xi_{cc}^{+}$  & $2.17$    & $14.6$  & $3.90$  & $28.8$ & $1.46$ & $9.57$\\
$\Xi_{cc}^{*++}\rightarrow\Xi_{cc}^{++}$  & $2.79$  & $16.7$  & $7.21$  & $23.5$ & $1.19$ & $22.0$\\
$\Omega_{cc}^{*+}\rightarrow\Omega_{cc}^{+}$  & $1.60$   & $6.93$  & $0.82$  & $2.11$ & $1.22$  & $9.45$\\
$\Xi_{bb}^{*-}\rightarrow\Xi_{bb}^{-}$  & $0.0268$  & $0.24$ & $0.21$ & $0.059$ & $0.0215$ & $5.17$\\
$\Xi_{bb}^{*0}\rightarrow\Xi_{bb}^{0}$  & $0.137$  & $1.19$ & $0.98$ & $0.31$ & $0.113$ & $31.1$\\
$\Omega_{bb}^{*-}\rightarrow\Omega_{bb}^{-}$  & $0.0148$ & $0.08$ & $0.04$ & $0.0226$ & $0.0131$ & $5.08$\\
$\Omega_{ccb}^{*+}\rightarrow\Omega_{ccb}^{+}$  & $0.0096$  & $\cdot\cdot\cdot$ & $\cdot\cdot\cdot$  & $\cdot\cdot\cdot$  & $\cdot\cdot\cdot$  & $\cdot\cdot\cdot$ \\
$\Omega_{cbb}^{*0}\rightarrow\Omega_{cbb}^{0}$  & $0.0130$  & $\cdot\cdot\cdot$ & $\cdot\cdot\cdot$  & $\cdot\cdot\cdot$  & $\cdot\cdot\cdot$  & $\cdot\cdot\cdot$ \\
\hline\hline  
\end{tabular}
\end{table*}

We see that for singly heavy baryons our predictions are similar to
the corresponding results obtained using other approaches. The exception
is the transition $\Sigma_{c}^{*+}\rightarrow\Sigma_{c}^{+}$. In
this case the transition moment is very sensitive to the choice of
the model, and the accuracy of the predicted values is not quite clear.
Our prediction for this transition practically coincides with the
result obtained using hypercentral approach \cite{MPV09}, while other
models predict lower values. For the
doubly charmed baryons our predictions are similar to the effective
mass and charge scheme (EM\&C) results, and lower than $\chi$QM or
$\chi$PT estimates. For the doubly bottom and triply heavy baryons
we have not found other predictions. The comparison with NR results
shows that our predictions are somewhat lower. This is usual, expected
behavior.
Analyzing results presented in Tables~\ref{t7.3},
\ref{t7.4} we see that the state mixing effect is especially large
for the transitions $\Xi_{c}^{\prime\,0}\rightarrow\Xi_{c}^{0}$ and
$\Xi_{c}^{*+}\rightarrow\Xi_{c}^{\prime\,+}$. In these cases the
values of the mixed transition moments are approximately four times
smaller than the unmixed ones. The mixed transition moment
 $\mu(\Xi_{c}^{*0}\rightarrow\Xi_{c}^{0})$
is $\approx20\%$ larger, the transition moment
 $\mu(\Xi_{b}^{\prime\,-}\rightarrow\Xi_{b}^{-})$
is $\approx20\%$ smaller, and the remaining moments do not undergo large
changes. In order to estimate the state mixing effect on the transition
moments of $\Xi_{cb}$ and $\Omega_{cb}$ baryons in Table~\ref{t7.6}
the unmixed moments in the two spin coupling schemes $(qc)b$ and $(cb)q$
are given. The unmixed results calculated in the $(cb)q$ basis are
also compared with recent predictions obtained in the framework of
heavy hadron chiral perturbation theory \cite{LMLZ18}. We see that
the effect of the mixing is appreciable, especially for the transitions
$\Xi_{cb}^{*0}\rightarrow\Xi_{cb}^{\prime\,0}$ , $\Xi_{cb}^{\prime\,+}\rightarrow\Xi_{cb}^{+}$,
and $\Omega_{cb}^{*0}\rightarrow\Omega_{cb}^{\prime\,0}$. 

As the last step, we have used the obtained above transition moments to
calculate the corresponding decay widths. In the
calculations the values of the photon momentum given
in Sec.~\ref{sec_hyp} were used.  The results are collected
in Tables~\ref{t7.9}--\ref{t7.14}. The predictions are compared
with other available results given by various approaches. In the case of $\Sigma_{Q}$
baryons our predictions are more or less
compatible with the potential model (PM, Hyp) results, but the agreement
is not very good. The predictions obtained using other approaches
differ from each other. For the $\Xi_{Q}$ 
baryons again all predictions are different. Note that the decay widths
$\Gamma(\Xi_{c}^{\prime\,0}\rightarrow\Xi_{c}^{0}\,\gamma)$ and $\Gamma(\Xi_{c}^{*+}\rightarrow\Xi_{c}^{\prime\,+}\,\gamma)$
in the physical basis are order of magnitude smaller than corresponding
unmixed values. Available lattice QCD estimates $\Gamma(\Omega_{c}^{*0}\rightarrow\Omega_{c}^{0}\,\gamma)$,
$\Gamma(\Xi_{c}^{\prime\,0}\rightarrow\Xi_{c}^{0}\,\gamma)$, and
$\Gamma(\Xi_{c}^{\prime\,+}\rightarrow\Xi_{c}^{+}\,\gamma)$ are considerably
lower than other predictions. For the decay $\Xi_{c}^{\prime\,0}\rightarrow\Xi_{c}^{0}\,\gamma$
they practically even did not find any signal. 

The case of doubly heavy $\Xi_{cb}$ and $\Omega_{cb}$ baryons is the nice exception for
us, because we can compare our predictions with
the mixed estimates obtained using other approaches. These are the predictions obtained in nonrelativistic
potential model \cite{AHN10b} and in relativistic three quark model
\cite{BFGIKLO10}. Models are rather different and provide to some
extent different results. Our predictions for the transitions $\Xi_{cb}^{*0}\rightarrow\Xi_{cb}^{\prime\,0}$,
$\Xi_{cb}^{*0}\rightarrow\Xi_{cb}^{0}$, $\Xi_{cb}^{*+}\rightarrow\Xi_{cb}^{\prime\,+}$,
$\Omega_{cb}^{*0}\rightarrow\Omega_{cb}^{\prime\,0}$, and $\Omega_{cb}^{\prime\,0}\rightarrow\Omega_{cb}^{0}$
are somewhere between their results, for the transitions $\Xi_{cb}^{*+}\rightarrow\Xi_{cb}^{+}$, $\Xi_{cb}^{\prime\,+}\rightarrow\Xi_{cb}^{+}$, and $\Omega_{cb}^{*0}\rightarrow\Omega_{cb}^{0}$
are larger than both, and for the transition $\Xi_{cb}^{\prime\,0}\rightarrow\Xi_{cb}^{0}$
our result is close to the PM prediction. For the sake of comparison
we also included in Table~\ref{t7.13} the unmixed results obtained
in $(cb)q$ basis using potential model and heavy hadron chiral perturbation
theory. We see that $\chi$PT predictions are approximately order of
magnitude larger than PM results. We have no our own predictions for the decay widths in this case because we have no good estimates for the hyperfine mass splittings
of unmixed states in $(cb)q$ basis. 

For the transitions of doubly heavy baryons $\Xi_{QQ}^{*}\rightarrow\Xi_{QQ}$
and $\Omega_{QQ}^{*}\rightarrow\Omega_{QQ}$ our results are smaller
than corresponding results obtained using other approaches. The models
are different, and we can expect different predictions. But in one
case (i.e. the RQM \cite{BFGIKLO10}) the main reason of the discrepancies
is clear. In this model the hyperfine splittings $\Delta\Omega_{cc}$
and $\Delta\Omega_{bb}$ are $\approx1.2$ times, $\Delta\Xi_{bb}$ $\approx1.4$
times, and $\Delta\Xi_{cc}$ even $\approx2.7$ times larger than ours.
Simple rescaling brings RQM results close to ours. For comparison, these rescaled values are also included in Table~\ref{t7.14} (labeled as Resc).

The decay widths for the transitions $\Xi_{cc}^{*+}\rightarrow\Xi_{cc}^{+}$ and $\Xi_{cc}^{*++}\rightarrow\Xi_{cc}^{++}$
were also analyzed in the framework of chiral quark model in Ref.~\cite{HM06b}.
Using their results one can obtain the following estimates

\begin{align}
\Gamma(\Xi_{cc}^{*+}\rightarrow\Xi_{cc}^{+}\,\gamma)  & \approx(2.6-3.5)\,\mathrm{keV},\\
\Gamma(\Xi_{cc}^{*++}\rightarrow\Xi_{cc}^{++}\,\gamma)  & \approx(2.3-3.4)\,\mathrm{keV}.
\end{align}

These estimates are compatible with our predictions.

\section{Discussion and summary}  \label{sec_sum}

The main purpose of this work was to obtain better estimates for the
magnetic moments and magnetic dipole decay widths of heavy ground-state
hadrons. For this end several improvements were implemented.
One such improvement is the new prescription proposed in our previous work
\cite{S16a} to deal with c.m.m. corrections for the magnetic observables 
calculated in the bag model framework. Usually c.m.m. corrections
are applied to the whole observable, e.g. magnetic moment of the hadron.
On the other hand, the expression (\ref{mbag 037}) is defined at
the quark level. This choice allows us to overcome the difficulty
present in the case of heavy-light hadrons, where the usual approach
can not ensure reliable predictions. The point is that in the heavy-light
systems the size of the c.m.m. correction is governed mainly
by heavy quarks, but it is applied to the whole observable (all quarks) in contradiction with the expectations from the HQS.
Eq. (\ref{mbag 037}) solves this problem because the magnetic observables
of heavy and light quarks are scaled independently. In addition, in
the present paper the new procedure, seemingly more suitable for the description
of the magnetic properties of heavy hadrons, is used to determine
the model parameters ($C_{L}$ and $C_{H}$) responsible for the scaling
of quark magnetic observables.

In order to reduce the uncertainties in the calculation of M1 decay
widths the experimental values of the hadron masses (if available)
were used. When the experimental data are absent we resort to the theoretical estimates.  Some efforts were undertaken to make these estimates more reliable. 
For this purpose we have employed the semiempirical approach based
on the quark model relations using as input the experimental
data, lattice QCD predictions, and theoretical predictions
obtained in the AL1 potential model. So, strongly speaking, 
the decay widths obtained in our work are
not pure bag model predictions. We have no intent to test the bag model
facilities. The advantages and drawbacks of this model are well known.
Instead, our strategy was to combine various methods in order to obtain
as good predictions as possible. As a byproduct we also obtain an
estimate for the mass of the $\Xi_{cc}^{*}$ baryon $M(\Xi_{cc}^{*})=3695\pm5$ MeV. 

The essential ingredient in our analysis of the baryons containing
three differently flavored quarks is the proper treatment of the state
mixing problem. As we have seen above, if one seeks to get the reliable
predictions for the magnetic moments and M1 decay widths of these baryons the effect
of the hyperfine mixing must be taken into account (see also Refs.~\cite{FLNC81,AHN10b,BFGIKLO10}). 

It is a difficult task to obtain the meaningful error estimates in any model, especially when a model is extended to a new region. In the heavy meson sector one has the experimental
data (not very accurate) only for two M1 decays $J/\psi\rightarrow\eta_{c}\gamma$
and $D^{*+}\rightarrow D^{+}\gamma$. We have set the parameter $C_{H}$
to reproduce the PDG average of $\Gamma(J/\psi\rightarrow\eta_{c}\gamma)$.
Therefore, the only one experimental result remains for the comparison.
The agreement is very good, but this is not enough to trust all other
predictions. Also one can check if the predictions in the cases when
nonrelativistic approximation is expected to hold (M1 transition moment 
$\mu(\Upsilon\rightarrow\eta_{b}\gamma)$, magnetic moment of triply heavy baryon
 $\Omega_{bbb}^{*-}$\,) agree with the estimates obtained using NR approach.
We see that the agreement is again good. 

Having in mind that the accuracy of the model predictions can not
be higher than the accuracy of the data used in the fitting of the
model parameters we can obtain a crude estimate of possible errors.
Such estimate for the magnetic moments of the up, down, strange, and
bottom quarks is $\approx8\%$, for the charmed quark $\approx15\%$. 
Because the magnetic moment of the bottom quark is small the uncertainty in its value does not affect other predictions very much.
The main source of possible errors is the uncertainty in the value
of the magnetic moment of charmed quark, but the substantial improvement in this field at the present time seems to be hardly possible. 
Though these estimates come from the meson sector, we optimistically expect similar uncertainties to hold in the heavy baryon sector, too. To estimate 
the accuracy of the predictions for the magnetic properties of particular
hadrons in each case more detailed analysis is necessary.

To conclude, we have performed a comprehensive analysis of the magnetic
properties (magnetic moments, magnetic dipole decay widths) of 
the ground-state heavy hadrons. The agreement of the predictions with the 
available experimental data and with some (but not all)
theoretical predictions is good. To our knowledge, some of our predictions, such as magnetic moments of neutral heavy mesons ($D^{*0}$, $B^{*0}$, $B_{s}^{*0}$),
and the decay widths of triply heavy baryons $\Gamma(\Omega_{ccb}^{*+}\rightarrow\Omega_{ccb}^{+}\,\gamma)$,
$\Gamma(\Omega_{cbb}^{*0}\rightarrow\Omega_{cbb}^{0}\,\gamma)$ still are the only available theoretical estimates.


\begin{thebibliography}{399}
\bibitem{KR10} E. Klempt and J. M. Richard, \href{https://doi.org/10.1103/RevModPhys.82.1095}{Rev. Mod. Phys. \textbf{82}, 1095 (2010)}.

\bibitem{BeA11a} N. Brambilla \textit{et al}., \href{https://doi.org/10.1140/epjc/s10052-010-1534-9}{Eur. Phys. J. C \textbf{71}, 1534 (2011)}.

\bibitem{C15} H.-Y. Cheng, \href{https://doi.org/10.1007/s11467-015-0483-z}{Front. Phys. \textbf{10}, 101406 (2015)}.

\bibitem{CCLLZ17} H.-X. Chen, W. Chen, X. Liu, Y.-R. Liu, and S.-L.
Zhu, \href{https://doi.org/10.1088/1361-6633/aa6420}{Rept. Prog. Phys. \textbf{80}, 076201 (2017)}.

\bibitem{AeA17b} R. Aaij \textit{et al}. (LHCb Collaboration), \href{https://doi.org/10.1103/PhysRevLett.119.112001}{Phys. Rev. Lett. \textbf{119}, 112001 (2017)}.

\bibitem{KKP94}  J.  G. K\"{o}rner, M. Kr\"{a}mer, and D. Pirjol, 
\href{https://doi.org/10.1016/0146-6410(94)90053-1}
{Prog. Part. Nucl. Phys. \textbf{33}, 787 (1994)}.

\bibitem{KR14} M. Karliner and J. L. Rosner, \href{https://doi.org/10.1103/PhysRevD.90.094007}{Phys. Rev. D \textbf{90}, 094007 (2014)}.

\bibitem{S96d} B. Silvestre-Brac, \href{https://doi.org/10.1007/s006010050028}{Few-Body Syst. \textbf{20}, 1 (1996)}.

\bibitem{S96e} B. Silvestre-Brac, \href{https://doi.org/10.1016/0146-6410(96)00030-0}{Prog. Part. Nucl. Phys. \textbf{36}, 263 (1996)}.

\bibitem{EFGM02} D. Ebert, R. N. Faustov, V. O. Galkin, and A. P.
Martynenko, \href{https://doi.org/10.1103/PhysRevD.66.014008}{Phys. Rev. D \textbf{66}, 014008 (2002)}.

\bibitem{NeA13} Y. Namekawa \textit{et al}. (PACS-CS Collaboration),
\href{https://doi.org/10.1103/PhysRevD.87.094512}{Phys. Rev. D \textbf{87}, 094512 (2013)}.

\bibitem{BDMO14} Z. S. Brown, W. Detmold, S. Meinel, and K. Orginos,
\href{https://doi.org/10.1103/PhysRevD.90.094507}{Phys. Rev. D \textbf{90}, 094507 (2014)}.

\bibitem{AK17} C. Alexandrou and C. Kallidonis, \href{https://doi.org/10.1103/PhysRevD.96.034511}{Phys. Rev. D \textbf{96}, 034511 (2017)}.

\bibitem{S16a}  V. \v{S}imonis, 
\href{https://doi.org/10.1140/epja/i2016-16090-5}
{Eur. Phys. J. A \textbf{52}, 90 (2016)}.

\bibitem{L78a} H. J. Lipkin, \href{https://doi.org/10.1016/0370-2693(78)90689-5}{Phys. Lett. \textbf{74}, 399 (1978)}.

\bibitem{R81b} J. M. Richard, \href{https://doi.org/10.1016/0370-2693(81)90618-3}{Phys. Lett. B \textbf{100}, 515 (1981)}.

\bibitem{BCN81} R. K. Bhaduri, L. E. Cohler, and Y. Nogami, \href{https://doi.org/10.1007/BF02827441}{Nuov. Cim. A \textbf{65}, 376 (1981)}.

\bibitem{BSS02} F. Brau, C. Semay, and B. Silvestre-Brac, \href{https://doi.org/10.1103/PhysRevC.66.055202}{Phys. Rev. C \textbf{66}, 055202 (2002)}.

\bibitem{CI86} S. Capstick and N. Isgur, \href{https://doi.org/10.1103/PhysRevD.34.2809}{Phys. Rev. D \textbf{34}, 2809 (1986)}.

\bibitem{ECAKR09} G. Eichmann, I.  C. Clo\"{e}t, R. Alkofer, A. Krassnigg, and C.  D. Roberts, 
\href{https://doi.org/10.1103/PhysRevC.79.012202}
{Phys. Rev. C \textbf{79}, 012202(R) (2009)}.

\bibitem{RCCR11} H.  L.  L. Roberts, L. Chang, I.  C.  Clo\"{e}t, and C. D. Roberts, 
\href{https://doi.org/10.1007/s00601-011-0225-x}
{Few-Body Syst. \textbf{51}, 1 (2011)}.

\bibitem{S03a} A. Samsonov, \href{http://dx.doi.org/10.1088/1126-6708/2003/12/061}{J. High Energy Phys. \textbf{12}, 061 (2003)}.

\bibitem{AOS09}  T.  M. Aliev, A. \"{O}zpineci, and M. Savc\i ,
\href{https://doi.org/10.1016/j.physletb.2009.06.073}
{Phys. Lett. B \textbf{678}, 470 (2009)}. 

\bibitem{CGNSS95a}  F. Cardarelli, I. Grach, I. Narodetskii, G. Salm\`{e}, and S. Simula, 
\href{https://doi.org/10.1016/0370-2693(95)00230-I}
{Phys. Lett. B \textbf{349}, 393 (1995)}.

\bibitem{MS02} D. Melikhov and S. Simula, \href{https://doi.org/10.1103/PhysRevD.65.094043}{Phys. Rev. D \textbf{65}, 094043 (2002)}.

\bibitem{J03} W. Jaus, \href{https://doi.org/10.1103/PhysRevD.67.094010}{Phys. Rev. D \textbf{67}, 094010 (2003)}.

\bibitem{CJ04} H.-M. Choi and C.-R. Ji, \href{https://doi.org/10.1103/PhysRevD.70.053015}{Phys. Rev. D \textbf{70}, 053015 (2004)}.

\bibitem{ddMF14} J. P. B. C. de Melo, A. N. da Silva, C. S. Mello, and T. Frederico, 
\href{https://doi.org/10.1051/epjconf/20147303017}{EPJ Web Conf. \textbf{73}, 03017 (2014)}.

\bibitem{MddF15} C. S. Mello, A. N. da Silva, J. P. B. C.
de Melo, and T. Frederico, \href{https://doi.org/10.1007/s00601-015-0961-4}{Few-Body Syst. \textbf{56}, 509 (2015)}.

\bibitem{dT17} J. P. B. C. de Melo and K. Tsushima, \href{https://doi.org/10.1007/s00601-017-1233-2}{Few-Body Syst. \textbf{58}, 82 (2017)}.

\bibitem{SD17a} B.-D. Sun and Y.-B. Dong, \href{https://doi.org/10.1103/PhysRevD.96.036019}{Phys. Rev. D \textbf{96}, 036019 (2017)}.

\bibitem{HP99} F. T. Hawes and M. A. Pichowsky, \href{https://doi.org/10.1103/PhysRevC.59.1743}{Phys. Rev. C \textbf{59}, 1743 (1999)}.

\bibitem{BM08} M. S. Bhagwat and P. Maris, \href{https://doi.org/10.1103/PhysRevC.77.025203}{Phys. Rev. C \textbf{77}, 025203 (2008)}.

\bibitem{RBGRW11}  H.  L.  L. Roberts, A. Bashir, L.  X. Guti\'{e}rrez-Guerrero, C.  D. Roberts, and D.  J. Wilson, 
\href{https://doi.org/10.1103/PhysRevC.83.065206}
{Phys. Rev. C \textbf{83}, 065206 (2011)}.

\bibitem{PSRRSW13} M. Pitschmann, C.-Y. Seng, M. J. Ramsey-Musolf,
C. D. Roberts, S. M. Schmidt, and D. J. Wilson, \href{https://doi.org/10.1103/PhysRevC.87.015205}{Phys. Rev. C \textbf{87}, 015205 (2013)}.

\bibitem{BS13d} A. M. Badalian and Yu. A. Simonov, \href{https://doi.org/10.1103/PhysRevD.87.074012}{Phys. Rev. D \textbf{87}, 074012 (2013)}.

\bibitem{DEGM14} D. Djukanovic, E. Epelbaum, J. Gegelia, and U.-G.
Mei\ss ner, \href{https://doi.org/10.1016/j.physletb.2014.01.001}{Phys. Lett. B \textbf{730}, 115 (2014)}.

\bibitem{AW97} W. Andersen and W. Wilcox, \href{https://doi.org/10.1006/aphy.1996.5648}{Ann. Phys. \textbf{255}, 34 (1997)}.

\bibitem{LMW08} F. X. Lee, S. Moerschbacher, and W. Wilcox, \href{https://doi.org/10.1103/PhysRevD.78.094502}{Phys. Rev. D \textbf{78}, 094502 (2008)}.

\bibitem{HKLLWZ07} J. N. Hedditch, W. Kamleh, B. G. Lasscock,
D. B. Leinweber, A. G. Williams, and J. M. Zanotti, \href{https://doi.org/10.1103/PhysRevD.75.094504}{Phys. Rev. D \textbf{75}, 094504 (2007)}.

\bibitem{OKLMM15a} B. J. Owen, W. Kamleh, D. B. Leinweber, M. S.
Mahbub, and B. J. Menadue, \href{https://doi.org/10.1103/PhysRevD.91.074503}{Phys. Rev. D \textbf{91}, 074503 (2015)}.

\bibitem{SDE15} C. J. Shultz, J. J. Dudek, and R. G. Edwards,
\href{https://doi.org/10.1103/PhysRevD.91.114501}{Phys. Rev. D \textbf{91}, 114501 (2015)}.

\bibitem{LST17} E. V. Lushevskaya, O. E. Solovjeva, and O. V.
Teryaev, \href{https://doi.org/10.1007/JHEP09(2017)142}{J. High Energy Phys. \textbf{09}, 142 (2017)}.

\bibitem{GS15a}  D.  G. Guidi\~{n}o and G.  T. S\'{a}nchez, 
\href{https://doi.org/10.1142/S0217751X15501146}
{Int. J. Mod. Phys. A \textbf{30}, 1550114 (2015)}. 

\bibitem{BS80} S. K. Bose and L. P. Singh, \href{https://doi.org/10.1103/PhysRevD.22.773}{Phys. Rev. D \textbf{22}, 773 (1980)}.

\bibitem{LCD15} Y.-L. Luan, X.-L. Chen, and W.-Z. Deng, \href{https://doi.org/10.1088/1674-1137/39/11/113103}{Chin. Phys. C  \textbf{39}, 113103 (2015)}.

\bibitem{CBCT15} 
M.  E. Carrillo-Serrano, W. Bentz, I.  C. Clo\"{e}t, and A.  W. Thomas, 
\href{https://doi.org/10.1103/PhysRevC.92.015212}
{Phys. Rev. C \textbf{92}, 015212 (2015)}.

\bibitem{BS14} E. P. Biernat and W. Schweiger, \href{https://doi.org/10.1103/PhysRevC.89.055205}{Phys. Rev. C \textbf{89}, 055205 (2014)}.

\bibitem{KPT18} A. F. Krutov, R. G. Polezhaev, and V. E. Troitsky,
\href{https://doi.org/10.1103/PhysRevD.97.033007}
{Phys. Rev. D \textbf{97}, 033007 (2018)}.

\bibitem{ABL78} T. Applelquist, R. M. Barnett, and K. Lane, \href{https://doi.org/10.1146/annurev.ns.28.120178.002131}{Annu. Rev. Nucl. Part. Sci. \textbf{28}, 387 (1978)}.

\bibitem{HI82} C. Hayne and N. Isgur, \href{https://doi.org/10.1103/PhysRevD.25.1944}{Phys. Rev. D \textbf{25}, 1944 (1982)}.

\bibitem{KRQ87} W. Kwong, J. L. Rosner, and C. Quigg, \href{https://doi.org/10.1146/annurev.ns.37.120187.001545}{Ann. Rev. Nucl. Part. Sci. \textbf{37}, 325 (1987)}.

\bibitem{ST87} E. Sucipto and R. L. Thews, \href{https://doi.org/10.1103/PhysRevD.36.2074}{Phys. Rev. D \textbf{36}, 2074 (1987)}.

\bibitem{KX92} A. N. Kamal and Q. P. Xu, \href{https://doi.org/10.1016/0370-2693(92)90455-D}{Phys. Lett. B \textbf{284}, 421 (1992)}.

\bibitem{R13} J. L. Rosner, \href{https://doi.org/10.1103/PhysRevD.88.034034}{Phys. Rev. D \textbf{88}, 034034 (2013)}.

\bibitem{GYS14}  P. Guo, T. Y\'{e}pez-Mart\'{\i}nez, and A.  P. Szczepaniak, 
\href{https://doi.org/10.1103/PhysRevD.89.116005}
{Phys. Rev. D \textbf{89}, 116005 (2014)}.

\bibitem{CDN93} P. Colangelo, F. De Fazio, and G. Nardulli, \href{https://doi.org/10.1016/0370-2693(93)91043-M}{Phys. Lett. B \textbf{316}, 555 (1993)}.

\bibitem{FR94} Fayyazuddin and Riazuddin, \href{https://doi.org/10.1016/0370-2693(94)91467-2}{Phys. Lett. B \textbf{337}, 189 (1994)}.

\bibitem{JMS95} P. Jain, A. Momen, and J. Schechter, \href{https://doi.org/10.1142/S0217751X95001182}{Int. J. Mod. Phys. A \textbf{10}, 2467 (1995)}.

\bibitem{OX94} P. J. O'Donnell and Q. P. Xu, \href{https://doi.org/10.1016/0370-2693(94)00975-9}{Phys. Lett. B \textbf{336}, 113 (1994)}.

\bibitem{CGNSS95b}  F. Cardarelli, I. Grach, I. Narodetskii, G. Salm\`{e}, and S. Simula,
\href{https://doi.org/10.1016/0370-2693(95)01058-X}
{Phys. Lett. B \textbf{359}, 1 (1995)}.

\bibitem{J96b} W. Jaus, \href{https://doi.org/10.1103/PhysRevD.53.1349}{Phys. Rev. D \textbf{53}, 1349 (1996)}.

\bibitem{J99} W. Jaus, \href{https://doi.org/10.1103/PhysRevD.60.054026}{Phys. Rev. D \textbf{60}, 054026 (1999)}.

\bibitem{CJ99} H.-M. Choi and C.-R. Ji, \href{https://doi.org/10.1103/PhysRevD.59.074015}{Phys. Rev. D \textbf{59}, 074015 (1999)}.

\bibitem{C07} H.-M. Choi, \href{https://doi.org/10.1103/PhysRevD.75.073016}{Phys. Rev. D \textbf{75}, 073016 (2007)}.

\bibitem{CJ09} H.-M.  Choi and  C.-R.  Ji, 
\href{https://doi.org/10.1103/PhysRevD.80.054016}{Phys.  Rev.  D  \textbf{80}, 054016  (2009)}.

\bibitem{HW07} C.-W. Hwang and Z.-T Wei, \href{https://doi.org/10.1088/0954-3899/34/4/008}{J. Phys. G \textbf{34}, 687 (2007)}.

\bibitem{HDDT78} R. H. Hackman, N. G. Deshpande, D. A. Dicus,
and V. L. Teplitz, \href{https://doi.org/10.1103/PhysRevD.18.2537}{Phys. Rev. D \textbf{18}, 2537 (1978)}.

\bibitem{IDS82} D. Izatt, C. DeTar, and M. Stephenson, \href{https://doi.org/10.1016/0550-3213(82)90347-9}{Nucl. Phys. B \textbf{199}, 269 (1982)}.

\bibitem{WMM85} W. Wilcox, O. V. Maxwell, and K. A. Milton, \href{https://doi.org/10.1103/PhysRevD.31.1081}{Phys. Rev. D \textbf{31}, 1081 (1985)}.

\bibitem{SM86a} P. Singer and G. A. Miller, \href{https://doi.org/10.1103/PhysRevD.33.141}{Phys. Rev. D \textbf{33}, 141 (1986)}.

\bibitem{MS88} G. A. Miller and P. Singer, \href{https://doi.org/10.1103/PhysRevD.37.2564}{Phys. Rev. D \textbf{37}, 2564 (1988)}.

\bibitem{SM89} P. Singer and G. A. Miller, \href{https://doi.org/10.1103/PhysRevD.39.825}{Phys. Rev. D \textbf{39}, 825 (1989)}.

\bibitem{ZCT93} Y. S. Zhong, T. S. Cheng, and A. W. Thomas,
\href{https://doi.org/10.1016/0375-9474(93)90261-U}{Nucl. Phys. A \textbf{559}, 579 (1993)}.

\bibitem{OH99}  A.  H. \"{O}rsland and H. H\"{o}gaasen, 
\href{https://doi.org/10.1007/s100520050044}
{Eur. Phys. J. C \textbf{9}, 503 (1999)}.

\bibitem{EGKLY80a} E. Eichten, K. Gottfried, T. Kinoshita, K. D.
Lane, and T.-M. Yan, \href{https://doi.org/10.1103/PhysRevD.21.203}{Phys. Rev. D \textbf{21}, 203 (1980)}.

\bibitem{EQ94} E. J. Eichten and C. Quigg, \href{https://doi.org/10.1103/PhysRevD.49.5845}{Phys. Rev. D \textbf{49}, 5845 (1994)}.

\bibitem{EGMR08} E. Eichten, S. Godfrey, H. Mahlke, and J. L. Rosner,
\href{https://doi.org/10.1103/RevModPhys.80.1161}{Rev. Mod. Phys. \textbf{80}, 1161 (2008)}.

\bibitem{BJ82} N. Barik and S. N. Jena, \href{https://doi.org/10.1103/PhysRevD.26.618}{Phys. Rev. D \textbf{26}, 618 (1982)}.

\bibitem{GOS84} H. Grotch, D. A. Owen, and K. J. Sebastian, \href{https://doi.org/10.1103/PhysRevD.30.1924}{Phys. Rev. D \textbf{30}, 1924 (1984)}.

\bibitem{ZSG91} X. Zhang, K. J. Sebastian, and H. Grotch, \href{https://doi.org/10.1103/PhysRevD.44.1606}{Phys. Rev. D \textbf{44}, 1606 (1991)}.

\bibitem{GKLT95a} S. S. Gershtein, V. V. Kiselev, A. K. Likhoded,
and A. V. Tkabladze, \href{https://doi.org/10.1103/PhysRevD.51.3613}{Phys. Rev. D \textbf{51}, 3613 (1995)}.

\bibitem{F99d} L. P. Fulcher, \href{https://doi.org/10.1103/PhysRevD.60.074006}{Phys. Rev. D \textbf{60}, 074006 (1999)}.

\bibitem{BSG02} R. Bonnaz, B. Silvestre-Brac, and C. Gignoux, \href{https://doi.org/10.1007/s10050-002-8765-6}{Eur. Phys. J. A \textbf{13}, 363 (2002)}.

\bibitem{BEH03} W. A. Bardeen, E. J. Eichten, and C. T. Hill,
\href{https://doi.org/10.1103/PhysRevD.68.054024}{Phys. Rev. D \textbf{68}, 054024 (2003)}.

\bibitem{GVGV03} P. Gonz\'{a}lez, A. Valcarce, H. Garcilazo, and J. Vijande, 
\href{https://doi.org/10.1103/PhysRevD.68.034007}
{Phys. Rev. D \textbf{68}, 034007 (2003)}.

\bibitem{CS05} F. E. Close and E. S. Swanson, \href{https://doi.org/10.1103/PhysRevD.72.094004}{Phys. Rev. D \textbf{72}, 094004 (2005)}.

\bibitem{LS07} O. Lakhina and E. S. Swanson, \href{https://doi.org/10.1016/j.physletb.2007.01.075}{Phys. Lett. B \textbf{650}, 159 (2007)}.

\bibitem{PPV10} A. Parmar, B. Patel, and P. C. Vinodkumar, \href{https://doi.org/10.1016/j.nuclphysa.2010.08.016}{Nucl. Phys. A \textbf{848}, 299 (2010)}.

\bibitem{BVM11} Bhagyesh, K. B. Vijaya Kumar, and A. P. Monterio,
\href{https://doi.org/10.1088/0954-3899/38/8/085001}{J. Phys. G \textbf{38}, 085001 (2011)}.

\bibitem{MBV17a} A. P. Monteiro, M. Bhat, and K. B. Vijaya Kumar,
\href{https://doi.org/10.1142/S0217751X1750021X}{Int. J. Mod. Phys. A \textbf{32}, 1750021 (2017)}.

\bibitem{CYC12} L. Cao, Y.-C. Yang, and H. Chen, \href{https://doi.org/10.1007/s00601-012-0478-z}{Few-Body Syst. \textbf{53}, 327 (2012)}.

\bibitem{SOEF16} J. Segovia, P. G. Ortega, D. R. Entem, and F.
Fern\'{a}ndez, \href{https://doi.org/10.1103/PhysRevD.93.074027}{Phys. Rev. D \textbf{93}, 074027 (2016)}.

\bibitem{DLGZ17a} W.-J. Deng, H. Liu, L.-C. Gui, and X.-H. Zhong,
\href{https://doi.org/10.1103/PhysRevD.95.034026}{Phys. Rev. D \textbf{95}, 034026 (2017)}.

\bibitem{DLGZ17b} W.-J. Deng, H. Liu, L.-C. Gui, and X.-H. Zhong,
\href{https://doi.org/10.1103/PhysRevD.95.074002}{Phys. Rev. D \textbf{95}, 074002 (2017)}.

\bibitem{ASMA17} N. Akbar, M. A. Sultan, B. Masud, and F. Akram,
\href{https://doi.org/10.1103/PhysRevD.95.074018}{Phys. Rev. D \textbf{95}, 074018 (2017)}.

\bibitem{ER93} E. Elaaoud and Riazuddin, \href{https://doi.org/10.1103/PhysRevD.47.1026}{Phys. Rev. D \textbf{47}, 1026 (1993)}.

\bibitem{IV95} M. A. Ivanov and Yu. M. Valit, \href{https://doi.org/10.1007/BF01553990}{Z. Phys. C \textbf{67}, 633 (1995)}.

\bibitem{GR01b} J. L. Goity and W. Roberts, \href{https://doi.org/10.1103/PhysRevD.64.094007}{Phys. Rev. D \textbf{64}, 094007 (2001)}.

\bibitem{EFG02b} D. Ebert, R. N. Faustov, and V. O. Galkin, \href{https://doi.org/10.1016/S0370-2693(02)01939-1}{Phys. Lett. B \textbf{537}, 241 (2002)}.

\bibitem{EFG03a} D. Ebert, R. N. Faustov, and V. O. Galkin, \href{https://doi.org/10.1103/PhysRevD.67.014027}{Phys. Rev. D \textbf{67}, 014027 (2003)}.

\bibitem{GI85} S. Godfrey and N. Isgur, \href{https://doi.org/10.1103/PhysRevD.32.189}{Phys. Rev. D \textbf{32}, 189 (1985)}.

\bibitem{G04} S. Godfrey, \href{https://doi.org/10.1103/PhysRevD.70.054017}{Phys. Rev. D \textbf{70}, 054017 (2004)}.

\bibitem{BGS05} T. Barnes, S. Godfrey, and E. S. Swanson, \href{https://doi.org/10.1103/PhysRevD.72.054026}{Phys. Rev. D \textbf{72}, 054026 (2005)}.

\bibitem{GM15} S. Godfrey and K. Moats, \href{https://doi.org/10.1103/PhysRevD.92.054034}{Phys. Rev. D \textbf{92}, 054034 (2015)}.

\bibitem{GM16} S. Godfrey and K. Moats, \href{https://doi.org/10.1103/PhysRevD.93.034035}{Phys. Rev. D \textbf{93}, 034035 (2016)}.

\bibitem{GMS16} S. Godfrey, K. Moats, and E. S. Swanson, \href{https://doi.org/10.1103/PhysRevD.94.054025}{Phys. Rev. D \textbf{94}, 054025 (2016)}.

\bibitem{RR07} S. F. Radford and W. W. Repko, \href{https://doi.org/10.1103/PhysRevD.75.074031}{Phys. Rev. D \textbf{75}, 074031 (2007)}.

\bibitem{RRS09} S. F. Radford, W. W. Repko, and M. J. Saelim,
\href{https://doi.org/10.1103/PhysRevD.80.034012}{Phys. Rev. D \textbf{80}, 034012 (2009)}.

\bibitem{GRR17} N. Green, W. W. Repko, and S. F. Radford, \href{https://doi.org/10.1016/j.nuclphysa.2016.11.006}{Nucl. Phys. A \textbf{958}, 71 (2017)}.

\bibitem{DR11} N. Devlani and A. K. Rai, \href{https://doi.org/10.1103/PhysRevD.84.074030}{Phys. Rev. D \textbf{84}, 074030 (2011)}.

\bibitem{DR12} N. Devlani and A. K. Rai, \href{https://doi.org/10.1140/epja/i2012-12104-8}{Eur. Phys. J. A \textbf{48}, 104 (2012)}.

\bibitem{DR13} N. Devlani and A. K. Rai, \href{https://doi.org/10.1007/s10773-013-1494-6}{Int. J. Theor. Phys. \textbf{52}, 2196 (2013)}.

\bibitem{DKR14} N. Devlani, V. Kher, and A. K. Rai, \href{https://doi.org/10.1140/epja/i2014-14154-2}{Eur. Phys. J. A \textbf{50}, 154 (2014)}.

\bibitem{DS87} F. Daghighian and D. Silverman, \href{https://doi.org/10.1103/PhysRevD.36.3401}{Phys. Rev. D \textbf{36}, 3401 (1987)}.

\bibitem{FM93} Fayyazuddin and O. H. Mobarek, \href{https://doi.org/10.1103/PhysRevD.48.1220}{Phys. Rev. D \textbf{48}, 1220 (1993)}.

\bibitem{CDN94} P. Colangelo, F. De Fazio, and G. Nardulli, \href{https://doi.org/10.1016/0370-2693(94)90607-6}{Phys. Lett. B \textbf{334}, 175 (1994)}.

\bibitem{A94a} M. Avila, \href{https://doi.org/10.1103/PhysRevD.49.309}{Phys. Rev. D \textbf{49}, 309 (1994)}.

\bibitem{DFS00} Y. B. Dong, A. Faessler, and K. Shimizu, \href{https://doi.org/10.1016/S0375-9474(99)00822-2}{Nucl. Phys. A \textbf{671}, 380 (2000)}.

\bibitem{CK01c}  H. \c{C}iftci and H. Koru, 
\href{https://doi.org/10.1142/S0217732301004959}
{Mod. Phys. Lett. A \textbf{16}, 1785 (2001)}.

\bibitem{BD94} N. Barik and P. C. Dash, \href{https://doi.org/10.1103/PhysRevD.49.299}{Phys. Rev. D \textbf{49}, 299 (1994)}.

\bibitem{JPT02} S. N. Jena, S. Panda, and T. C. Tripathy, \href{https://doi.org/10.1016/S0375-9474(01)01299-4}{Nucl. Phys. A \textbf{699}, 649 (2002)}.

\bibitem{JMPS10a} S. N. Jena, M. K. Muni, H. R. Pattnaik, and
K. P. Sahu, \href{https://doi.org/10.1142/S0217751X10048214}{Int. J. Mod. Phys. A \textbf{25}, 2063 (2010)}.

\bibitem{JMMP10} S. N. Jena, H. H. Muni, P. K. Mohapatra, and
P. Panda, \href{https://doi.org/10.1139/P10-034}{Can. J. Phys. A \textbf{88}, 517 (2010)}.

\bibitem{SPV14} M. Shah, B. Patel, and P. C. Vinodkumar, \href{https://doi.org/10.1103/PhysRevD.90.014009}{Phys. Rev. D \textbf{90}, 014009 (2014)}.

\bibitem{SPV16a} M. Shah, B. Patel, and P. C. Vinodkumar, \href{https://doi.org/10.1140/epjc/s10052-016-3875-5}{Eur. Phys. J. C \textbf{76}, 36 (2016)}.

\bibitem{SPV16b} M. Shah, B. Patel, and P. C. Vinodkumar, \href{https://doi.org/10.1103/PhysRevD.93.094028}{Phys. Rev. D \textbf{93}, 094028 (2016)}.

\bibitem{PSDR15} J. N. Pandya, N. R. Soni, N. Devlani, and A. K.
Rai, \href{https://doi.org/10.1088/1674-1137/39/12/123101}{Chin. Phys. C \textbf{39}, 123101 (2015)}.

\bibitem{PDKPB16} M. Priyadarsini, P. C. Dash, S. Kar, S. P.
Patra, and N. Barik, \href{https://doi.org/10.1103/PhysRevD.94.113011}{Phys. Rev. D \textbf{94}, 113011 (2016)}.

\bibitem{PDKPB17} S. Patnaik, P. C. Dash, S. Kar, S. P. Patra,
and N. Barik, \href{https://doi.org/10.1103/PhysRevD.96.116010}{Phys. Rev. D \textbf{96}, 116010 (2017)}.

\bibitem{MBV17b} A. P. Monteiro, M. Bhat, and K. B. Vijaya Kumar,
\href{https://doi.org/10.1103/PhysRevD.95.054016}{Phys. Rev. D \textbf{95}, 054016 (2017)}.

\bibitem{BGS17} A. Barducci, R. Giachetti, and E. Sorace, \href{https://doi.org/10.1103/PhysRevD.95.054022}{Phys. Rev. D \textbf{95}, 054022 (2017)}.

\bibitem{DHJ93} Y.-B. Dai, C.-S. Huang, and H.-Y. Jin, \href{https://doi.org/10.1007/BF01560051}{Z. Phys. C \textbf{60}, 527 (1993)}.

\bibitem{MRMP95} C. R. M\"{u}nz, J. Resag, B. C. Metsch, and H. R.
Petry, \href{https://doi.org/10.1103/PhysRevC.52.2110}{Phys. Rev. C \textbf{52}, 2110 (1995)}.

\bibitem{KRMMP00} M. Koll, R. Ricken, D. Merten, B. Metsch, and H.
Petry, \href{https://doi.org/10.1007/PL00013675}{Eur. Phys. J. A \textbf{9}, 73 (2000)}.

\bibitem{IKR99} M. A. Ivanov, Yu. L. Kalinovsky, and C. D.
Roberts, \href{https://doi.org/10.1103/PhysRevD.60.034018}{Phys. Rev. D \textbf{60}, 034018 (1999)}.

\bibitem{ASV05} A. Abd El-Hady, J. R. Spence, and J. P. Vary,
\href{https://doi.org/10.1103/PhysRevD.71.034006}{Phys. Rev. D \textbf{71}, 034006 (2005)}.

\bibitem{KWLC10} H.-W. Ke, G.-L. Wang, X.-Q. Li, and C.-H. Chang,
\href{https://doi.org/10.1007/s11433-010-4151-6}{Sci. China Phys. Mech. Astron. \textbf{53}, 2025 (2010)}.

\bibitem{LNR00}  T.  A. L\"{a}hde, C.  J. Nyf\"{a}lt, and D.  O. Riska, 
\href{https://doi.org/10.1016/S0375-9474(00)00154-8}
{Nucl. Phys. A \textbf{674}, 141 (2000)}.

\bibitem{L03b}  T.  A. L\"{a}hde, 
\href{https://doi.org/10.1016/S0375-9474(02)01362-3}
{Nucl. Phys. A \textbf{714}, 183 (2003)}.

\bibitem{BeA04b} N. Brambilla \textit{et al}., \href{https://arxiv.org/abs/hep-ph/0412158}{arXiv:hep-ph/0412158}.

\bibitem{BJV06} N. Brambilla, Y. Jia, and A. Vairo, \href{https://doi.org/10.1103/PhysRevD.73.054005}{Phys. Rev. D \textbf{73}, 054005 (2006)}.

\bibitem{LPW12} F. J. Llanes-Estrada, O. I. Pavlova, and R. Williams
\href{https://doi.org/10.1140/epjc/s10052-012-2019-9}{Eur. Phys. J. C \textbf{72}, 2019 (2012)}.

\bibitem{P12} A. Pineda, \href{https://doi.org/10.1016/j.ppnp.2012.01.038}{Prog. Part. Nucl. Phys. \textbf{67}, 735 (2012)}.

\bibitem{PS13} A. Pineda and J. Segovia, \href{https://doi.org/10.1103/PhysRevD.87.074024}{Phys. Rev. D \textbf{87}, 074024 (2013)}.

\bibitem{S80d} M. A. Shifman, \href{https://doi.org/10.1007/BF01421576}{Z. Phys. C \textbf{4}, 345 (1980)}; 
Erratum, \href{https://doi.org/10.1007/BF01557779}{Z. Phys. C \textbf{6}, 282 (1980)}.

\bibitem{K80a} A. Yu. Khodjamirian, \href{https://doi.org/10.1016/0370-2693(80)90974-0}{Phys. Lett. B \textbf{90}, 460 (1980)}.

\bibitem{A84} T. M. Aliyev, \href{https://doi.org/10.1007/BF01421766}{Z. Phys. C \textbf{26}, 275 (1984)}.

\bibitem{AIP94b} T. M. Aliev, E. Iltan, and N. K. Pak, \href{https://doi.org/10.1016/0370-2693(94)90606-8}{Phys. Lett. B \textbf{334}, 169 (1994)}.

\bibitem{ADIP96} T. M. Aliev, D. A. Demir, E. Iltan, and N. K.
Pak, \href{https://doi.org/10.1103/PhysRevD.54.857}{Phys. Rev. D \textbf{54}, 857 (1996)}.

\bibitem{EK85} V. L. Eletsky and Ya. I. Kogan, \href{https://doi.org/10.1007/BF01550263}{Z. Phys. C \textbf{28}, 155 (1985)}.

\bibitem{BR85b} V. A. Beilin and A. V. Radyushkin, \href{https://doi.org/10.1016/0550-3213(85)90310-4}{Nucl. Phys. B \textbf{260}, 61 (1985)}.

\bibitem{DN96b} H. G. Dosch and S. Narison, \href{https://doi.org/10.1016/0370-2693(95)01417-9}{Phys. Lett. B \textbf{368}, 163 (1996)}.

\bibitem{ZHY97b} S.-L. Zhu, W-Y. P. Hwang, and Z.-S. Yang, \href{https://doi.org/10.1142/S0217732397003150}{Mod. Phys. Lett. A \textbf{12}, 3027 (1997)}.

\bibitem{V08} M. B. Voloshin, \href{https://doi.org/10.1016/j.ppnp.2008.02.001}{Progr. Part. Nucl. Phys. \textbf{61}, 455 (2008)}.

\bibitem{W13b} Z.-G. Wang, \href{https://doi.org/10.1140/epjc/s10052-013-2559-7}{Eur. Phys. J. C \textbf{73}, 2559 (2013)}.

\bibitem{ADMNS07} V. V. Anisovich, L. G Dakhno, M. A. Matveev,
V. A. Nikonov, and A. V. Sarantsev, \href{https://doi.org/10.1134/S1063778807010097}{Phys. Atom. Nucl. \textbf{70}, 63 (2007)}.

\bibitem{DET09} J. J. Dudek, R. G. Edwards, and C. E. Thomas,
\href{https://doi.org/10.1103/PhysRevD.79.094504}{Phys. Rev. D \textbf{79}, 094504 (2009)}.

\bibitem{CeA11b} Y. Chen \textit{et al}. (CLQCD Collaboration), \href{https://doi.org/10.1103/PhysRevD.84.034503}{Phys. Rev. D \textbf{84}, 034503 (2011)}.

\bibitem{BH11} D. Be\v{c}irevi\'{c} and B. Haas, \href{https://doi.org/10.1140/epjc/s10052-011-1734-y}{Eur. Phys. J. C \textbf{71}, 1734 (2011)}.

\bibitem{BS13c}  D. Be\v{c}irevi\'{c} and F. Sanfilippo, 
\href{https://doi.org/10.1007/JHEP01(2013)028}
{J. High Energy Phys. \textbf{01}, 028 (2013)}.

\bibitem{DeA12a} G. C. Donald \textit{et al}. (HPQCD Collaboration),
\href{https://doi.org/10.1103/PhysRevD.86.094501}{Phys. Rev. D \textbf{86}, 094501 (2012)}.

\bibitem{DDKL14} G. C. Donald, C. T. H. Davies, J. Koponen,
and G. P. Lepage (HPQCD Collaboration), \href{https://doi.org/10.1103/PhysRevLett.112.212002}{Phys. Rev. Lett. \textbf{112}, 212002 (2014)}.

\bibitem{OKLMM15b} B. J. Owen, W. Kamleh, D. B. Leinweber, M. S.
Mahbub, and B. J. Menadue, \href{https://doi.org/10.1103/PhysRevD.92.034513}{Phys. Rev. D \textbf{92}, 034513 (2015)}.

\bibitem{LW16} N. Li and Y.-J. Wu, \href{https://doi.org/10.1142/S0217751X16501098}{Int. J. Mod. Phys. A \textbf{31}, 1650109 (2016)}.

\bibitem{CCLLYY93} H.-Y. Cheng, C.-Y. Cheung, G.-L. Lin, Y. C.
Lin, T.-M. Yan, and H.-L. Yu, \href{https://doi.org/10.1103/PhysRevD.47.1030}{Phys. Rev. D \textbf{47}, 1030 (1993)}.

\bibitem{CCLLYY94a} H.-Y. Cheng, C.-Y. Cheung, G.-L. Lin, Y. C.
Lin, T.-M. Yan, and H.-L. Yu, \href{https://doi.org/10.1103/PhysRevD.49.5857}{Phys. Rev. D \textbf{49}, 5857 (1994)}; Erratum,
\href{https://doi.org/10.1103/PhysRevD.55.5851}{Phys. Rev. D \textbf{55}, 5851 (1997)}.

\bibitem{CDDGFN97} R. Casalbuoni, A. Deandrea, N. Di Bartolomeo,
R. Gatto, F. Feruglio and G. Nardulli, \href{https://doi.org/10.1016/S0370-1573(96)00027-0}{Phys. Rep. \textbf{281}, 145 (1997)}.

\bibitem{DDGNP98} A. Deandrea, N. Di Bartolomeo, R. Gatto, G. Nardulli,
and A. D. Polosa, \href{https://doi.org/10.1103/PhysRevD.58.034004}{Phys. Rev. D \textbf{58}, 034004 (1998)}.

\bibitem{NW00} M. A. Nobes and R. M. Woloshyn, \href{https://doi.org/10.1088/0954-3899/26/7/308}{J. Phys. G \textbf{26}, 1079 (2000)}.

\bibitem{LZ11} G. Li and Q. Zhao, \href{https://doi.org/10.1103/PhysRevD.84.074005}{Phys. Rev. D \textbf{84}, 074005 (2011)}.

\bibitem{CH14} C.-Y. Cheung and C.-W. Hwang, \href{https://doi.org/10.1007/JHEP04(2014)177}{J. High Energy Phys. \textbf{04}, 177 (2014)}.

\bibitem{BBHMR93}  V. Bernard, A.  H. Blin, B. Hiller, U.-G. Mei\ss ner, and M.  C. Ruivo, 
\href{https://doi.org/10.1016/0370-2693(93)91122-4}
{Phys. Lett. B \textbf{305}, 163 (1993)}.

\bibitem{IT95} A. N. Ivanov, and N. I. Troitskaya, \href{https://doi.org/10.1016/0370-2693(94)01589-5}{Phys. Lett. B \textbf{345}, 175 (1995)}.

\bibitem{DCD14} H.-B. Deng, X.-L. Chen, and W.-Z. Deng, \href{https://doi.org/10.1088/1674-1137/38/1/013103}{Chin. Phys. C \textbf{38}, 013103 (2014)}.

\bibitem{FLNC81} J. Franklin, D. B. Lichtenberg, W. Namgung, and
D. Carydas, \href{https://doi.org/10.1103/PhysRevD.24.2910}{Phys. Rev. D \textbf{24}, 2910 (1981)}.

\bibitem{DL96} R. Delbourgo and D. Liu, \href{https://doi.org/10.1103/PhysRevD.53.6576}{Phys. Rev. D \textbf{53}, 6576 (1996)}.

\bibitem{AHNV07b}  C. Albertus, E. Hern\'{a}ndez, J. Nieves, and J.  M. Verde-Velasco, 
\href{https://doi.org/10.1140/epja/i2007-10364-y}
{Eur. Phys. J. A \textbf{32}, 183 (2007)}; Erratum,
\href{https://doi.org/10.1140/epja/i2008-10547-0}{Eur. Phys. J. A \textbf{36}, 119 (2008)}.

\bibitem{KDV05} S. Kumar, R. Dhir, and R. C. Verma, \href{https://doi.org/10.1088/0954-3899/31/2/006}{J. Phys. G \textbf{31}, 141 (2005)}.

\bibitem{DV09} R. Dhir and R. C. Verma, \href{https://doi.org/10.1140/epja/i2009-10872-8}{Eur. Phys. J. A \textbf{42}, 243 (2009)}.

\bibitem{DKV13} R. Dhir, C. S. Kim, and R. C. Verma, \href{https://doi.org/10.1103/PhysRevD.88.094002}{Phys. Rev. D \textbf{88}, 094002 (2013)}.

\bibitem{MPV08} A. Majethiya, B. Patel, and P. C. Vinodkumar, \href{https://doi.org/10.1140/epja/i2008-10674-6}{Eur. Phys. J. A \textbf{38}, 307 (2008)}.

\bibitem{PRV08a} B. Patel, A. K. Rai, and P. C. Vinodkumar, \href{https://doi.org/10.1088/0954-3899/35/6/065001}{J. Phys. G \textbf{35}, 065001 (2008)}.

\bibitem{PMV09} B. Patel, A. Majethiya, and P. C. Vinodkumar, \href{https://doi.org/10.1007/s12043-009-0061-4}{Pramana \textbf{72}, 679 (2009)}.

\bibitem{GRQR14} Z. Ghalenovi, A. A. Rajabi, S.-x. Qin, and D. H.
Rischke, \href{https://doi.org/10.1142/S0217732314501065}{Mod. Phys. Lett. A \textbf{29}, 1450106 (2014)}.

\bibitem{STRV16b} Z. Shah, K. Thakkar, A. K. Rai, and P. C. Vinodkumar,
\href{https://doi.org/10.1088/1674-1137/40/12/123102}{Chin. Phys. C \textbf{40}, 123102 (2016)}.

\bibitem{STR16} Z. Shah, K. Thakkar, and A. K. Rai, \href{https://doi.org/10.1140/epjc/s10052-016-4379-z}{Eur. Phys. J. C \textbf{76}, 530 (2016)}.

\bibitem{SR17a} Z. Shah and A. K. Rai, \href{https://doi.org/10.1140/epjc/s10052-017-4688-x}{Eur. Phys. J. C \textbf{77}, 129 (2017)}.

\bibitem{MTV16} A. Majethiya, K. Thakkar, and P. C. Vinodkumar,
\href{https://doi.org/10.1016/j.cjph.2016.06.008}{Chin. J Phys. \textbf{54}, 495 (2016)}.

\bibitem{TMV16} K. Thakkar, A. Majethiya, and P. C. Vinodkumar,
\href{https://doi.org/10.1140/epjp/i2016-16339-4}{Eur. Phys. J. Plus \textbf{131}, 339 (2016)}.

\bibitem{JR04}  B. Juli\'{a}-D\'{\i}az and D.  O. Riska, 
\href{https://doi.org/10.1016/j.nuclphysa.2004.03.078}
{Nucl. Phys. A \textbf{739}, 69 (2004)}.

\bibitem{FGIKLNP06}  
A. Faessler, T. Gutsche, M.  A. Ivanov, J.  G. K\"{o}rner, V.  E. Lyubovitskij, D. Nicmorus, and K. Pumsa-ard, 
\href{https://doi.org/10.1103/PhysRevD.73.094013}
{Phys. Rev. D \textbf{73}, 094013 (2006)}.

\bibitem{GSP16} A. N. Gadaria, N. R. Soni, and J. N. Pandya,
\href{http://inspirehep.net/record/1603524?ln=en}
{DAE Symp. Nucl. Phys. \textbf{61}, 698 (2016)}.

\bibitem{BD83a} N. Barik and M. Das, \href{https://doi.org/10.1103/PhysRevD.28.2823}{Phys. Rev. D \textbf{28}, 2823 (1983)}.

\bibitem{JR86} S. N. Jena and D. P. Rath, \href{https://doi.org/10.1103/PhysRevD.34.196}{Phys. Rev. D \textbf{34}, 196 (1986)}.

\bibitem{JB95} S. N. Jena and M. R. Behera, \href{https://doi.org/10.1007/BF02847612}{Pramana \textbf{44} 357 (1995)}.  

\bibitem{SDCG10} N. Sharma, H. Dahiya, P. K. Chatley, and M. Gupta,
\href{https://doi.org/10.1103/PhysRevD.81.073001}{Phys. Rev. D \textbf{81}, 073001 (2010)}.

\bibitem{OMRS91} Y. Oh, D.-P. Min, M. Rho, and N. N. Scoccola,
\href{https://doi.org/10.1016/0375-9474(91)90458-I}{Nucl. Phys. A \textbf{534}, 493 (1991)}.

\bibitem{SW04} S. Scholl and H. Weigel, \href{https://doi.org/10.1016/j.nuclphysa.2004.01.132}{Nucl. Phys. A \textbf{735}, 163 (2004)}.

\bibitem{S94a} M. J. Savage, \href{https://doi.org/10.1016/0370-2693(94)91326-9}{Phys. Lett. B \textbf{326}, 303 (1994)}.

\bibitem{LMLZ17a} H.-S. Li, L. Meng, Z.-W. Liu, and S.-L. Zhu, \href{https://doi.org/10.1103/PhysRevD.96.076011}{Phys. Rev. D \textbf{96}, 076011 (2017)}.

\bibitem{MLLZ17} L. Meng, H.-S. Li, Z.-W. Liu, and S.-L. Zhu, \href{https://doi.org/10.1140/epjc/s10052-017-5447-8}{Eur. Phys. J. C \textbf{77}, 869 (2017)}.

\bibitem{ZHY97a} S.-L. Zhu, W-Y. P. Hwang, and Z.-S. Yang, \href{https://doi.org/10.1103/PhysRevD.56.7273}{Phys. Rev. D \textbf{56}, 7273 (1997)}.

\bibitem{AOS02b} T. M. Aliev, A. \"{O}zpineci, and M. Savc\i , \href{https://doi.org/10.1103/PhysRevD.65.056008}{Phys. Rev. D \textbf{65}, 056008 (2002)}.

\bibitem{AAO08} T. M. Aliev, K. Azizi, and A. Ozpineci, \href{https://doi.org/10.1103/PhysRevD.77.114006}{Phys. Rev. D \textbf{77}, 114006 (2008)}.

\bibitem{AAO09a} T. M. Aliev, K. Azizi, and A. Ozpineci, \href{https://doi.org/10.1016/j.nuclphysb.2008.09.018}{Nucl. Phys. B \textbf{808}, 137 (2009)}.

\bibitem{ABS15a} T. M. Aliev, T. Barakat, and M. Savc\i , \href{https://doi.org/10.1103/PhysRevD.91.116008}{Phys. Rev. D \textbf{91}, 116008 (2015)}.

\bibitem{CEIOT13} K. U. Can, G. Erkol, B. Isildak, M. Oka, and T. T.
Takahashi, \href{https://doi.org/10.1016/j.physletb.2013.09.024}{Phys. Lett. B \textbf{726}, 703 (2013)}.

\bibitem{CEIOT14b} K. U. Can, G. Erkol, B. Isildak, M. Oka, and T. T.
Takahashi, \href{https://doi.org/10.1007/JHEP05(2014)125}{J. High Energy Phys. \textbf{05}, 125 (2014)}.

\bibitem{CEOT15} K. U. Can, G. Erkol, M. Oka, and T. T. Takahashi,
\href{https://doi.org/10.1103/PhysRevD.92.114515}{Phys. Rev. D \textbf{92}, 114515 (2015)}.

\bibitem{BS13a} A. Bernotas and V. \v{S}imonis, \href{https://doi.org/10.3952/lithjphys.53202}{Lith. J. Phys. \textbf{53}, 84 (2013)}.

\bibitem{C97e} H.-Y. Cheng, \href{https://doi.org/10.1016/S0370-2693(97)00305-5}{Phys. Lett. B \textbf{399}, 281 (1997)}.

\bibitem{HM06b} J. Hu and T. Mehen, \href{https://doi.org/10.1103/PhysRevD.73.054003}{Phys. Rev. D \textbf{73}, 054003 (2006)}.

\bibitem{BPS00} M.  C. Ba\~{n}uls, A. Pich, and I. Scimemi, 
\href{https://doi.org/10.1103/PhysRevD.61.094009}
{Phys. Rev. D \textbf{61}, 094009 (2000)}.

\bibitem{JCZ15} N. Jiang, X.-L. Chen, and S.-L. Zhu, \href{https://doi.org/10.1103/PhysRevD.92.054017}{Phys. Rev. D \textbf{92}, 054017 (2015)}.

\bibitem{LMLZ18} H.-S. Li, L. Meng, Z.-W. Liu, and S.-L. Zhu, \href{https://doi.org/10.1016/j.physletb.2017.12.031}{Phys. Lett. B \textbf{777}, 169 (2018)}.

\bibitem{TKO01}  S. Tawfiq, J.  G. K\"{o}rner, and P.  J. O'Donnell, 
\href{https://doi.org/10.1103/PhysRevD.63.034005}
{Phys. Rev. D \textbf{63}, 034005 (2001)}.

\bibitem{XWLZZ17} L.-Y. Xiao, K.-L. Wang, Q.-F L\"{u}, X.-H. Zhong, and S.-L. Zhu,  
\href{https://doi.org/10.1103/PhysRevD.96.094005}
{Phys. Rev. D \textbf{96}, 094005 (2017)}.

\bibitem{WYZZ17} K.-L. Wang, Y.-X. Yao, X.-H. Zhong, and Q. Zhao,
\href{https://doi.org/10.1103/PhysRevD.96.116016}{Phys. Rev. D \textbf{96}, 116016 (2017)}.

\bibitem{DDSV94} J. Dey, M. Dey, V. Shevchenko, and P. Volkovitsky,
\href{https://doi.org/10.1016/0370-2693(94)91466-4}{Phys. Lett. B \textbf{337}, 185 (1994)}.

\bibitem{MPV09} A. Majethiya, B. Patel, and P. C. Vinodkumar, \href{https://doi.org/10.1140/epja/i2009-10880-8}{Eur. Phys. J. A \textbf{42}, 213 (2009)}.

\bibitem{AHN10b}  C. Albertus, E. Hern\'{a}ndez, and J. Nieves, 
\href{https://doi.org/10.1016/j.physletb.2010.05.042}
{Phys. Lett. B \textbf{690}, 265 (2010)}.

\bibitem{IKLR99b}  
M.  A. Ivanov, J.  G. K\"{o}rner, V.  E. Lyubovitskij, and A.  G. Rusetsky, 
\href{https://doi.org/10.1103/PhysRevD.60.094002}
{Phys. Rev. D \textbf{60}, 094002 (1999)}.

\bibitem{BFGIKLO10}  T. Branz, A. Faessler, T. Gutsche, M.  A. Ivanov, J.  G. K\"{o}rner, V.  E. Lyubovitskij, and B. Oexl, 
\href{https://doi.org/10.1103/PhysRevD.81.114036}
{Phys. Rev. D \textbf{81}, 114036 (2010)}.

\bibitem{LWXZ17} Q.-F. L\"u, K.-L. Wang, L.-Y. Xiao, and X.-H. Zhong, 
\href{https://doi.org/10.1103/PhysRevD.96.114006}
{Phys. Rev. D \textbf{96}, 114006 (2017)}.

\bibitem{AOS02a}  T.  M. Aliev, A. \"{O}zpineci, and M. Savc\i , 
\href{https://doi.org/10.1103/PhysRevD.65.096004}
{Phys. Rev. D \textbf{65}, 096004 (2002)}.

\bibitem{AAO09b} T. M. Aliev, K. Azizi, and A. Ozpineci, \href{https://doi.org/10.1103/PhysRevD.79.056005}{Phys. Rev. D \textbf{79}, 056005 (2009)}.

\bibitem{AAS14b} T. M. Aliev, K. Azizi, and M. Savc\i , \href{https://doi.org/10.1103/PhysRevD.89.053005}{Phys. Rev. D \textbf{89}, 053005 (2014)}.

\bibitem{AAS15b} T. M. Aliev, K. Azizi, and H. Sundu, \href{https://doi.org/10.1140/epjc/s10052-014-3229-0}{Eur. Phys. J. C \textbf{75}, 14 (2015)}.

\bibitem{ABS16} T. M. Aliev, T. Barakat, and M. Savc\i , \href{https://doi.org/10.1103/PhysRevD.93.056007}{Phys. Rev. D \textbf{93}, 056007 (2016)}.

\bibitem{ASZ12} T. M. Aliev, M. Savc\i , and V. S. Zamiralov,
\href{https://doi.org/10.1142/S021773231250054X}{Mod. Phys. Lett. A \textbf{27}, 1250054 (2012)}.

\bibitem{ZD99} S.-L. Zhu and Y.-B. Dai, \href{https://doi.org/10.1103/PhysRevD.59.114015}{Phys. Rev. D \textbf{59} 114015 (1999)}.

\bibitem{W10a} Z.-G. Wang, \href{https://doi.org/10.1103/PhysRevD.81.036002}{Phys. Rev. D \textbf{81}, 036002 (2010)}.

\bibitem{W10b} Z.-G. Wang, \href{https://doi.org/10.1140/epja/i2010-11004-3}{Eur. Phys. J. A \textbf{44}, 105 (2010)}.

\bibitem{BCEO15} H. Bahtiyar, K. U. Can, G. Erkol, and M. Oka,
\href{https://doi.org/10.1016/j.physletb.2015.06.006}{Phys. Lett. B \textbf{747}, 281 (2015)}.

\bibitem{BCEOT17} H. Bahtiyar, K. U. Can, G. Erkol, M. Oka, and
T. T. Takahashi, \href{https://doi.org/10.1016/j.physletb.2017.06.022}{Phys. Lett. B \textbf{772}, 121 (2017)}.

\bibitem{CCCLZ18} E.-L. Cui, H.-X. Chen, W. Chen, X. Liu, and S.-L. Zhu, 
\href{https://doi.org/10.1103/PhysRevD.97.034018}
{Phys. Rev. D \textbf{97}, 034018 (2018)}.

\bibitem{BS13b} A. Bernotas and V. \v{S}imonis, \href{https://doi.org/10.1103/PhysRevD.87.074016}{Phys. Rev. D \textbf{87}, 074016 (2013)}.

\bibitem{BS04b} A. Bernotas and V. \v{S}imonis, \href{https://doi.org/10.1016/j.nuclphysa.2004.05.017}{Nucl. Phys. A \textbf{741}, 179 (2004)}.

\bibitem{CJJTW74} A. Chodos, R. L. Jaffe, K. Johnson, C. B. Thorn,
and V. F. Weisskopf, \href{https://doi.org/10.1103/PhysRevD.9.3471}{Phys. Rev. D \textbf{9}, 3471 (1974)}.

\bibitem{CJJT74} A. Chodos, R. L. Jaffe, K. Johnson, and C. B.
Thorn, \href{https://doi.org/10.1103/PhysRevD.10.2599}{Phys. Rev. D \textbf{10}, 2599 (1974)}.

\bibitem{DJJK75} T. DeGrand, R. L. Jaffe, K. Johnson, and J. Kiskis,
\href{https://doi.org/10.1103/PhysRevD.12.2060}{Phys. Rev. D \textbf{12}, 2060 (1975)}.

\bibitem{T84b} A. W. Thomas, \href{https://doi.org/10.1007/978-1-4613-9892-9_1}{Adv. Nucl. Phys. \textbf{13}, 1 (1984)}.

\bibitem{DJ80} J. F. Donoghue and K. Johnson, \href{https://doi.org/10.1103/PhysRevD.21.1975}{Phys. Rev. D \textbf{21}, 1975 (1980)}.

\bibitem{RP08} W. Roberts and M. Pervin, \href{https://doi.org/10.1142/S0217751X08041219}{Int. J. Mod. Phys. A \textbf{23}, 2817 (2008)}.

\bibitem{AOZ11} T. M. Aliev, A. Ozpineci, and V. Zamiralov, \href{https://doi.org/10.1103/PhysRevD.83.016008}{Phys. Rev. D \textbf{83}, 016008 (2011)}.

\bibitem{AAS12b} T. M. Aliev, K. Azizi, and M. Savc\i , \href{https://doi.org/10.1016/j.physletb.2012.07.033}{Phys. Lett. B \textbf{715}, 149 (2012)}.

\bibitem{BS08a} A. Bernotas and V. \v{S}imonis, \href{https://doi.org/10.3952/lithjphys.48202}{Lith. J. Phys. \textbf{48}, 127 (2008)}.

\bibitem{AeA00a} A. Ali Khan \textit{et al}., \href{https://doi.org/10.1103/PhysRevD.62.054505}{Phys. Rev. D \textbf{62}, 054505 (2000)}.

\bibitem{MLW02a} N. Mathur, R. Lewis, and R. M. Woloshyn, \href{https://doi.org/10.1103/PhysRevD.66.014502}{Phys. Rev. D \textbf{66}, 014502 (2002)}.

\bibitem{LW09} R. Lewis and R. M. Woloshyn, \href{https://doi.org/10.1103/PhysRevD.79.014502}{Phys. Rev. D \textbf{79}, 014502 (2009)}.

\bibitem{FMT03} J. M. Flynn, F. Mescia, and A. S. B. Tariq,
\href{https://doi.org/10.1088/1126-6708/2003/07/066}{J. High Energy Phys. \textbf{07}, 066 (2003)}.

\bibitem{ACCDGP12} C. Alexandrou, J. Carbonell, D. Christaras, V.
Drach, M. Gravina, and M. Papinutto, \href{https://doi.org/10.1103/PhysRevD.86.114501}{Phys. Rev. D \textbf{86}, 114501 (2012)}.

\bibitem{ADJKK14} C. Alexandrou, V. Drach, K. Jansen, C. Kallidonis,
and G. Koutsou, \href{https://doi.org/10.1103/PhysRevD.90.074501}{Phys. Rev. D \textbf{90}, 074501 (2014)}.

\bibitem{B15b} T. Burch, \href{https://arxiv.org/abs/1502.00675}{arXiv:1502.00675}.

\bibitem{PCB15} P. P\'{e}rez-Rubio, S. Collins, and G.  S. Bali, 
\href{https://doi.org/10.1103/PhysRevD.92.034504}
{Phys. Rev. D \textbf{92}, 034504 (2015)}.

\bibitem{ZH08b} J.-R. Zhang and M.-Q. Huang, \href{https://doi.org/10.1103/PhysRevD.78.094007}{Phys. Rev. D \textbf{78}, 094007 (2008)}.

\bibitem{AN10b} R. M. Albuquerque and S. Narison, \href{https://doi.org/10.1016/j.physletb.2010.09.051}{Phys. Lett. B \textbf{694}, 217 (2010)}.

\bibitem{W10d} Z.-G. Wang, \href{https://doi.org/10.1140/epja/i2010-11004-3}{Eur. Phys. J. A \textbf{45}, 267 (2010)}.

\bibitem{W10e} Z.-G. Wang, \href{https://doi.org/10.1140/epjc/s10052-010-1357-8}{Eur. Phys. J. C \textbf{68}, 459 (2010)}.

\bibitem{AAS12a} T. M. Aliev, K. Azizi, and M. Savc\i , \href{https://doi.org/10.1016/j.nuclphysa.2012.09.009}{Nucl. Phys. A \textbf{895}, 59 (2012)}.

\bibitem{AAS13b} T. M. Aliev, K. Azizi, and M. Savc\i , \href{https://doi.org/10.1088/0954-3899/40/6/065003}{J. Phys. G \textbf{40}, 065003 (2013)}.

\bibitem{RLP95} R. Roncaglia, D. B. Lichtenberg, and E. Predazzi,
\href{https://doi.org/10.1103/PhysRevD.52.1722}{Phys. Rev. D \textbf{52}, 1722 (1995)}.

\bibitem{LRP96} D. B. Lichtenberg, R. Roncaglia, and E. Predazzi,
\href{https://doi.org/10.1103/PhysRevD.53.6678}{Phys. Rev. D \textbf{53}, 6678 (1996)}.

\bibitem{BVR05}  N. Brambilla, A. Vairo, and T. R\"{o}sch, 
\href{https://doi.org/10.1103/PhysRevD.72.034021}
{Phys. Rev. D \textbf{72}, 034021 (2005)}.

\bibitem{ER12} B. Eakins and W. Roberts, \href{https://doi.org/10.1142/S0217751X1250039X}{Int. J. Mod. Phys. A \textbf{27}, 1250039 (2012)}.

\bibitem{AHN10a} C. Albertus, E. Hern\'{a}ndez, and J. Nieves, \href{https://doi.org/10.1016/j.physletb.2009.11.048}{Phys. Lett. B \textbf{683}, 21 (2010)}.

\bibitem{IMMW00} C. Itoh, T. Minamikawa, K. Miura, and T. Watanabe,
\href{https://doi.org/10.1103/PhysRevD.61.057502}{Phys. Rev. D \textbf{61}, 057502 (2000)}.

\bibitem{KLPS02} V. V. Kiselev, A. K. Likhoded, O. N. Pakhomova,
and V. A. Saleev, \href{https://doi.org/10.1103/PhysRevD.66.034030}{Phys. Rev. D \textbf{66}, 034030 (2002)}.

\bibitem{L09} A. K. Likhoded, \href{https://doi.org/10.1134/S1063778809030181}{Phys. Atom. Nuclei \textbf{72}, 529 (2009)}.

\bibitem{VGV08} A. Valcarce, H. Garcilazo, and J. Vijande, \href{https://doi.org/10.1140/epja/i2008-10616-4}{Eur. Phys. J. A \textbf{37}, 217 (2008)}.

\bibitem{YHHOS15} T. Yoshida, E. Hiyama, A. Hosaka, M. Oka, and K.
Sadato, \href{https://doi.org/10.1103/PhysRevD.92.114029}{Phys. Rev. D \textbf{92}, 114029 (2015)}.

\bibitem{CDR98} F. Coester, K. Dannbom, and D. O. Riska, \href{https://doi.org/10.1016/S0375-9474(98)00142-0}{Nucl. Phys. A \textbf{634}, 335 (1998)}.

\bibitem{MMMP06} S. Migura, D. Merten, B. Metsch and H.-R. Petry,
\href{https://doi.org/10.1140/epja/i2006-10017-9}{Eur. Phys. J. A \textbf{28}, 41 (2006)}.

\bibitem{M08b} A. P. Martynenko, \href{https://doi.org/10.1016/j.physletb.2008.04.030}{Phys. Lett. B \textbf{663}, 317 (2008)}.

\bibitem{G09} F. Giannuzzi, \href{https://doi.org/10.1103/PhysRevD.79.094002}{Phys. Rev. D \textbf{79}, 094002 (2009)}.

\bibitem{AAS13a} T. M. Aliev, K. Azizi, and M. Savc\i , \href{https://doi.org/10.1007/JHEP04(2013)042}{J. High Energy Phys. \textbf{04}, 42 (2013)}.

\bibitem{AAS14a} T. M. Aliev, K. Azizi, and M. Savc\i , \href{https://doi.org/10.1088/0954-3899/41/6/065003}{J. Phys. G \textbf{41}, 065003 (2014)}.

\bibitem{FHN12} J.  M. Flynn, E. Hern\'{a}ndez, and J. Nieves,
\href{https://doi.org/10.1103/PhysRevD.85.014012}
{Phys. Rev. D \textbf{85}, 014012 (2012)}.

\bibitem{AAHN04}  C. Albertus, J.  E. Amaro, E. Hern\'{a}ndez, and J. Nieves,
\href{https://doi.org/10.1016/j.nuclphysa.2004.04.114}
{Nucl. Phys. A \textbf{740}, 333 (2004)}.

\bibitem{PDG17}  C. Patrignani \textit{et al}. (Particle Data Group), 
\href{https://doi.org/10.1088/1674-1137/40/10/100001}
{Chin. Phys. C \textbf{40}, 100001 (2016)}; and
\href{http://pdg.lbl.gov/2017/listings/contents_listings.html}
{2017 update}.

\bibitem{PEMP15} M. Padmanath, R. G. Edwards, N. Mathur, and M.
Peardon (Hadron Spectrum Collaboration), \href{https://doi.org/10.1103/PhysRevD.91.094502}{Phys. Rev. D \textbf{91}, 094502 (2015)}.

\bibitem{LT88} G. P. Lepage and B. A. Thacker, 
\href{https://doi.org/10.1016/0920-5632(88)90102-8}{Nucl. Phys. B (Proc. Suppl.) \textbf{4}, 199 (1988)}.

\bibitem{IW91} N. Isgur and M. B. Wise, \href{https://doi.org/10.1103/PhysRevLett.66.1130}{Phys. Rev. Lett. \textbf{66}, 1130 (1991)}.

\bibitem{FI92} J. M. Flynn and N. Isgur, \href{https://doi.org/10.1088/0954-3899/18/10/004}{J. Phys. G \textbf{18}, 1627 (1992)}.

\bibitem{EQ17} E. J. Eichten and C. Quigg, \href{https://doi.org/10.1103/PhysRevLett.119.202002}{Phys. Rev. Lett. \textbf{119}, 202002 (2017)}.

\bibitem{M97a} T. Mannel, \href{https://doi.org/10.1088/0034-4885/60/10/003}{Rep. Prog. Phys. \textbf{60}, 1113 (1997)}.

\bibitem{MW00} A. V. Manohar and M. B. Wise, \textit{Heavy Quark
Physics} (Cambridge University Press, Cambridge, 2000).

\bibitem{J08} E. E. Jenkins, \href{https://doi.org/10.1103/PhysRevD.77.034012}{Phys. Rev. D \textbf{77}, 034012 (2008)}.

\bibitem{WQGW17} Z.-Y. Wang, J.-J. Qi, X.-H. Guo, and K.-W. Wei,
\href{https://doi.org/10.1088/1674-1137/41/9/093103}{Chin. Phys. C \textbf{41}, 093103 (2017)}.

\bibitem{EFG11b} D. Ebert, R. N. Faustov, and V. O. Galkin, \href{https://doi.org/10.1103/PhysRevD.84.014025}{Phys. Rev. D \textbf{84},  014025 (2011)}.

\bibitem{RY94} B. Rosenstein and H.-L. Yu, \href{https://doi.org/10.1103/PhysRevD.49.4949}{Phys. Rev. D \textbf{49}, 4949 (1994)}.

\bibitem{ACOV04} T. J. Allen, T. Coleman, M. G. Olsson, and S.
Veseli, \href{https://doi.org/10.1103/PhysRevD.69.074010}{Phys. Rev. D \textbf{69}, 074010 (2004)}.

\bibitem{MDFHL12} C. McNeile, C. T. H. Davies, E. Follana, K.
Hornbostel, and G. P. Lepage (HPQCD Collaboration), \href{https://doi.org/10.1103/PhysRevD.86.074503}{Phys. Rev. D \textbf{86}, 074503 (2012)}.

\bibitem{PPSS04a} A. A. Penin, A. Pineda, V. A. Smirnov, and M. Steinhauser, 
\href{https://doi.org/10.1016/j.physletb.2004.04.066}{Phys. Lett. B \textbf{593}, 124 (2004)}; 
Erratum, \href{https://doi.org/10.1016/j.physletb.2009.05.036}{Phys. Lett. B \textbf{677}, 343 (2009)};
Erratum, \href{https://doi.org/10.1016/j.physletb.2009.12.035}{Phys. Lett. B \textbf{683}, 358 (2010)}.


\bibitem{W13a} Z.-G. Wang, \href{https://doi.org/10.1140/epja/i2013-13131-7}{Eur. Phys. J. A \textbf{49}, 131 (2013)}.

\bibitem{GeA10} E. B. Gregory \textit{et al}., \href{https://doi.org/10.1103/PhysRevLett.104.022001}{Phys. Rev. Lett. \textbf{104}, 022001 (2010)}.

\bibitem{DDHH12} R. J. Dowdall, C. T. H. Davies, T. C. Hammant,
and R. R. Horgan (HPQCD Collaboration), \href{https://doi.org/10.1103/PhysRevD.86.094510}{Phys. Rev. D \textbf{86}, 094510 (2012)}.

\bibitem{WLW15} M. Wurtz, R. Lewis, and R. M. Woloshyn, \href{https://doi.org/10.1103/PhysRevD.92.054504}{Phys. Rev. D \textbf{92}, 054504 (2015)}.

\bibitem{RDLP95} R. Roncaglia, A. Dzierba, D. B. Lichtenberg, and
E. Predazzi, \href{https://doi.org/10.1103/PhysRevD.51.1248}{Phys. Rev. D \textbf{51}, 1248 (1995)}.

\bibitem{CO96} Y.-Q. Chen and R. J. Oakes, \href{https://doi.org/10.1103/PhysRevD.53.5051}{Phys. Rev. D \textbf{53}, 5051 (1996)}.

\bibitem{CIKM97} S. J. Collins, T. D. Imbo, B. A. King, and
E. C. Martell, \href{https://doi.org/10.1016/S0370-2693(96)01514-6}{Phys. Lett. B \textbf{393}, 155 (1997)}.

\bibitem{MZ98a} L. Motyka and K. Zalewski, \href{https://doi.org/10.1007/s100529800743}
{Eur. Phys. J. C \textbf{4}, 107 (1998)}.

\bibitem{PV09} B. Patel and P. C. Vinodkumar, \href{https://doi.org/10.1088/0954-3899/36/3/035003}{J. Phys. G \textbf{36}, 035003 (2009)}.

\bibitem{Cv88} H. W. Crater and P. Van Alstine, \href{https://doi.org/10.1103/PhysRevD.37.1982}{Phys. Rev. D \textbf{37}, 1982 (1988)}.

\bibitem{IMMW92} C. Itoh, T. Minamikawa, K. Miura, and T. Watanabe,
\href{https://doi.org/10.1007/BF02731983}{Nuovo Cim. A \textbf{105}, 1539 (1992)}.

\bibitem{ZVR95} J. Zeng, J. W. Van Orden, and W. Roberts, \href{https://doi.org/10.1103/PhysRevD.52.5229}{Phys. Rev. D \textbf{52}, 5229 (1995)}.

\bibitem{GJ96} S. N. Gupta and J. M. Johnson, \href{https://doi.org/10.1103/PhysRevD.53.312}{Phys. Rev. D \textbf{53}, 312 (1996)}.

\bibitem{BBD10a} A. M. Badalian, B. L. G. Bakker, and I. V.
Danilkin, \href{https://doi.org/10.1103/PhysRevD.81.071502}{Phys. Rev. D \textbf{81}, 071502(R) (2010)}; Erratum,
\href{https://doi.org/10.1103/PhysRevD.81.099902}{Phys. Rev. D \textbf{81}, 099902 (2010)}.

\bibitem{EFG11a} D. Ebert, R. N. Faustov, and V. O. Galkin, \href{https://doi.org/10.1140/epjc/s10052-011-1825-9}{Eur. Phys. J. C \textbf{71}, 1825 (2011)}.

\bibitem{ALV99} A. A. El-Hady, M. A. K. Lodhi, and J. P. Vary,
\href{https://doi.org/10.1103/PhysRevD.59.094001}{Phys. Rev. D \textbf{59}, 094001 (1999)}.

\bibitem{BP00} M. Baldicchi and G. M. Prosperi, \href{https://doi.org/10.1103/PhysRevD.62.114024}{Phys. Rev. D \textbf{62}, 114024 (2000)}.

\bibitem{IS05b} S. M. Ikhdair and R. Sever, \href{https://doi.org/10.1142/S0217751X05021294}{Int. J. Mod. Phys. A \textbf{20}, 6509 (2005)}.

\bibitem{GHK16} M. G\'{o}mez-Rocha, T. Hilger, and A. Krassnigg, 
\href{https://doi.org/10.1103/PhysRevD.93.074010}
{Phys. Rev. D \textbf{93}, 074010 (2016)}.

\bibitem{MNSMMT93} S. N. Mukherjee, R. Nag, S. Sanyal, T. Morii,
J. Morishita, and M. Tsuge, \href{https://doi.org/10.1016/0370-1573(93)90010-B}{Phys. Rept. \textbf{231}, 201 (1993)}.

\bibitem{KL08} M. Karliner and H. J. Lipkin, \href{https://doi.org/10.1016/j.physletb.2008.01.023}{Phys. Lett. B \textbf{660}, 539 (2008)}.

\bibitem{B16} V. G. Baryshevsky, \href{https://doi.org/10.1016/j.physletb.2016.04.025}{Phys. Lett. B \textbf{757}, 426 (2016)}.

\bibitem{FeA17} A. S. Fomin \textit{et al}., \href{https://doi.org/10.1007/JHEP08(2017)120}{J. High Energy Phys. \textbf{08}, 120 (2017)}.

\bibitem{PSTWCDO17} A. Parreno, M. J. Savage, B. C. Tiburzi,
J. Wilhelm, E. Chang, W. Detmold, and K. Orginos, \href{https://arxiv.org/abs/1709.01564}{arXiv:1709.01564}.

\end{thebibliography}
\end{document}